\documentclass[12pt,a4paper,oneside]{article}

\usepackage{graphicx,amssymb,amsmath}
\usepackage{setspace} 
\usepackage[linkcolor={blue},citecolor={blue},colorlinks=true,linktocpage,urlcolor={blue}]
{hyperref}
\usepackage{cite}
\usepackage[margin=2cm]{geometry}
\usepackage[dvipsnames]{xcolor}
\definecolor{darkblue}{rgb}{0,0,0.54}
\usepackage{tikz}
\usetikzlibrary{patterns}
\usetikzlibrary{calc}
\usepackage{multirow} 
\usepackage{appendix}
\usepackage{bbm}
\usepackage{import}
\usepackage{tensor}
\usepackage{pdfpages}
\usepackage{cancel} 
\usepackage{slashed} 
\usepackage{xparse} 
\usepackage{tabu} 
\usepackage{adjustbox} 
\usepackage{simpler-wick}
\usepackage[bb=boondox]{mathalfa}
\usepackage{dsfont}
\usepackage[bottom]{footmisc}

\numberwithin{equation}{section} 
\usepackage{empheq} 

\graphicspath{{Figures/}} 
\usepackage{caption}
\usepackage{subcaption}

\newcommand{\cE}{\mathcal{E}}

\newcommand{\Tr}{\text{Tr}}

\DeclareMathOperator{\tr}{tr}

\renewcommand{\Im}{\operatorname{Im}}

\newcommand{\mean}{\overline}
\newcommand{\ns}{\rho_N}
\newcommand{\es}{\rho_E}
\newcommand{\Smicro}{\textbf{S}}
\DeclareMathOperator{\ext}{ext}
\usepackage{titling} 
\usepackage[affil-it]{authblk} 

\usepackage{tocloft} 
\title{Negativity Spectra in Random Tensor Networks and Holography}
\author[a]{Jonah Kudler-Flam,}
\author[b]{Vladimir Narovlansky,}
\author[b,c]{Shinsei Ryu}

\affil[a]{Kadanoff Center for Theoretical Physics, University of Chicago, Chicago, IL~60637, USA}
\affil[b]{Princeton Center for Theoretical Science,
Princeton University, Princeton, NJ 08544, USA}
\affil[c]{Department of Physics, Princeton University, Princeton, NJ 08544, USA}
\date{\small \texttt{jkudlerflam@uchicago.edu, narovlansky@princeton.edu, shinseir@princeton.edu}}

\makeatletter 
\let\@fnsymbol\@arabic
\makeatother

\begin{document}

\begin{titlingpage}
    \maketitle
    \begin{abstract}
                 Negativity is a measure of entanglement that can be used both in pure and mixed states. The negativity spectrum is the spectrum of eigenvalues of the partially transposed density matrix, and characterizes the degree and ``phase'' of entanglement. For pure states, it is simply determined by the entanglement spectrum. We use a diagrammatic method complemented by a modification of the Ford-Fulkerson algorithm to find
                 the negativity spectrum in general random tensor networks with large bond dimensions. In holography, these describe the entanglement of fixed-area states. It was found that many fixed-area states have a negativity spectrum given by a semi-circle. More generally, we find new negativity spectra that appear in random tensor networks, as well as in phase transitions in holographic states, wormholes, and holographic states with bulk matter. The smallest random tensor network is the same as a micro-canonical version of Jackiw-Teitelboim (JT) gravity decorated with end-of-the-world branes. We consider the semi-classical negativity of Hawking radiation and find that contributions from islands should be included. We verify this in the JT gravity model, showing the Euclidean wormhole origin of these contributions.
    \end{abstract}
\end{titlingpage}

\tableofcontents
\pagebreak

\section{Introduction} \label{sec:intro}

Quantum entanglement has proven to be a key concept in modern quantum many-body physics. 
For example, the (von Neumann) entanglement entropy 
of ground states
can be used to distinguish 
different quantum phases and phase transitions
\cite{ Fradkin:2013sab, zeng2018quantum}.
It is also an important diagnostic for
problems in quantum dynamics, 
such as thermalization 
in isolated systems and
quantum information scrambling
\cite{Calabrese_2016, Nandkishore_2015}.
Furthermore, 
quantum entanglement is expected to 
be behind 
the mechanism for emergent geometry in holographic duality \cite{2017LNP...931.....R}.

While the vast majority of the literature focuses on the entanglement entropy as a measure of quantum entanglement, entanglement entropy has a well-known caveat when applied to mixed quantum states, such as thermal states (Gibbs states) and reduced density matrices obtained by tracing out a part of the total system. For these mixed states, the entanglement entropy is no longer a measure of correlations, let alone quantum correlations. 


Negativity
\cite{Zyczkowski:1998yd, 
Vidal:2002zz, 
Peres:1996dw, 1999JMOp...46..145E, 2005PhRvL..95i0503P, 2000PhRvL..84.2726S, 1996PhLA..223....1H}
is a measure of entanglement that is well defined not only in pure states, but in mixed states as well. 
Consider a Hilbert space that is a tensor product of two Hilbert spaces $\mathcal{H} _A \otimes \mathcal{H} _B$, with a state given by a bipartite density matrix $\rho $. The partial transpose (say with respect to $B$) of $\rho $ in the orthonormal basis composed of $|i\rangle _A$ of $\mathcal{H} _A$ and $|j\rangle _B$ of $\mathcal{H} _B$ is defined to have matrix elements
\begin{equation}
\rho  ^{T_B} _{i_A j_B,k_A l_B} = \rho _{i_A l_B,k_A j_B} ,
\end{equation}
which is a Hermitian matrix with unit trace. While the eigenvalues of $\rho $ are non-negative, this is not necessarily the case for its partial transpose. Negativity is based on this property and quantifies the amount of entanglement by the negative eigenvalues of $\rho ^{T_B} $. Denoting the eigenvalues of $\rho ^{T_B}$ by $\lambda _i$, the \emph{negativity} is defined by
\begin{equation}
\mathcal{N} :=\frac{||\rho ^{T_B} ||_1-1}{2} = \sum _{\lambda _i<0} |\lambda _i|,
\end{equation}
where $|| \cdot ||_1$ is the trace norm (sum of absolute values of eigenvalues), and similarly the \emph{logarithmic negativity} is
\begin{equation} \label{eq:log_negativity_def}
\mathcal{E} =\log ||\rho  ^{T_B} ||_1.
\end{equation}
Indeed \cite{Peres:1996dw}, a separable (unentangled) state can be written as
\begin{equation} \label{eq:separable_state}
\rho  = \sum _a p_a \rho _a^{(A)} \otimes \rho _a^{(B)} 
\end{equation}
with $p_a \ge 0$, $\sum _a p_a=1$, and $\rho _a^{(A)} $, $\rho _a^{(B)} $ are density matrices, so that \eqref{eq:separable_state} is indeed a density matrix. Such states satisfy Bell's inequality. For these states, $\rho ^{T_B} =\sum _a p_a \rho _a^{(A)} \otimes \left( \rho _a^{(B)} \right) ^T$, where $\left( \rho _a^{(B)} \right) ^T$ is also a density matrix, so $\rho ^{T_B} $ has no negative eigenvalues. This shows that a necessary condition for a state to be unentangled is to have vanishing negativity.\footnote{It is not a sufficient condition. However, the entanglement that goes undetected by the negativity may not be useful as negativity places an upper bound on the distillable entanglement \cite{Vidal:2002zz}.} Note that it is known that determining if a density matrix is separable is generally NP hard \cite{gurvits2003classical,gharibian2008strong}.
These definitions of negativity do not depend on the choice of bases for $\mathcal{H} _{A}, \mathcal{H} _B$, and are invariant under interchanging $A$ and $B$.
In the rest of the paper, when we say the \textit{negativity}, we will always mean the logarithmic negativity \eqref{eq:log_negativity_def}.

A finer measure of entanglement is given by the \emph{negativity spectrum} which is the spectrum of the partially transposed density matrix. This is analogous to the \textit{entanglement spectrum} which has been studied extensively,
e.g., in the context of topological phases 
of matter
\cite{Ryu_2006,  PhysRevLett.101.010504, Pollmann_2010}.
The negativity spectrum has been 
studied in 
many-body quantum systems
--- see e.g., Refs.\
\cite{Ruggiero_2,
Mbeng_2017,
Shapourian_2019,
Inamura_2020,
Shapourian:2020mkc}.

The basic objects we will study for negativity are the moments
\begin{equation}
m_k := \tr \left( \rho ^{T_B} \right) ^k .
\end{equation}
For even $k$, the moments are given in terms of the absolute values of the eigenvalues by $ \sum _i |\lambda _i|^k$, so we can obtain the logarithmic negativity by an analytic continuation of the even moments
\begin{equation}
\mathcal{E} = \lim _{k \text{ even} \to 1} \log m_{k}.
\end{equation}
An important object that can be extracted from the moments is the resolvent, which is defined by
\begin{equation}
R(z) = \tr \left( z-\rho ^{T_B} \right) ^{-1} = \sum _{k=0} ^{\infty } \frac{m_k}{z^{k+1} } .
\end{equation}
Indeed, the negativity spectrum $\ns$ (density of eigenvalues) can be obtained from it using the Stieltjes transformation
\begin{equation} \label{eq:neg_spectrum_from_resolvent}
\ns (\lambda )=-\frac{1}{\pi } \lim _{\epsilon  \to 0^+} \Im R(\lambda +i\epsilon ).
\end{equation}
The negativity can be extracted from the spectrum without relying on analytic continuation
\begin{align}
    \mathcal{E} = \log \left[\int d\lambda\, \rho_N(\lambda)|\lambda| \right]. 
\end{align}

\begin{figure}[]
\centering
\includegraphics[width=0.5\textwidth]{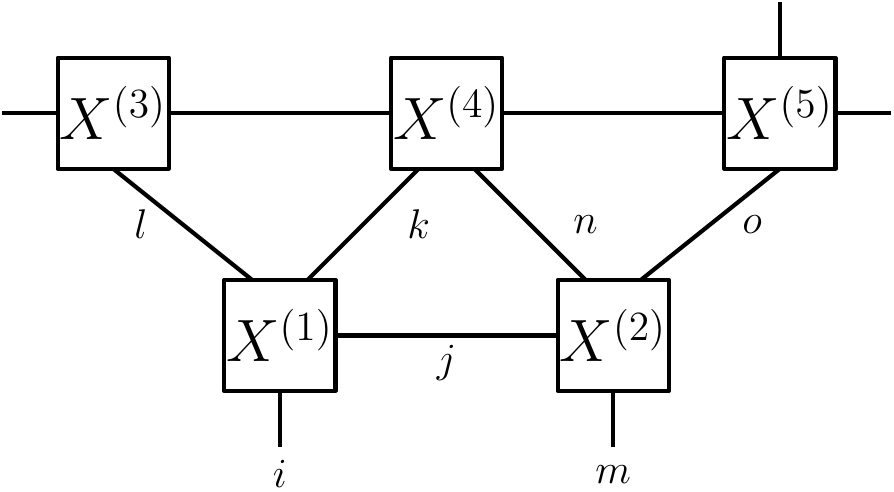}
\caption{An example of a general random tensor network.}
\label{fig:gen_tensor_ntw}
\end{figure}

In this paper we will use a diagrammatic approach to study negativity and negativity spectra in general random tensor networks \cite{2010JPhA...43A5303C,Hayden:2016cfa}. We define the tensor networks that we study as follows. These are unoriented graphs consisting of internal vertices, and edges connecting the vertices. We include external edges that attach to a single 
vertex. 
An example is shown in Fig.~\ref{fig:gen_tensor_ntw}. A 
vertex corresponds to a tensor $X^{(i)} $, where we will take the different tensors to be independent, complex Gaussian random variables. The edges correspond to Hilbert spaces. An edge connecting two vertices represents a contraction of the corresponding tensor indices, while external edges stand for free indices that form the resulting state. A tensor network, such as the one shown in Fig.~\ref{fig:gen_tensor_ntw}, prepares a state
\begin{align}
|\psi\rangle = \mathcal{N}^{-1/2 }\sum X^{(1)} _{ijkl} X^{(2)} _{mjno} \cdots |i\rangle |m\rangle \cdots ,
\end{align}
where $\mathcal{N}$ is the normalization.\footnote{Usually random tensor networks are expressed in terms of projected Haar random states. An argument for the equivalence of the two descriptions can be found in \cite{2021arXiv210800011K}.}

Each Hilbert space corresponding to an edge has a dimension $L_i$. We will assume here that all
\begin{equation}
    L_i=N^{w_i} d_i
\end{equation}
with $N \gg 1$ and $d_i,w_i>0$ finite.
We take the variances of the tensors such that
\begin{equation} \label{eq:variance_general}
    \prod _i\text{Var}\left( X^{(i)} \right) = \prod_i \mean{ \left( X^{(i)} \right)^2}=  \prod _{\text{edges}} L_i^{-1},
\end{equation}
where we denote an average over the random tensors by an overline, and their mean value is taken to be vanishing.
The density matrix corresponding to the state above is
\begin{equation}
\rho = |\psi\rangle \langle \psi| = \mathcal{N}^{-1}\sum X^{(1)} _{ijkl} X^{(2)} _{mjno} \cdots |i\rangle |m\rangle \cdots  X^{(1)*} _{i'j'k'l'} X^{(2)*} _{m'j'n'o'} \cdots \langle i'| \langle m'| \cdots .
\end{equation}
With the choice of variances above, the expectation value of the normalization is 1, and the fluctuations around it are small in the large-$N$ limit. Therefore, we can drop it at leading order.\footnote{For example, considering an averaged moment of the density matrix, the normalization is a singlet that equals one at leading order.} When discussing $m_k$, $\cE$, $R$, and $\ns$ for a random tensor network, these quantities will always stand for their averaged values.

We will start (Section \ref{sec:1_tensor}) with the case of a Haar random state, which corresponds to a single tensor. The negativity spectrum for this case was studied in \cite{2010arXiv1011.0275A,2011arXiv1105.2556B,2013JMP....54d2202F,Shapourian:2020mkc} for the bulk of the different phases. 
We then move to the case of a two-tensor network (Section \ref{sec:2_tensor}).
We use it as a relatively simple example where we demonstrate the different ingredients of our analysis. We explain the diagrammatic method that we use and that also applies to general tensor networks, its large-$N$ structure, and its relation both to random tensors and permutations. We will classify the rich phase diagram for the two-tensor network and check these results with finite size numerics. The details are given in Appendix \ref{sec:2_tensor_phases}.

Pure states are analyzed in Section \ref{sec:pure_states}. The negativity spectrum for pure states is explicitly determined in terms of the entanglement spectrum. We analyze it for the two-tensor network in several phases and write a non-trivial Schwinger-Dyson equation for general large Hilbert space dimensions.

In Section \ref{sec:gen_network} we analyze general random tensor networks, and express the negativity moments using a solution to a flow network problem.
In \cite{Dong:2021clv} the negativity of holographic states in contractible topologies with unique extremal Ryu-Takayanagi surfaces was found. Here we consider general random tensor networks with large bond dimensions. In particular, we do not assume that there is a unique RT surface, and allow the Hilbert spaces associated to edges to have different dimensions. We find a wide range of negativity spectra, beyond the semi-circle distribution.
We may also have different states on the same tensor network and ask how different they are, and this was studied recently in \cite{2021arXiv210800011K}.

In Sections \ref{sec:west_coast_1_tensor}-\ref{sec:islands}, we discuss the implications of the results in prior sections for holographic systems.
In Section \ref{sec:west_coast_1_tensor}, we recall the model of \cite{2019arXiv191111977P} involving Jackiw-Teitelboim gravity in the micro-canonical ensemble, a toy model for a black hole and its radiation. We note that negativity in this model, when we consider subsystems of the radiation and the black hole system, is described by the smallest random tensor network, namely the one-tensor network of Section \ref{sec:1_tensor}.

A holographic setting that is described by a larger class of random tensor networks is that of holographic fixed-area states, which are discussed in Section \ref{sec:holography}. 
The diagrammatic approach that we use is also useful in showing that the calculation of entanglement in a fixed-area state is given by that in a random tensor network. 
We discuss some examples of this. We also describe how more general random tensor networks are related to holography.

Assuming that we have access to part of the Hawking radiation of an evaporating black hole, it is interesting to look for an island formula that describes the negativity in such systems coupled to gravity. In Section \ref{sec:islands}, we only describe it semi-classically, rather than providing a formula analogous to the quantum extremal surface prescription. We do this at first by applying the random tensor network result to the doubly-holographic idea of \cite{Almheiri:2019hni}. We can think about this as holographic states where we fix the generalized entropy rather than the area. We then test it in the same simple model based on JT gravity \cite{2019arXiv191111977P}. This provides an argument for the replica wormhole origin of the island contributions to negativity in systems coupled to gravity.



\section{Haar random states} \label{sec:1_tensor}

\begin{figure}[]
\centering
\includegraphics[width=0.75\textwidth]{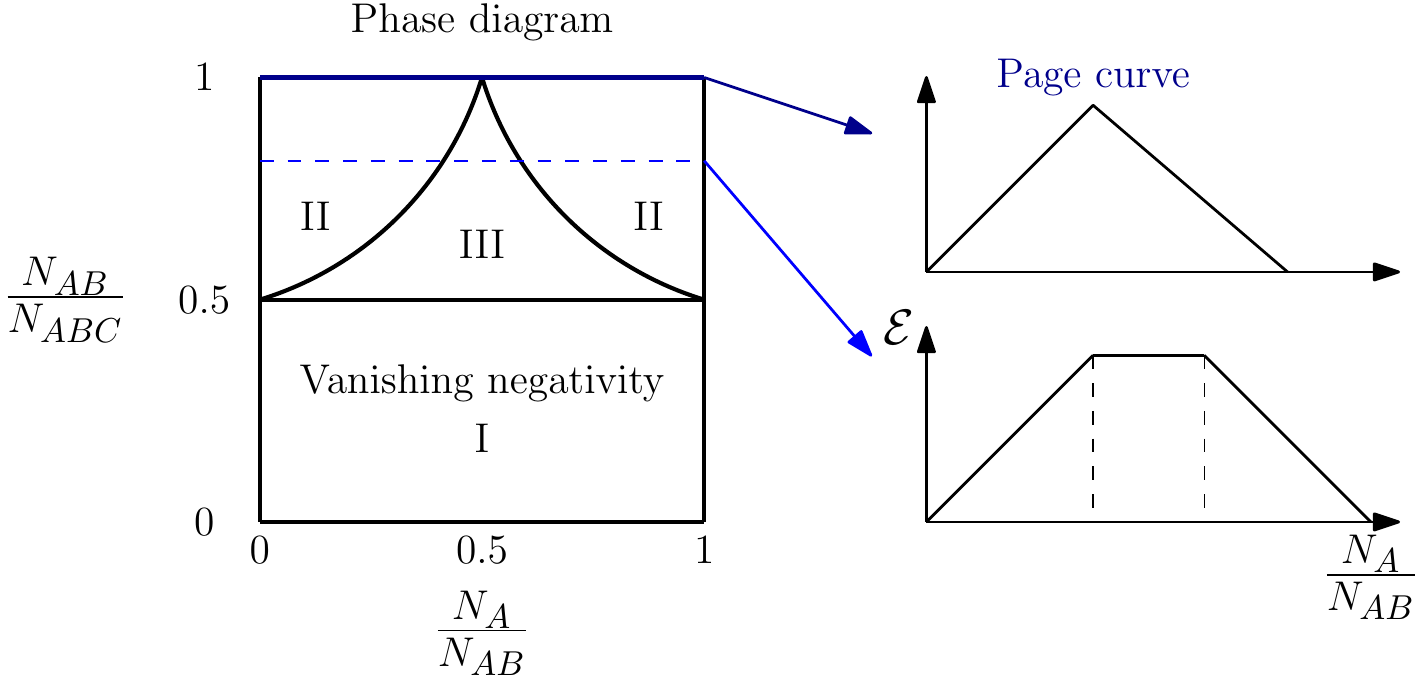}
\caption{Phase diagram for the entanglement in a random mixed state. For a given fixed $L_C$, the logarithmic negativity is shown on the right. In this figure, $L$ is the Hilbert space dimension, and $N=\log L$ is the number of qudits.
In the pure state limit, there is the Page curve for entanglement, being proportional to the size of the smaller subsystem $\min\{\log L_{A} ,\log L_{B} \}$. Away from this limit, there is a plateau in the Page curve in a region of the phase diagram where multipartite entanglement is present.
}
\label{fig:one_tensor_phase_diagram}
\end{figure}

Let us begin by considering the case of a single tensor $X$. The broad motivation here is to study mixed state entanglement in a random state (see \cite{Shapourian:2020mkc}). In order to do that, we start with a \emph{pure} Haar random state (as was used by Page \cite{Page:1993df})
\begin{equation} \label{eq:1_tensor_state}
|\Psi\rangle =\sum _{ij\alpha } X_{ij\alpha } |i\rangle _{A} |j\rangle _{B} |\alpha \rangle _C,
\end{equation}
where the Hilbert space is the tensor product of three Hilbert spaces, and we then trace out subsystem $C$. This leaves us with a random \emph{mixed} state for the $AB$ system, and we can study the negativity in order to learn about the entanglement between $A$ and $B$. In \cite{Shapourian:2020mkc}, a phase diagram for this entanglement structure was found, as a function of the sizes $L_{A},L_{B},L_C$ of the Hilbert spaces, and we reproduce it in Fig.~\ref{fig:one_tensor_phase_diagram}.

Let us first summarize the result for the phase diagram, and relate it to the more familiar entanglement entropy. There are three regions in the phase diagram:
\begin{itemize}
\item 
Region I ($L_C > L_A L_B$): the negativity vanishes, and the negativity spectrum is given by a semi-circle distribution with a support on the positive real axis.
\item 
Region II ($L_A > L_B L_C$ or $L_B > L_A L_C$): the negativity grows linearly with $\min\{ \log L_{A}, \log L_{B}\}$. The negativity spectrum is given by two copies of the Marchenko–Pastur distribution, one supported on the positive, and one on the negative real axis.
\item 
Region III ($L_{A} <L_B L_{C} $ and $L_{B} <L_A L_{C} $ and $L_{C} <L_A L_{B} $): the negativity does not change with the system size $L_{A}$ independently, only with $L_AL_B$. 
The negativity spectrum is a semi-circle with support on both positive and negative real axes.
\end{itemize}

Indeed, for an empty subsystem $C$ (top horizontal line in the phase diagram), this reduces to the random pure state studied by Page \cite{Page:1993df}. In this case, region III disappears, and the entanglement takes the form of the Page curve shown in Fig.~\ref{fig:one_tensor_phase_diagram}.
The unnormalized reduced density matrix is $X X ^{\dagger} $, where $X$ is considered as a matrix $X_{ij} $, which is known in random matrix theory as the Wishart ensemble.
Its entanglement spectrum is given by the Marchenko–Pastur distribution.

In the region in the phase diagram between phases II and III, we have a phase transition. The negativity spectrum within this phase transition was plotted numerically in \cite{Shapourian:2020mkc}. Here, in Section \ref{sec:wormhole_1_tensor}, we will find its analytic form and will compare to the numerical plot.


We will use a diagrammatic approach for negativity similar to \cite{Shapourian:2020mkc}, so we start by reviewing it for Haar random states. Every instance of the density matrix can be represented by drawing one line for every index, where we have one index for each Hilbert space factor or equivalently, a tensor network edge. This way, the matrix elements of the full density matrix are
\begin{equation}
\left(| \Psi\rangle \langle \Psi|\right) _{ij\alpha ,kl\beta } =
X_{ij\alpha } X^*_{kl\beta } =
    \tikz[baseline=-0.5ex]{
    \draw[densely dotted,thick] (0,-0.3)--(0,0.3);
    \draw (0,0.5) node[scale=0.8] {$i$};
    \draw (0.2,-0.3)--(0.2,0.3);
    \draw (0.2,0.5) node[scale=0.8] {$j$};
    \draw[dashed] (0.4,0.3)--(0.4,-0.3);
    \draw (0.4,0.5) node[scale=0.8] {$\alpha$};
    
    \draw[dashed] (1.3,0.3)--(1.3,-0.3);
    \draw (1.3,0.5) node[scale=0.8] {$\beta$};
    \draw (1.5,-0.3)--(1.5,0.3);
    \draw (1.5,0.5) node[scale=0.8] {$l$};
    \draw[densely dotted,thick] (1.7,-0.3)--(1.7,0.3);
    \draw (1.7,0.5) node[scale=0.8] {$k$};
    }
\end{equation}
where dotted, solid, and dashed lines correspond to subsystems $A,B$, and $C$ respectively. In these diagrams, we perform matrix operations in the lower part of the diagram, and ensemble averaging in the upper part.
This way, tracing over system $C$ gives us a mixed state for system $AB$ with matrix elements
\begin{equation}
\left( \rho \right) _{ij,kl} =
    \tikz[baseline=-0.5ex]{
    \draw[densely dotted,thick] (0,-0.3)--(0,0.3);
    \draw (0.2,-0.3)--(0.2,0.3);
    \draw[dashed] (0.4,0.3)--(0.4,-0.2)--(1.3,-0.2)--(1.3,0.2);
    
    \draw (1.5,-0.3)--(1.5,0.3);
    \draw[densely dotted,thick] (1.7,-0.3)--(1.7,0.3);
    },
\end{equation}
where from now on we omit the explicit indices. Then, for example for the purity, we should consider
\begin{equation}
\tr \rho ^2=
    \tikz[baseline=-0.5ex]{
    \draw[densely dotted,thick] (0,-0.8)--(0,0.3) node[pos=0](a1){} node[pos=1](a2){};
    \draw (0.2,-0.7)--(0.2,0.3) node[pos=0](b1){} node[pos=1](b2){};
    \draw[dashed] (0.4,0.3)--(0.4,-0.2)--(1.3,-0.2)--(1.3,0.2) node[pos=0](c2){} node[pos=1](cc2){};
    
    \draw (1.5,-0.4)--(1.5,0.3) node[pos=0](aa1){} node[pos=1](aa2){};
    \draw[densely dotted,thick] (1.7,-0.3)--(1.7,0.3) node[pos=0](bb1){} node[pos=1](b2){};

    \draw[densely dotted,thick] (2.5,-0.3)--(2.5,0.3) node[pos=0](d1){} node[pos=1](d2){};
    \draw (2.7,-0.4)--(2.7,0.3) node[pos=0](e1){} node[pos=1](e2){};
    \draw[dashed] (2.9,0.3)--(2.9,-0.2)--(3.8,-0.2)--(3.8,0.2) node[pos=0](f2){} node[pos=1](ff2){};
    
    \draw (4,-0.7)--(4,0.3) node[pos=0](dd1){} node[pos=1](dd2){};
    \draw[densely dotted,thick] (4.2,-0.8)--(4.2,0.3) node[pos=0](ee1){} node[pos=1](ee2){};
    
    \draw[densely dotted,thick] (1.7,-0.3)--(2.5,-0.3);
    \draw (1.5,-0.4)--(2.7,-0.4);
    \draw (0.2,-0.7)--(4,-0.7);
    \draw[densely dotted,thick] (0,-0.8)--(4.2,-0.8);
    } .
\end{equation}
For the negativity, we implement the partial transpose just by switching the indices
\begin{equation}
\rho ^{T_B} =
    \tikz[baseline=-0.5ex]{
    \draw[densely dotted,thick] (0,-0.3)--(0,0.3);
    \draw (0.2,-0.3)--(0.2,0.3);
    \draw[dashed] (0.4,0.3)--(0.4,-0.2)--(1.3,-0.2)--(1.3,0.2);
    
    \draw (1.5,-0.3)--(1.5,0.3);
    \draw[densely dotted,thick] (1.7,-0.3)--(1.7,0.3);
        \draw (0.2,-0.3)--(1.5,-0.6);
    \draw (0.2,-0.6)--(1.5,-0.3);
    } \,\,\,.
\end{equation}
For the averaging, we sum over all Wick contractions, going from the left end of one density matrix instance to the right end of another. There are two such contractions for 
the purity of the partially transposed density matrix $\mean{\tr \left( \rho ^{T_B} \right) ^2}$
\begin{equation}
    \tikz[baseline=-0.5ex]{
    \draw[densely dotted,thick] (0,-0.3)--(0,0.3);
    \draw (0.2,-0.3)--(0.2,0.3);
    \draw[dashed] (0.4,0.3)--(0.4,-0.2)--(1.3,-0.2)--(1.3,0.2);
    
    \draw (1.5,-0.3)--(1.5,0.3);
    \draw[densely dotted,thick] (1.7,-0.3)--(1.7,0.3);
    
    \draw (0.2,-0.3)--(1.5,-0.6);
    \draw (0.2,-0.6)--(1.5,-0.3);
    
    \draw[densely dotted,thick] (2.5,-0.3)--(2.5,0.3);
    \draw (2.7,-0.3)--(2.7,0.3);
    \draw[dashed] (2.9,0.3)--(2.9,-0.2)--(3.8,-0.2)--(3.8,0.2);
    
    \draw (4,-0.3)--(4,0.3);
    \draw[densely dotted,thick] (4.2,-0.3)--(4.2,0.3);
    
    \draw (2.7,-0.3)--(4,-0.6);
    \draw (2.7,-0.6)--(4,-0.3);
    \draw[densely dotted,thick] (1.7,-0.3)--(2.5,-0.3);
    \draw (1.5,-0.6)--(2.7,-0.6);
    \draw (0.2,-0.6)--(0.2,-0.8)--(4,-0.8)--(4,-0.6);
    \draw[densely dotted,thick] (0,-0.3)--(0,-0.9)--(4.2,-0.9)--(4.2,-0.3);
    
    \draw[densely dotted,thick] (1.7,0.3) arc (0:180:0.85);
    \draw(1.5,0.3) arc (0:180:0.65);
    \draw[dashed] (1.3,0.3) arc (0:180:0.45);
    \draw[densely dotted,thick] (4.2,0.3) arc (0:180:0.85);
    \draw (4,0.3) arc (0:180:0.65);
    \draw[dashed] (3.8,0.3) arc (0:180:0.45);
    } 
    \hspace{.5cm}
    + 
    \hspace{.5cm}
    \tikz[baseline=-0.5ex]{
    \draw[densely dotted,thick] (0,-0.3)--(0,0.3);
    \draw (0.2,-0.3)--(0.2,0.3);
    \draw[dashed] (0.4,0.3)--(0.4,-0.2)--(1.3,-0.2)--(1.3,0.2);
    
    \draw (1.5,-0.3)--(1.5,0.3);
    \draw[densely dotted,thick] (1.7,-0.3)--(1.7,0.3);
    
    \draw (0.2,-0.3)--(1.5,-0.6);
    \draw (0.2,-0.6)--(1.5,-0.3);
    
    \draw[densely dotted,thick] (2.5,-0.3)--(2.5,0.3);
    \draw (2.7,-0.3)--(2.7,0.3);
    \draw[dashed] (2.9,0.3)--(2.9,-0.2)--(3.8,-0.2)--(3.8,0.2);
    
    \draw (4,-0.3)--(4,0.3);
    \draw[densely dotted,thick] (4.2,-0.3)--(4.2,0.3);
    
    \draw (2.7,-0.3)--(4,-0.6);
    \draw (2.7,-0.6)--(4,-0.3);
    \draw[densely dotted,thick] (1.7,-0.3)--(2.5,-0.3);
    \draw (1.5,-0.6)--(2.7,-0.6);
    \draw (0.2,-0.6)--(0.2,-0.8)--(4,-0.8)--(4,-0.6);
    \draw[densely dotted,thick] (0,-0.3)--(0,-0.9)--(4.2,-0.9)--(4.2,-0.3);
    
    \draw[densely dotted,thick] (4.2,0.3) arc (0:180:2.1);
    \draw(4.0,0.3) arc (0:180:1.9);
    \draw[dashed] (3.8,0.3) arc (0:180:1.7);
    \draw[densely dotted,thick] (2.5,0.3) arc (0:180:0.4);
    \draw (2.7,0.3) arc (0:180:0.6);
    \draw[dashed] (2.9,0.3) arc (0:180:0.8);
    } 
    \hspace{.5cm}
    .
\end{equation}

\noindent
The value we assign to every diagram is given simply by
\begin{itemize}
\item 
A factor of $L_{A} $, $L_{B} $, or $L_C$ for every dotted, solid, or dashed loop respectively, from summing over the index.
\item 
For every contraction, we get from the variance a factor of $\frac{1}{L_{A} L_{B} L_C} $.
\end{itemize}

There are several methods to evaluate the moments. One method is to obtain relations between the moments using manipulations of the defining integrals, similarly to random matrix theory. We show how to derive such loop equations in the presence of the partial transpose in Appendix \ref{sec:loop_eq}, and use it for random Haar states. Another method is to write consistency relations for all diagrams representing the summation of all moments. This is done in the following subsection.

\subsection{Schwinger-Dyson equation}

For any given moment we consider, one should first determine which diagrams dominate. In the case at hand, this was done for the bulk of the three regions in the phase diagram of Fig.~\ref{fig:one_tensor_phase_diagram} in \cite{Shapourian:2020mkc}. Here, we will not repeat this analysis, as we later analyze this in detail for general    tensor networks. One of the ingredients is an 't Hooft-like large-$N$ expansion that we will use. In each of the three regions of the phase diagram, different sets of diagrams are dominant, but all of them are special cases of planar diagrams, e.g., for the dotted and the dashed lines. We will write a Schwinger-Dyson equation that sums all the planar diagrams.

The Schwinger-Dyson equation is written for the resolvent. Let us consider the same object as the resolvent, but without taking the trace, i.e., it is an operator acting on the Hilbert space of system $AB$, and denote it by a shaded disc. That is, it is the sum over all the diagrams mentioned above with the same factors of $z$ and the Hilbert space dimensions. It is related to the resolvent just by
\begin{equation}
R=
    \tikz[baseline=-0.5ex]{
    \node[circle,draw,pattern=north east lines] (a) at (1,0) {};
    \node[circle] (b) at (0,0) {};
    \node[circle] (c) at (2,0) {};
    \draw[densely dotted,thick] (b.10)--(a.170);
    \draw (b.350)+ (0.1,0)--(a.190);
    \draw[densely dotted,thick] (a.10)--(c.170);
    \draw (c.190)+(-0.1,0)--(a.350);
    \draw (b.350)+ (0.1,0) .. controls (0.6,-1) and (1.4,-1) .. ($(c.190)+(-0.1,0)$);
    \draw[densely dotted,thick] (b.10).. controls (0.6,-1.15) and (1.4,-1.15) .. (c.170);
    }
\end{equation}
and it satisfies the recursion relation
\begin{align}
    \label{eq:radiation-exp}
    \tikz[baseline=-0.5ex]{
    \draw[densely dotted,thick] (0.35,0.05)--(0.65,0.05);
    \draw[densely dotted,thick] (-0.35,0.05)--(-0.65,0.05);
    \draw (0.35,-0.05)--(0.65,-0.05);
    \draw (-0.35,-0.05)--(-0.65,-0.05);
    \draw[thick] (0,0) circle (10pt) node[anchor=center,pattern=north east lines,circle,minimum size=20pt] {};
    }
    =& \nonumber
\ \,
   \tikz[baseline=-0.5ex]{
    \draw[densely dotted,thick] (-0.4,0.05)--(0.35,0.05);
    \draw (-0.4,-0.05)--(0.35,-0.05);
    }
\ \,    
+\vcenter{\hbox{\includegraphics[scale=0.5]{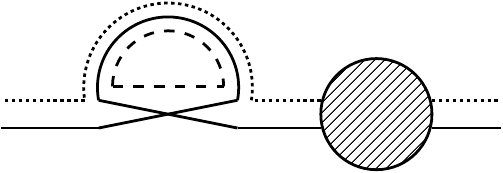}}}
+\vcenter{\hbox{\includegraphics[scale=0.5]{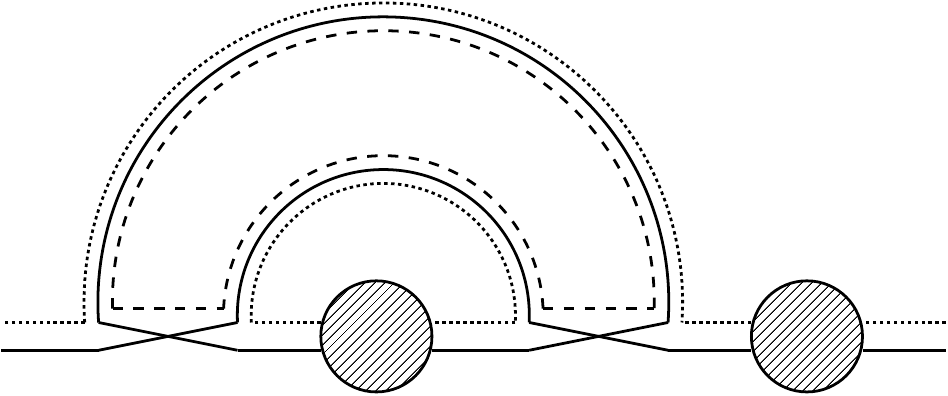}}} \\
&+\vcenter{\hbox{\includegraphics[scale=0.4]{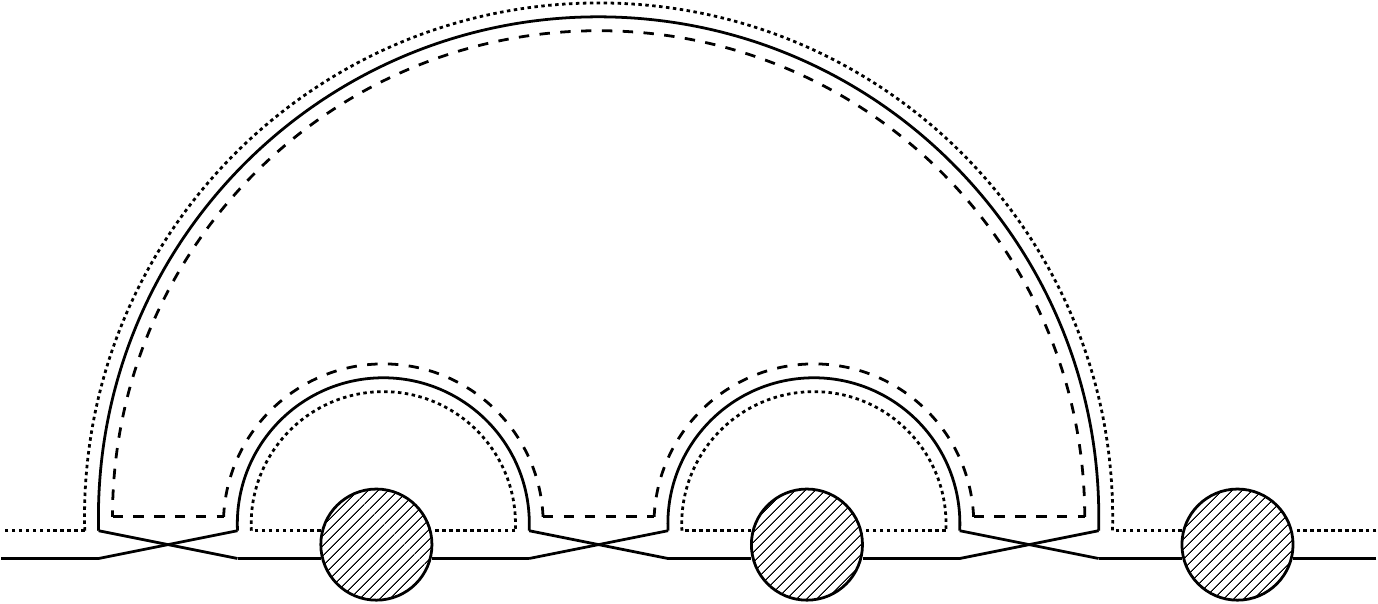}}}
+\vcenter{\hbox{\includegraphics[scale=0.4]{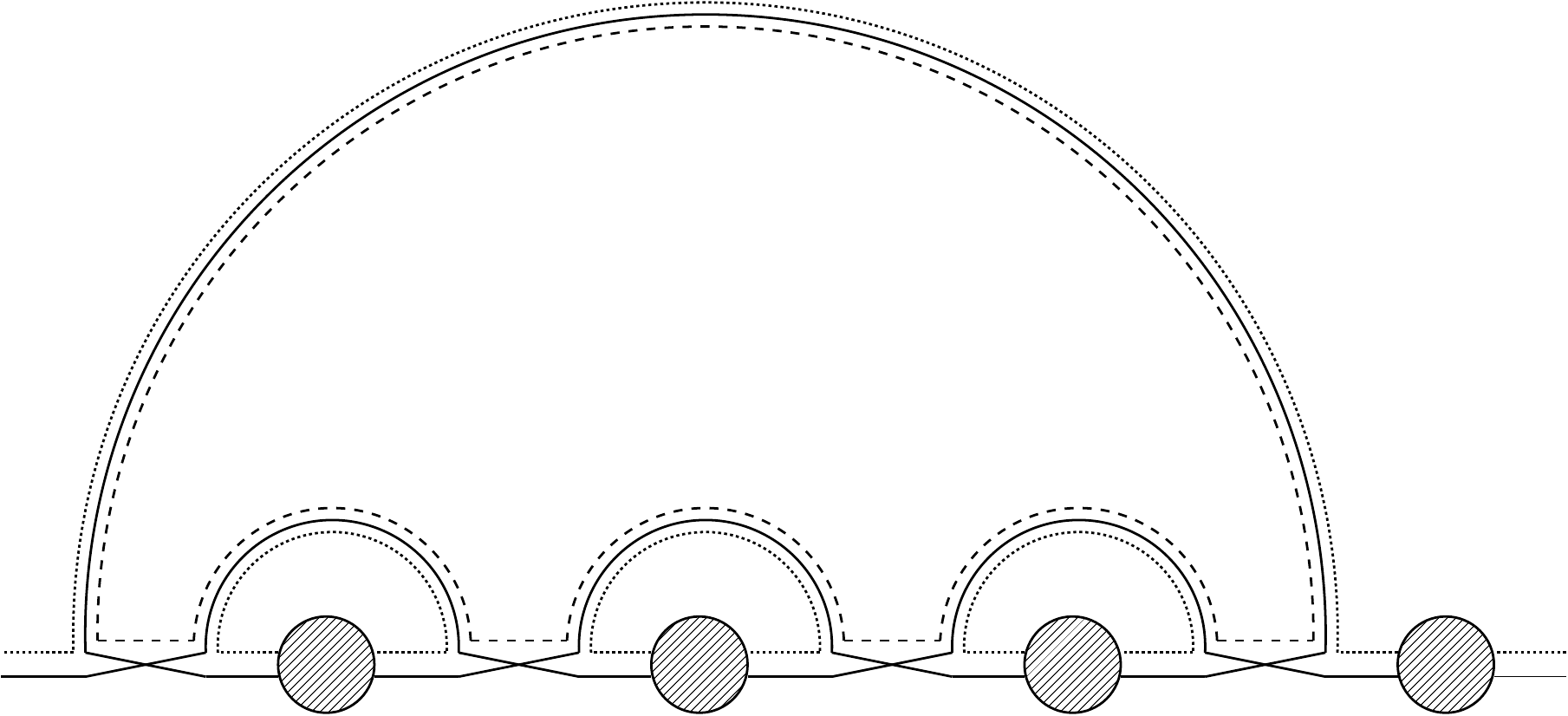}}} 
\end{align}
for planar diagrams.
The different terms are distinguished by which density matrices are contracted to the first one.
The main point to notice is that on the RHS, whenever there is an odd number of $R$'s, there is a single solid line, while for an even number, there are two solid lines. This gives the equation
\begin{equation}
\begin{split}
z R &= L_AL_B + \sum _{k=0} ^{\infty } \frac{R^{2k+1} }{L_C^{2k} L_{A} ^{2k+1} L_{B} ^{4k+1} } + \sum _{k=1} ^{\infty } \frac{R^{2k} }{L_C^{2k-1} L_{A} ^{2k} L_{B} ^{4k-2} } =\\
&=\frac{L_{A} L_{B} ^3 L_C^2(L_A^2+R)-L_AL_B R^2+L_{B} ^2L_C R^2}{L_C^2L_{A} ^2L_{B} ^4-R^2} .
\end{split}
\end{equation}
Completing the geometric sums, we find a cubic Schwinger-Dyson equation
\begin{equation}
zR^3+(L_{B} ^2L_C-L_AL_B)R^2+\left( L_{A} L_{B} ^3L_C^2-L_{A} ^2L_{B} ^4L_C^2 z\right) R+L_{A} ^3 L_{B} ^5 L_C^2 =0.
\end{equation}
This equation allows one to get the negativity spectrum in the different regimes. We will not perform that explicitly here, as our main focus lies in higher tensor networks.

\section{Two-tensor network} \label{sec:2_tensor}

The simplest case involving an internal structure of a tensor network is the case consisting of two tensors connected to each other. This allows us to ``tune an additional knob'' controlling the connectivity of the external Hilbert spaces. We will demonstrate our methods on this case, making the later generalization to an arbitrary tensor network easier.

Denoting the two tensors by $X$ and $Y$, the tensor network looks as follows
\begin{equation} \label{eq:two_tensor_network}
\tikz[baseline=-0.5ex]{
\path (0,0) node[rectangle,minimum size=1] (a) {} ++(2,0) node[rectangle,draw,minimum size=2] (X) {$X$} ++(3,0) node[rectangle,draw,minimum size=2] (Y) {$Y$} ++(2,0) node[rectangle,minimum size=1] (b) {};
\node (T1) at ($(X)+(0,-1.5)$) {};
\node (T2) at ($(Y)+(0,-1.5)$) {};
\draw (a)--(X) node[pos=0.5,above,scale=0.9] {$j \in \mathcal{H} _{A} $};
\draw (X)--(Y) node[pos=0.5,above,scale=0.9] {$k \in \mathcal{H} _{E_W} $};
\draw (Y)--(b) node[pos=0.5,above,scale=0.9] {$n \in \mathcal{H} _{B} $};
\draw (T1)--(X) node[pos=0.5,left,scale=0.9] {$i \in \mathcal{H}_{C_1} $};
\draw (T2)--(Y) node[pos=0.5,right,scale=0.9] {$m \in \mathcal{H}_{C_2} $};
} .
\end{equation}
We now explain the Hilbert space structure and the reason that there are two external edges for each tensor. The total external Hilbert space is the tensor product $\mathcal{H} _{C_1} \otimes \mathcal{H} _A \otimes \mathcal{H} _{C_2} \otimes \mathcal{H} _B$.
We have chosen the notation in a way that will correspond later to the holographic system that this tensor network represents.
The Hilbert space $\mathcal{H} _{E_W} $ is an internal space.
The two tensors are independent complex Gaussian random variables $X_{i j k} $ (with $i=1,\cdots ,L_{C_1} $, $j=1,\cdots ,L_A$, and $k=1,\cdots ,L_{E_W}$) and $Y_{m n k} $ (with $m=1, \cdots ,L_{C_2} $, $n=1, \cdots, L_B$ and again $k=1,\cdots ,L_{E_W}$). Their variance satisfies \eqref{eq:variance_general}, so we can take them to be
\begin{equation}
\begin{split}
& \mean{ X_{i_1j_1k_1} X^*_{i_2j_2k_2} } =\left( L_{C_1} L_A \sqrt{L_{E_W}}\right) ^{-1} \delta_{i_1i_2} \delta _{j_1j_2} \delta _{k_1k_2} ,\\
& \mean{ Y_{m_1n_1k_1} Y^*_{m_2n_2k_2} } =\left( L_{C_2} L_B \sqrt{L_{E_W}}\right) ^{-1} \delta_{m_1m_2} \delta _{n_1n_2} \delta _{k_1k_2} .
\end{split}
\end{equation}

The state that this tensor network prepares is
\begin{equation} \label{eq:2_tensor_state}
|\psi\rangle  = \sum_{i,j,m,n,k} X_{ijk} Y_{mnk} |i\rangle |j\rangle |m\rangle |n\rangle .
\end{equation}
While this is a pure state, we will trace over $\mathcal{H} _{C_1} $ and $\mathcal{H} _{C_2} $, inducing a mixed state on the system $AB$. We are interested in studying the entanglement between $A$ and $B$ using negativity. This means that we should study the moments of the partial transpose with respect to $A$ of the density matrix on $AB$
\begin{equation}
m_n = \mean{\tr \left( \rho ^{T_A} \right) ^n},
\end{equation}
where the density matrix on $AB$ is
\begin{equation}
\rho _{j_1n_1,j_2n_2} = \sum_{k,l} X_{i j_1k} Y_{m n_1k} X^*_{i j_2l} Y^*_{m n_2l} .
\end{equation}

\begin{figure}[]
\centering
\includegraphics[width=0.5\textwidth]{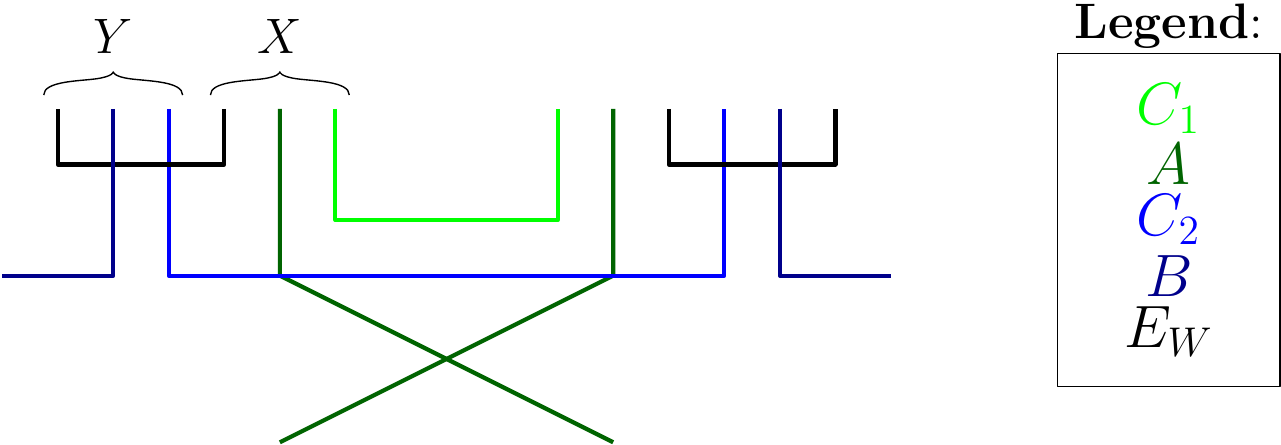}
\caption{The representation of the density matrix graphically.}
\label{fig:density_graphically}
\end{figure}

\begin{figure}[]
\centering
\includegraphics[width=0.8\textwidth]{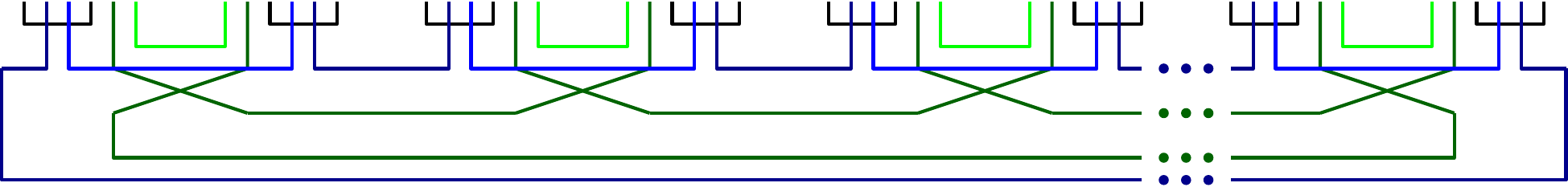}
\caption{The negativity moment is given by the sum over all contractions of this diagram. 
}
\label{fig:Zn_graphically}
\end{figure}

\begin{figure}[]
\centering
\includegraphics[width=0.6\textwidth]{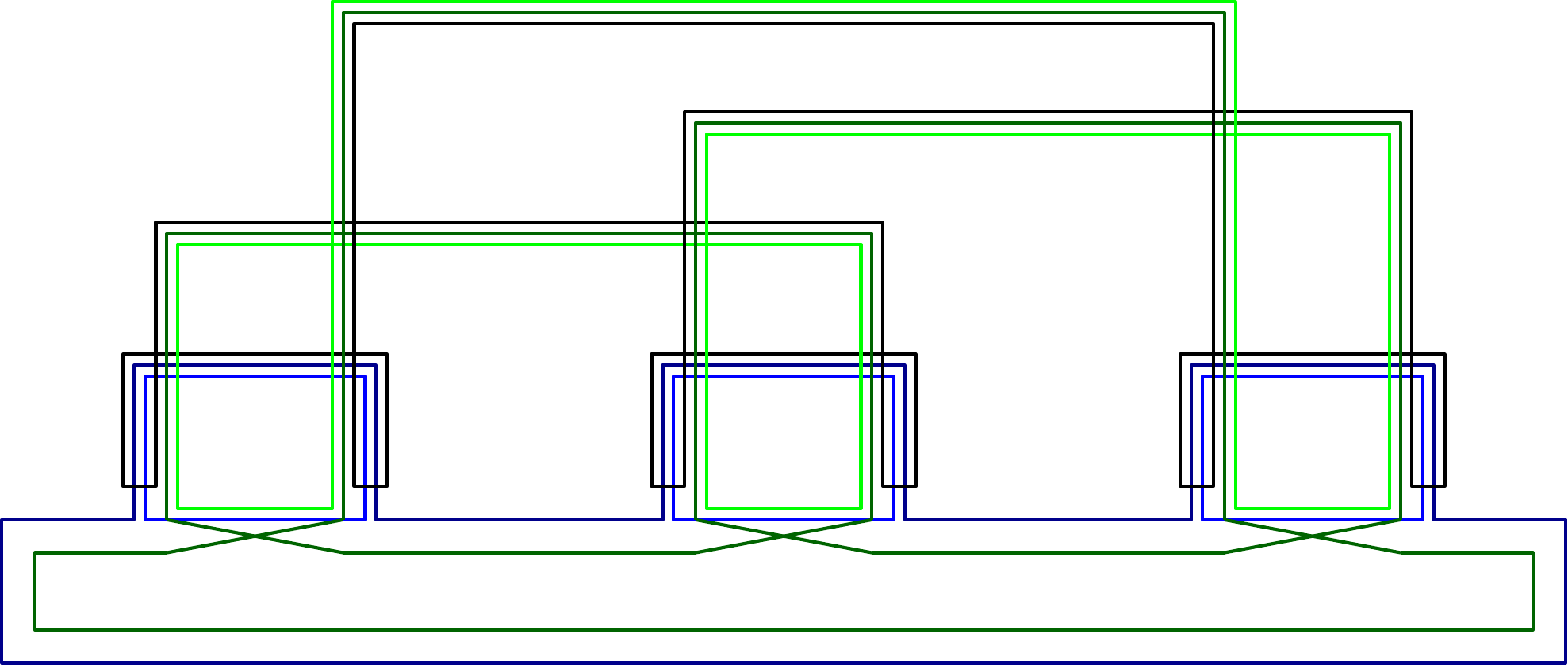}
\caption{A diagram contributing to the third negativity moment.
}
\label{fig:Zn_example}
\end{figure}

We introduce the following diagrammatic representation for calculating the moments $m_n$.
The basic building block corresponds to one instance of the partially transposed density matrix, and is shown in Fig.~\ref{fig:density_graphically}. We first connect $n$ copies of this basic ingredient, as shown in Fig.~\ref{fig:Zn_graphically}. The moment is then given by summing over all possible contractions. We group each of the six lines on each side of Fig.~\ref{fig:density_graphically} into two groups: the green group consisting of the two green and one black lines, corresponding to $X_{{\color{green}i} {\color{OliveGreen} j} k} $, and the blue group made of two blue and one black lines, corresponding to $Y_{{\color{blue}m} {\color{darkblue} n} k} $.\footnote{The colors are not essential but rather redundant, as they are determined according to their position. We include them for convenience.} These give us two \emph{multi-lines}. Then, the contractions are done so that left green multi-lines are contracted with right green multi-lines, and left blue multi-lines are contracted with right blue multi-lines. An example of an allowed contraction is shown in Fig.~\ref{fig:Zn_example}.

Each line corresponds to an index that we sum over in the corresponding Hilbert space. The Hilbert space associated with the light green line is of dimension $L_{C_1} $ and similarly for the rest (see the legend in Fig.~\ref{fig:density_graphically} for the mapping). Therefore, in each diagram, we will get $L_{C_1} $ to the power of the number of closed green lines, and the same for the rest.

To summarize, the moment is calculated by summing over all such diagrams, where each closed line is assigned the corresponding $L$ factor, each multi-line green contraction is assigned $\frac{1}{L_{C_1} L_A \sqrt{L_{E_W} }} $, and each multi-line blue contraction is assigned $\frac{1}{L_{C_2} L_B \sqrt{L_{E_W} }} $.

\subsection{'t Hooft-like large-$N$ expansion}

\begin{figure}[]
\centering
\includegraphics[width=0.3\textwidth]{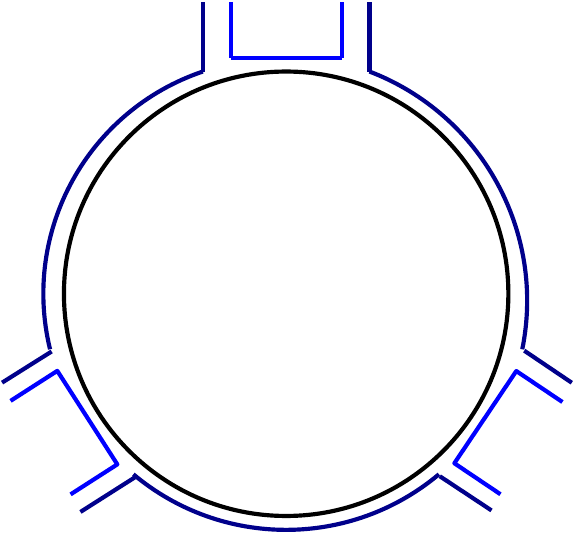}
\hspace{1cm}
\includegraphics[width=0.3\textwidth]{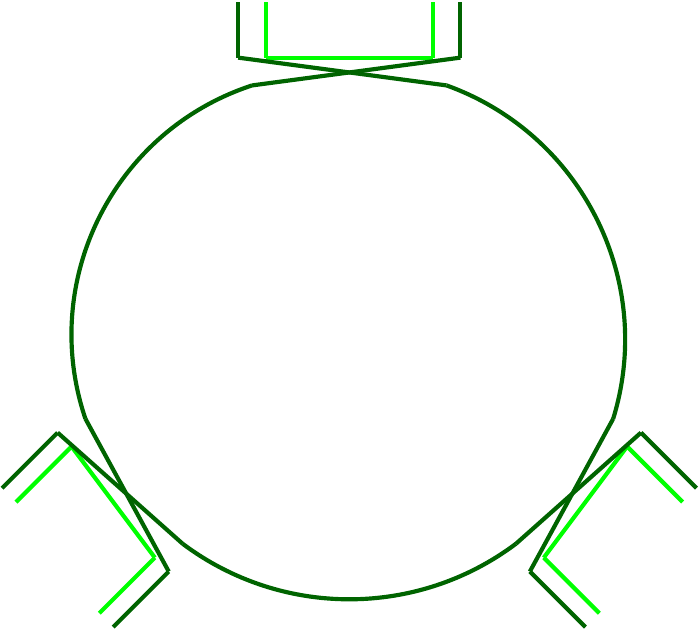}
\caption{large-$N$ counting in the blue (left) and green (right) parts for $n=3$.}
\label{fig:large_N_blue_and_green}
\end{figure}

We now show that these diagrams obey a topological expansion at large dimensions.
Suppose that $L_{C_1} ,L_{C_2} ,L_A,L_B \sim N \gg 1$ with $L_{E_W}$ fixed. The two blue lines of the diagram form a double line structure. To see this, we draw the diagram on a circle; an example with $n=3$ is shown in Fig.~\ref{fig:large_N_blue_and_green}. In this figure, the propagators contribute a factor of $1/N^{2n}$ and there is a factor of $N$ for every loop. The number of loops is $F-1$ (with $F$ being the number of faces), where the one is subtracted since there is one missing loop that was added in black at the center of the figure. We see that the diagram is made of $2n$ cubic large-$N$ vertices, and there are $3n$ edges. Therefore, each such diagram is assigned
\begin{equation}
\frac{1}{N^{2n} } N^{F-1} = N^{1-2g-n} ,
\end{equation}
where $g$ is the genus of the surface.

For the green part, a very similar analysis holds. An example is shown in Fig.~\ref{fig:large_N_blue_and_green}. We connect the LHS of each of the matrix insertions with the RHS of another matrix insertion. But then, if we exchange the two ends of each such insertion, we are in the same situation as in the blue part, with the same counting. Therefore, we once again get $N^{1-2g-n} $ where now $g$ is the genus of the surface obtained by this construction.


Together, we get that the large-$N$ counting of a general diagram is
\begin{equation}
N^{2-2g_A - 2g_B - n_A-n_B} ,
\end{equation}
where $g_A$ is the genus of the $A$ subsystem (green), $n_A$ the number of matrix insertions $XX ^{\dagger} $, and similarly for the $B$ part.

Note also that if we have two traces of only the (say) $B$ subsystem (only blue diagrams), then the connected part goes as $N^{-2g - n_1 - n_2} $ where $n_1,n_2$ are the number of matrices in each of the two traces. This is suppressed with respect to the disconnected contribution, which is $N^{2-2g_1-2g_2 - n_1-n_2} $.

\subsection{Permutations interpretation}

A particularly useful advantage of this diagrammatic method is that it makes it clear that there is an equivalence between random tensor networks and a problem of counting permutations. The permutations interpretation that we will show in a moment is immediately related to holographic entanglement in fixed-area states. This gives us another way to see how the entanglement structure in fixed-area states is described precisely by that in random tensor networks.

We will use permutations of $n$ elements. We denote by $\gamma =(1, 2, \cdots ,n)$ the cyclic permutation taking $i \to i+1$ and correspondingly $\gamma ^{-1} $ is its inverse. For any permutation $\tau $, we denote by $C(\tau )$ the number of cycles in the permutation. For example, $C\left( (1,2)(3,5)(4)\right) =3$.

In any diagram built by contractions of Fig.~\ref{fig:Zn_graphically}, let $\tau _1$ be the permutation of $n$ elements such that the right green multi-line of the $i$'th insertion of a density matrix is contracted to the left green multi-line of the $\tau _1(i)$'th insertion. Similarly $\tau _2$ is assigned for the blue counter-part. We then see that the partially transposed moment is given by
\begin{equation}
m_n = \frac{Z_n^{(PT)} }{Z_1^n}
\end{equation}
with the permutations partition function defined by
\begin{equation} \label{eq:2_tensor_perm_partition}
Z_{n} ^{(PT)} =\sum _{\tau _1,\tau _2 \in S_n} L_{C_1} ^{C(\tau _1)} L_A^{C(\gamma \circ \tau _1)} L_{C_2} ^{C(\tau _2)} L_B^{C(\gamma ^{-1} \circ \tau _2)} L_{E_W} ^{C(\tau _1^{-1} \circ \tau _2)} .
\end{equation}
Each element in the sum corresponds to a particular contraction.

Note that we can also calculate the unnormalized permutation partition function \eqref{eq:2_tensor_perm_partition} using averaged moments of the tensor network --- the only modification required is to take the tensors to have unit variance.

We can now simply prove the following familiar claim using this diagrammatic representation together with the large-$N$ expansion above:

\noindent
\emph{Claim} (the triangle inequality): for any two permutations $\sigma ,\tau $ of $n$ elements, we have that
$C(\tau )+C(\tau ^{-1} \circ \sigma )\le C(\sigma ) + n$.

In order to specify when the inequality is saturated, we will use the following terminology. Consider a permutation $c = (c_1,c_2,\cdots ,c_n)$ consisting of a single cycle. It induces an order $\prec$ on the $n$ elements $\{1,2, \cdots ,n\}$ according to the way that they appear in the cycle which is not necessarily the usual ordering. We will say that a permutation $\tau $ is non-crossing in $c$ if (1) any cycle of $\tau $ can be written as $(i_1,\cdots ,i_k)$ with $i_1\prec \cdots \prec i_k$ according to the order of $c$, and (2) there are no elements $a \prec x \prec b \prec y$ such that $a,b$ belong to one cycle of $\tau $ and $x,y$ belong to another cycle.\footnote{Note that the ordering of a cycle $(c_1,c_2,\cdots ,c_n)$ is in fact cyclic, so we could start it at any element. However, this definition does not depend on this, and we can use a linear order.} If we draw the nodes of $c$ on a line and represent cycles of $\tau $ by connecting arcs, this definition just says that the arcs do not cross. We will say in short that \emph{$\tau $ is non-crossing in cycles of $\sigma $} if the cycles of $\tau $ are contained in cycles of $\sigma $, and in each cycle of $\sigma $, $\tau $ is non-crossing.
For the simplest case $c=\gamma=(1,2, \cdots ,n)$, we will simply say that a permutation is non-crossing.\footnote{Note that in this case we can just think about the permutation as a partition, but there are two possible orientations, so we fixed the orientation to be that of $\gamma $.}
With these definitions, the inequality in the claim is saturated iff $\tau $ is non-crossing in $\sigma $.

\begin{figure}[]
\centering
\includegraphics[width=0.6\textwidth]{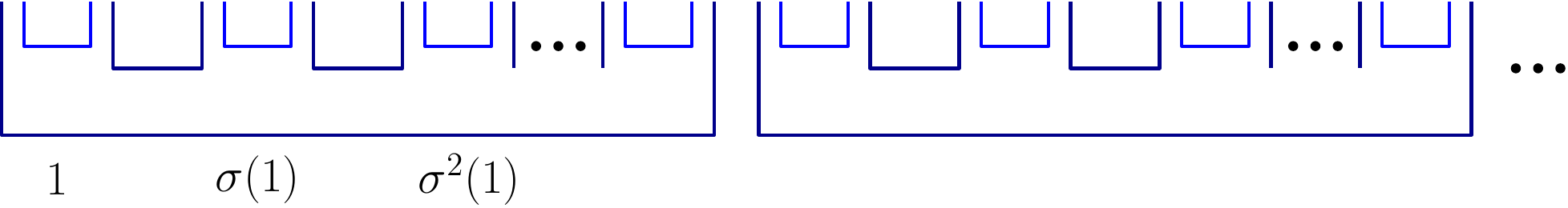}
\caption{Diagram used to prove the triangle inequality.}
\label{fig:triangle}
\end{figure}

The proof of this is simple using the tools above. We use a diagram consisting of the (say) blue part of what we had before (see Fig.~\ref{fig:triangle}). There is one block for each cycle of $\sigma $. For example, the first block has nodes $1,\sigma (1),\sigma ^2(1),\cdots $. The contractions are made according to $\tau $; we contract a right pair of lines of node $i$ with the left pair of node $\tau (i)$. The quantity $C(\tau )+C(\tau ^{-1} \circ \sigma )=C(\tau )+C(\sigma ^{-1} \circ \tau )$ is found by counting the number of closed lines.

In each connected component of size $n_i$, we saw that there is a genus expansion giving that the number of loops (closed lines) is $1-2g_i+n_i$.\footnote{Here we consider unnormalized ``traces'' where no value is assigned to a propagator, so we do not subtract $2n_i$.} As we saw, the disconnected piece dominates and is given by $\sum_i (1-2g_i+n_i) = C(\sigma )+n-2 \sum _i g_i$. This is indeed bounded by $C(\sigma )+n$ and is saturated by planar contractions, which are the permutations mentioned above. This concludes the proof of the claim.

Now, let us come back to the two-tensor network.
One way to think about the problem that we have is to replace the two tensors by a single tensor, and ask if we can apply the results of Section \ref{sec:1_tensor}. That is, by comparing \eqref{eq:1_tensor_state} to \eqref{eq:2_tensor_state} we replace
\begin{equation}
X^{\text{one-tensor}} _{(j)(n)(im)} \leftrightarrow \sum _k X_{ijk} Y_{mnk} .
\end{equation}
The right hand side is a random variable which has a variance
\begin{equation} \label{eq:variance_random_pure_state}
\sum_{k,l} \mean{ X_{i_1j_1k} Y_{m_1n_1k} X^*_{i_2j_2l} Y^*_{m_2n_2l} }  = \left(L_{C_1} L_{C_2} L_AL_B\right) ^{-1} \delta _{i_1i_2} \delta _{j_1j_2} \delta _{m_1m_2} \delta _{n_1n_2} ,
\end{equation}
with the $L_{E_W} $ dependence canceling. This is just the variance needed in Section \ref{sec:1_tensor}. The important difference from the one-tensor case is that this is not a Gaussian variable, so we cannot apply the results there. We expect that when $L_{E_W} $ is large, we may be able to apply these results. Indeed, this is easy to see using the permutations language.
In \eqref{eq:2_tensor_perm_partition} there are two terms that depend only on $\tau _1$, and similarly there are two terms for $\tau _2$. The only interaction between $\tau _1$ and $\tau _2$ is through the $E_W$ term.
For large $L_{E_W} $, \eqref{eq:2_tensor_perm_partition} will be dominated by $\tau _1=\tau _2$. Therefore, in the large $L_{E_W} $ limit, this two-tensor network reduces to the one-tensor network described in Section \ref{sec:1_tensor}.

\begin{figure}
     \centering
     \qquad\quad
     \begin{subfigure}[b]{0.13\textwidth}
         \centering
         \includegraphics[width=\textwidth]{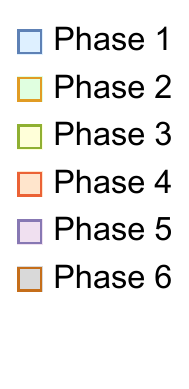}
     \end{subfigure}
     \qquad\qquad\quad\,\,\,\,\,\,\,
     \begin{subfigure}[b]{0.26\textwidth}
         \centering
         \includegraphics[width=\textwidth]{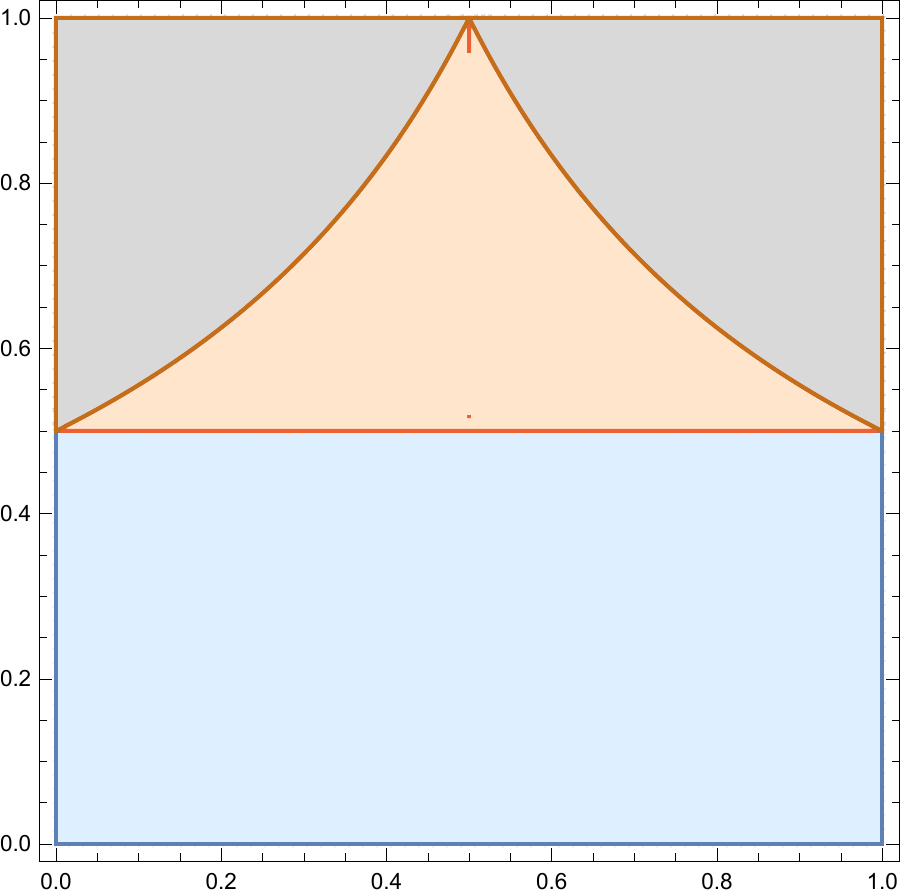}
         \caption{$w_{E_W}\ge 25$}
     \end{subfigure}
     \hfill
     \begin{subfigure}[b]{0.26\textwidth}
         \centering
         \includegraphics[width=\textwidth]{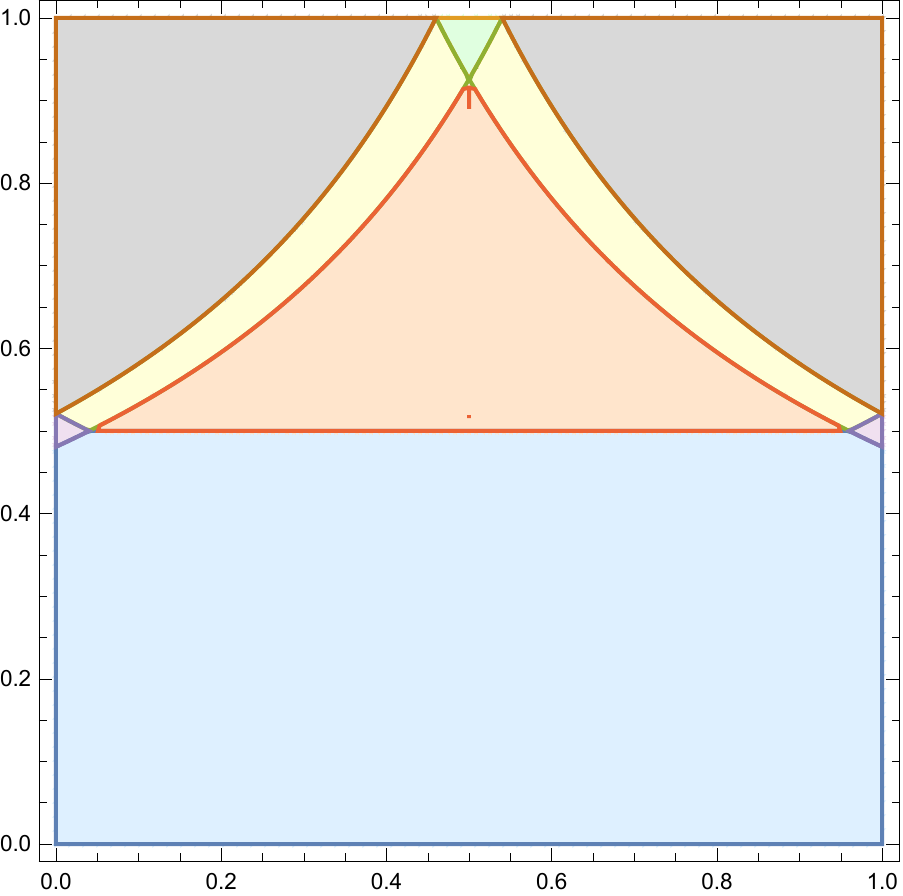}
         \caption{$w_{E_W}=23$}
     \end{subfigure}
          \hfill
     \begin{subfigure}[b]{0.26\textwidth}
         \centering
         \includegraphics[width=\textwidth]{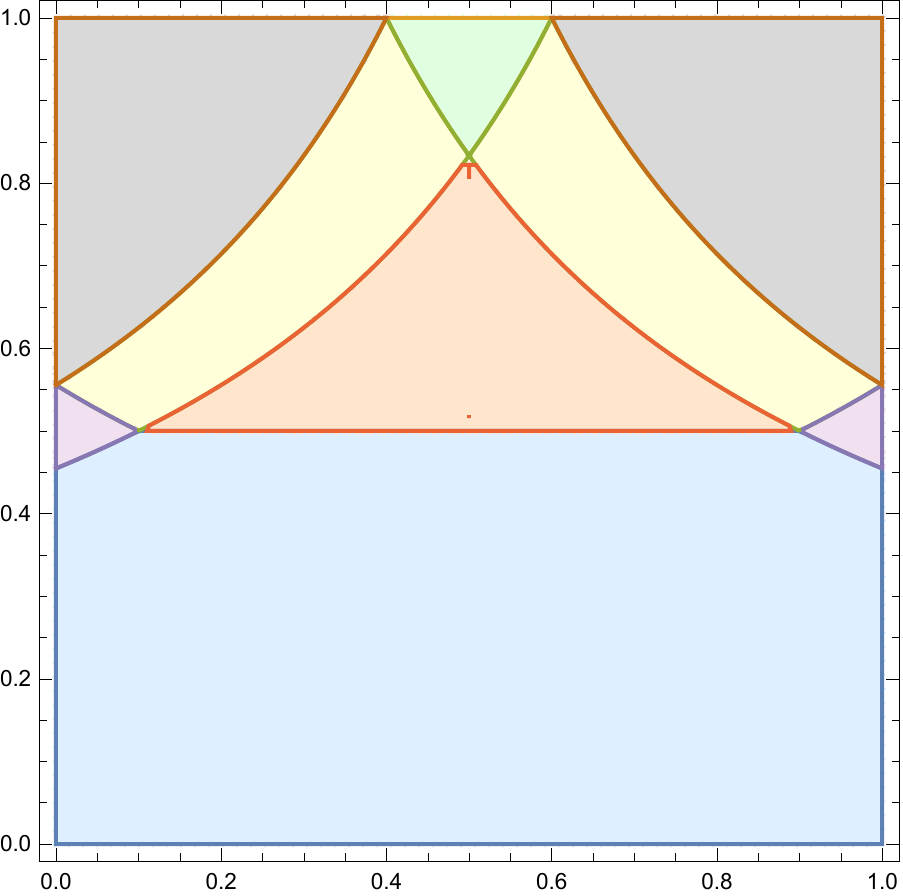}
         \caption{$w_{E_W}=20$}
     \end{subfigure}
          \hfill
     \begin{subfigure}[b]{0.26\textwidth}
         \centering
         \includegraphics[width=\textwidth]{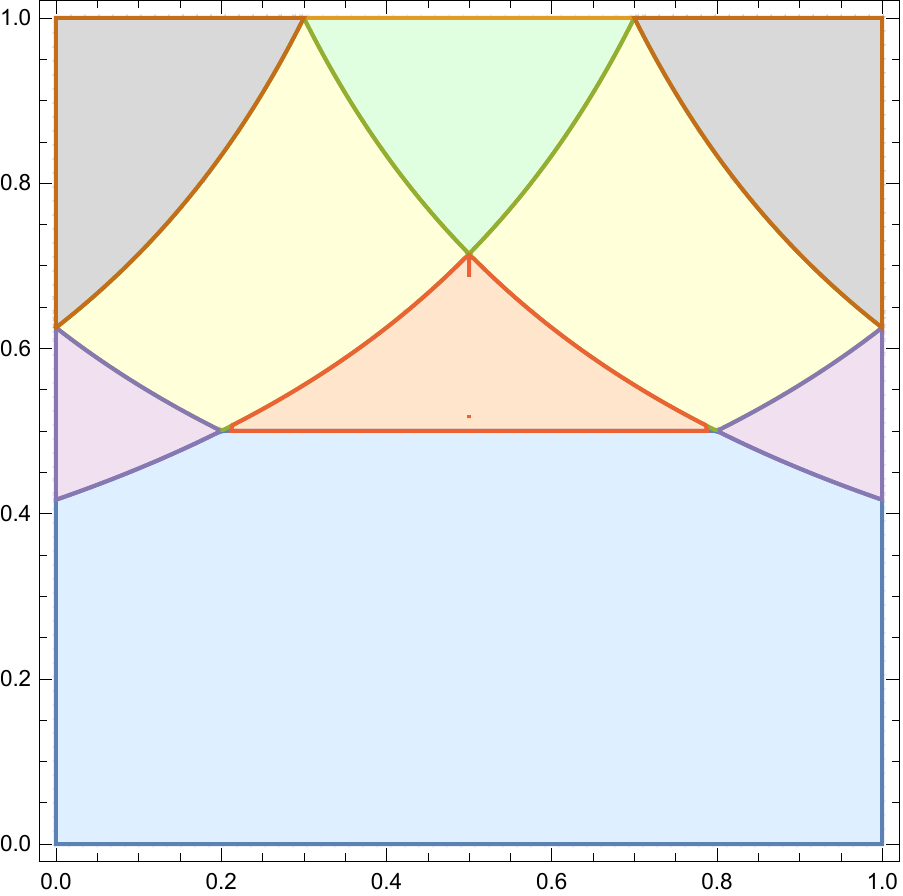}
         \caption{$w_{E_W}=15$}
     \end{subfigure}
          \hfill
     \begin{subfigure}[b]{0.26\textwidth}
         \centering
         \includegraphics[width=\textwidth]{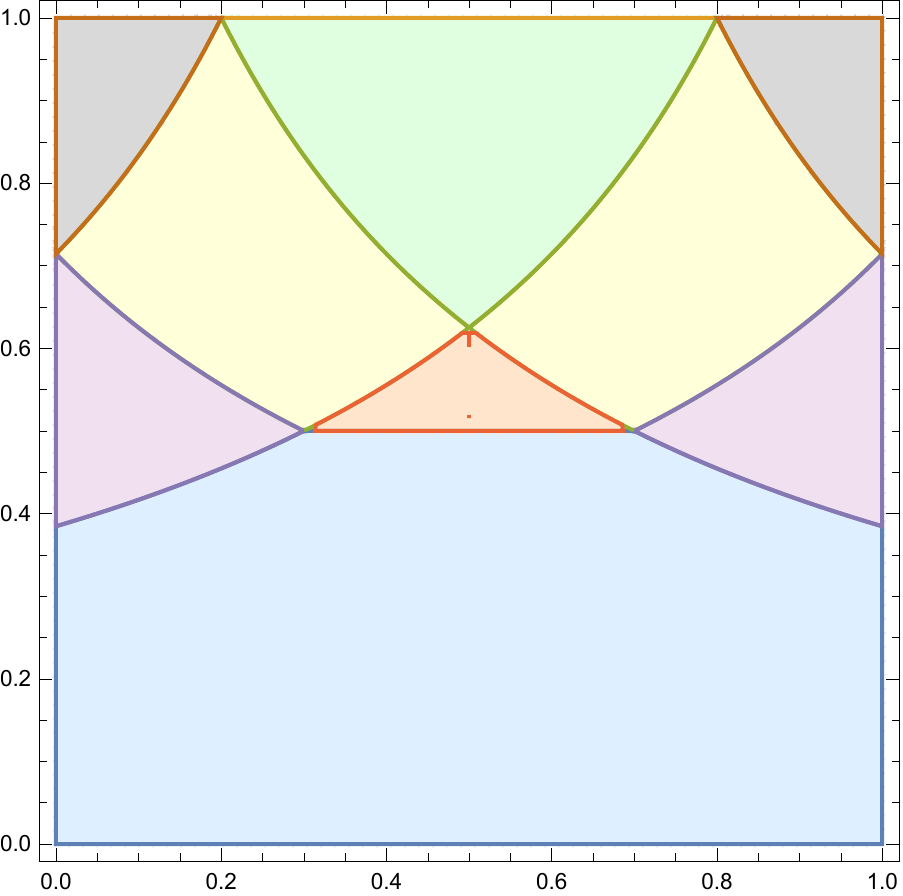}
         \caption{$w_{E_W}=10$}
     \end{subfigure}
          \hfill
     \begin{subfigure}[b]{0.26\textwidth}
         \centering
         \includegraphics[width=\textwidth]{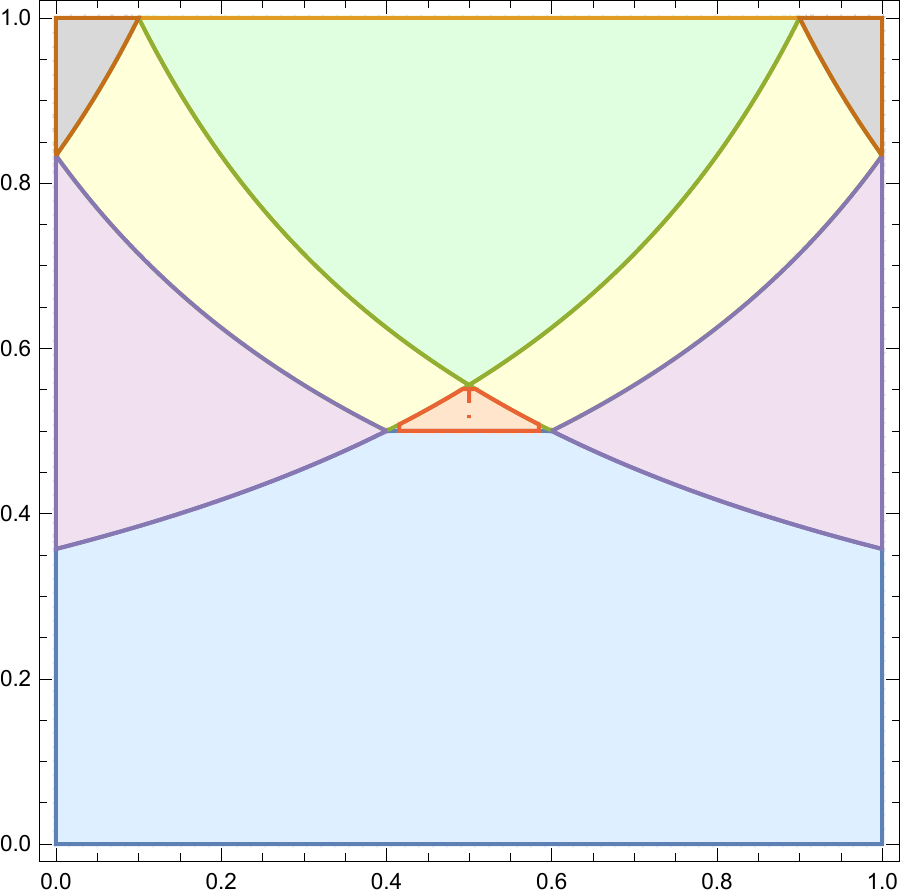}
         \caption{$w_{E_W}=5$}
     \end{subfigure}
          \hfill
     \begin{subfigure}[b]{0.26\textwidth}
         \centering
         \includegraphics[width=\textwidth]{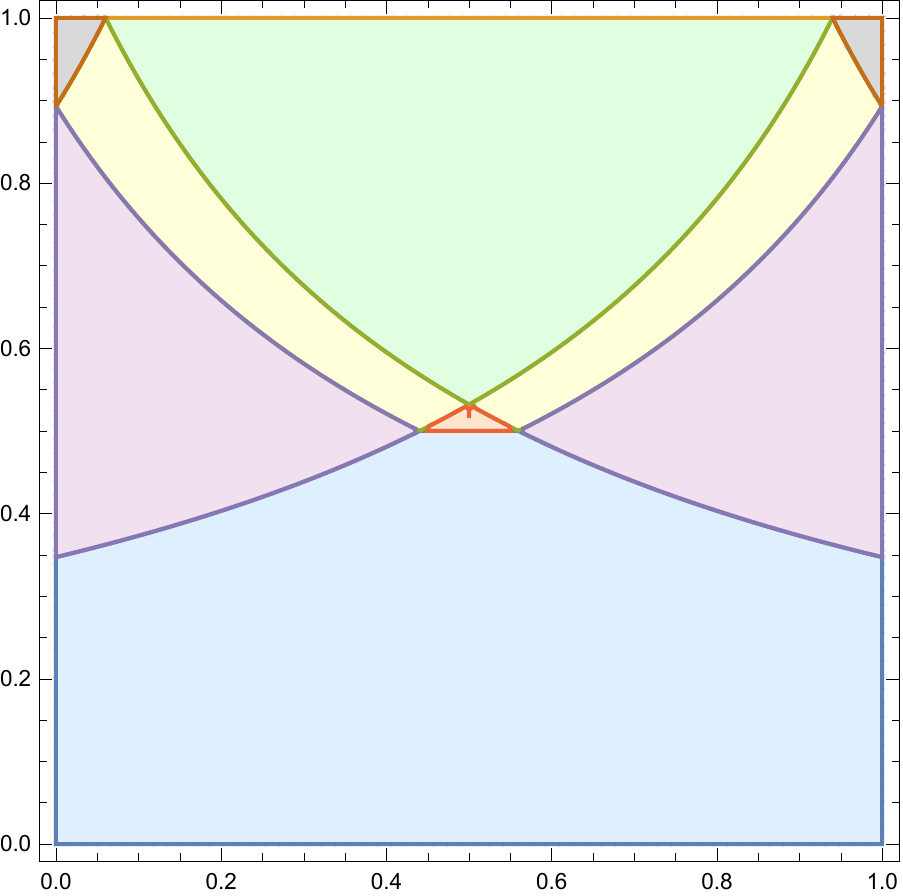}
         \caption{$w_{E_W}=3$}
     \end{subfigure}
          \hfill
     \begin{subfigure}[b]{0.26\textwidth}
         \centering
         \includegraphics[width=\textwidth]{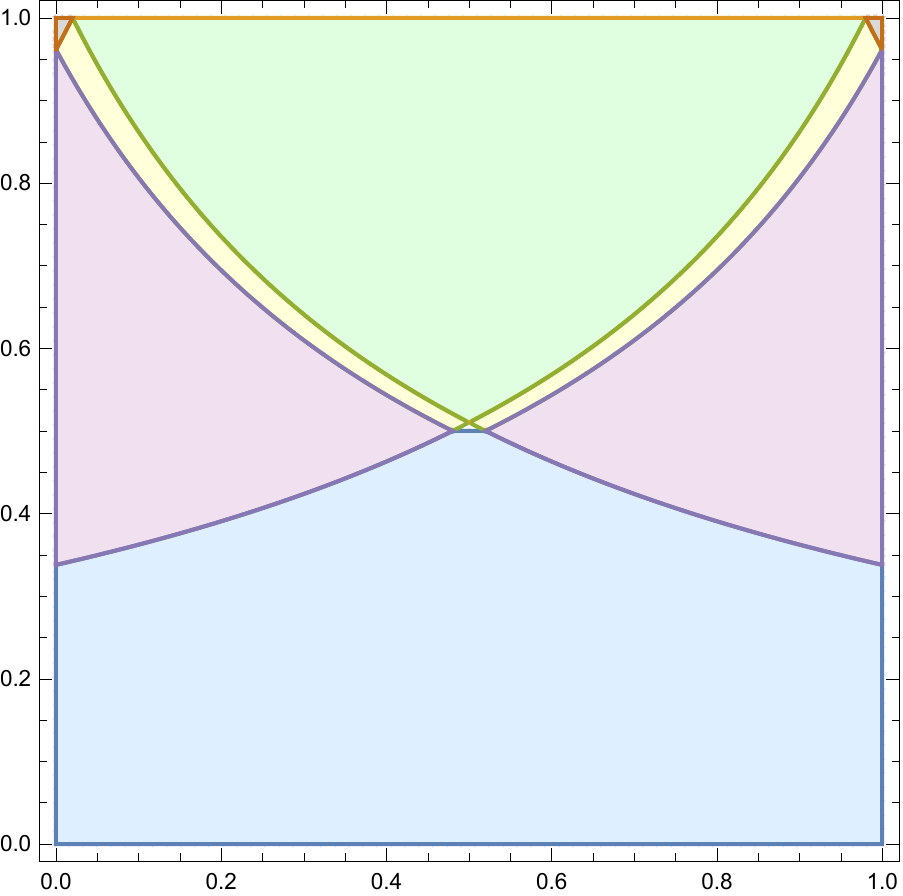}
         \caption{$w_{E_W}=1$}
     \end{subfigure}
    \caption{Plots of different slices of the four-dimensional phase diagram of the two-tensor network. We have taken $w_{C_1}=w_{C_2}$, keep $w_A+w_B=50$ fixed, and vary $w_{E_W}$. The horizontal axis is $\frac{w_A}{w_A+w_B}$ and the vertical axis is $\frac{w_A+w_B}{w_A+w_B+w_C}$ where $w_C=w_{C_1}+w_{C_2}$.}
    \label{fig:2_tensor_phases_plots}
\end{figure}

We can analyze the full phase structure of the two-tensor network. This is done in detail in Appendix \ref{sec:2_tensor_phases}, and the result is summarized in Table \ref{tab:2_tensor_phases}. In the table, we show the range of system sizes that correspond to each phase. In each phase, we indicate the dominant permutations in the large-$N$ limit, where we recall that $w_i$ is the power of $N$ for each $L_i$. As for the moments and the negativity spectrum, we can write them using the one-tensor case, and so we indicate the result using the corresponding phase in the one-tensor network. The precise parameters that should be used in the analogous one-tensor network can be found in Appendix \ref{sec:2_tensor_phases}. In one of the phases, the distribution is related to the corresponding one in the one-tensor case by a rescaling of both the eigenvalues and the distribution. Note, though, that as expected, in general the two-tensor network differs from the one-tensor network. It reduces to a one-tensor network for large enough $L_{E_W}$ as anticipated above. Indeed, in such a case, the two permutations are the same, as can be seen in the table. This happens for phases 1,4, and 6.
We also plot the different phases graphically in Fig.~\ref{fig:2_tensor_phases_plots}.

\begin{table}
\centering
\begin{tabular}{ |c|c|c|c|c| }
\hline
& Range & $\tau _1$ & $\tau _2$ & Analogous one-tensor \\
\hline
1 & \begin{tabular}{l}
$\begin{cases}
w_A<w_{C_1} \\
w_B<w_{C_2} +w_{E_W} \\
w_A+w_B < w_{C_1} +w_{C_2} 
\end{cases}$ 
or \\
$\begin{cases}
w_B<w_{C_2} \\
w_A<w_{C_1} +w_{E_W} \\
w_A+w_B < w_{C_1} +w_{C_2} 
\end{cases}$
\end{tabular}
 & id & id & Phase I \\
\hline
2 & $\begin{aligned}
& w_A> w_{C_1}+w_{E_W} \\
& w_B > w_{C_2} + w_{E_W} 
\end{aligned}$ & $\gamma ^{-1}$ & $\gamma $ & Phase II \\
\hline
\multirow{6}{*}{3} & $\begin{aligned}
& w_{E_W} <w_A+w_{C_1} \\
&w_{C_1} <w_A+w_{E_W} \\
& w_A < w_{C_1} +w_{E_W} \\
& w_B>w_{C_2} +w_{E_W}
\end{aligned}$ & $NC_2$ & $\gamma $ & \multirow{6}{*}{Rescaled phase III} \\
\cline{2-4}
& $\begin{aligned}
& w_{E_W} <w_B+w_{C_2} \\
&w_{C_2} <w_B+w_{E_W} \\
& w_B < w_{C_2} +w_{E_W} \\
& w_A>w_{C_1} +w_{E_W} 
\end{aligned}$ & $\gamma^{-1}$ & $NC_2$ & \\
\hline
\multirow{3}{*}{4} & $\begin{aligned}
& |w_{A} -w_{C_1}| <w_B-w_{C_2} <w_A+w_{C_1} \\
& w_{E_W} >w_B-w_{C_2} 
\end{aligned}$ & $NC_2$ & $=\tau_1$ & \multirow{3}{*}{Phase III} \\
\cline{2-4}
& $\begin{aligned}
& |w_{B} -w_{C_2} |<w_A-w_{C_1} <w_B+w_{C_2} \\
& w_{E_W} >w_A-w_{C_1} 
\end{aligned}$ & $NC_2$ & $=\tau_1$ & \\
\hline
\multirow{3}{*}{5} & $\begin{aligned}
& w_{C_1} > w_A+w_{E_W} \\
& w_B>w_{C_2} +w_{E_W} 
\end{aligned}$ & id & $\gamma$ & \multirow{3}{*}{Phase I} \\
\cline{2-4}
& $\begin{aligned}
& w_{C_2} > w_B+w_{E_W} \\
& w_A>w_{C_1} +w_{E_W} 
\end{aligned}$ & $\gamma^{-1}$ & id & \\
\hline
\multirow{3}{*}{6} & $\begin{aligned}
& w_{E_W} >w_B+w_{C_2} \\
& w_A > w_B+w_{C_1} +w_{C_2} 
\end{aligned}$ & $\gamma^{-1}$ & $=\tau_1$ & \multirow{3}{*}{Phase II} \\
\cline{2-4}
& $\begin{aligned}
& w_{E_W} >w_A+w_{C_1} \\
& w_B > w_A+w_{C_1} +w_{C_2} 
\end{aligned}$ & $\gamma$ & $=\tau_1$ & \\
\hline
\end{tabular}
\caption{Phases of a two-tensor network. We show the range of each phase corresponding to the intersection of the inequalities and the values of the permutations that give the dominant contribution. In these cases, the behavior in each phase can be mapped to a possibly rescaled phase of the one-tensor network, and this is shown in the table. A permutation is in $NC_2$ if it is non-crossing and consists of only 2-cycles, unless it is of odd size and in such a case it also has a single 1-cycle.}
\label{tab:2_tensor_phases}
\end{table}

\subsection{Numerics}

\begin{figure}
    \centering
    \includegraphics[width = .48\textwidth]{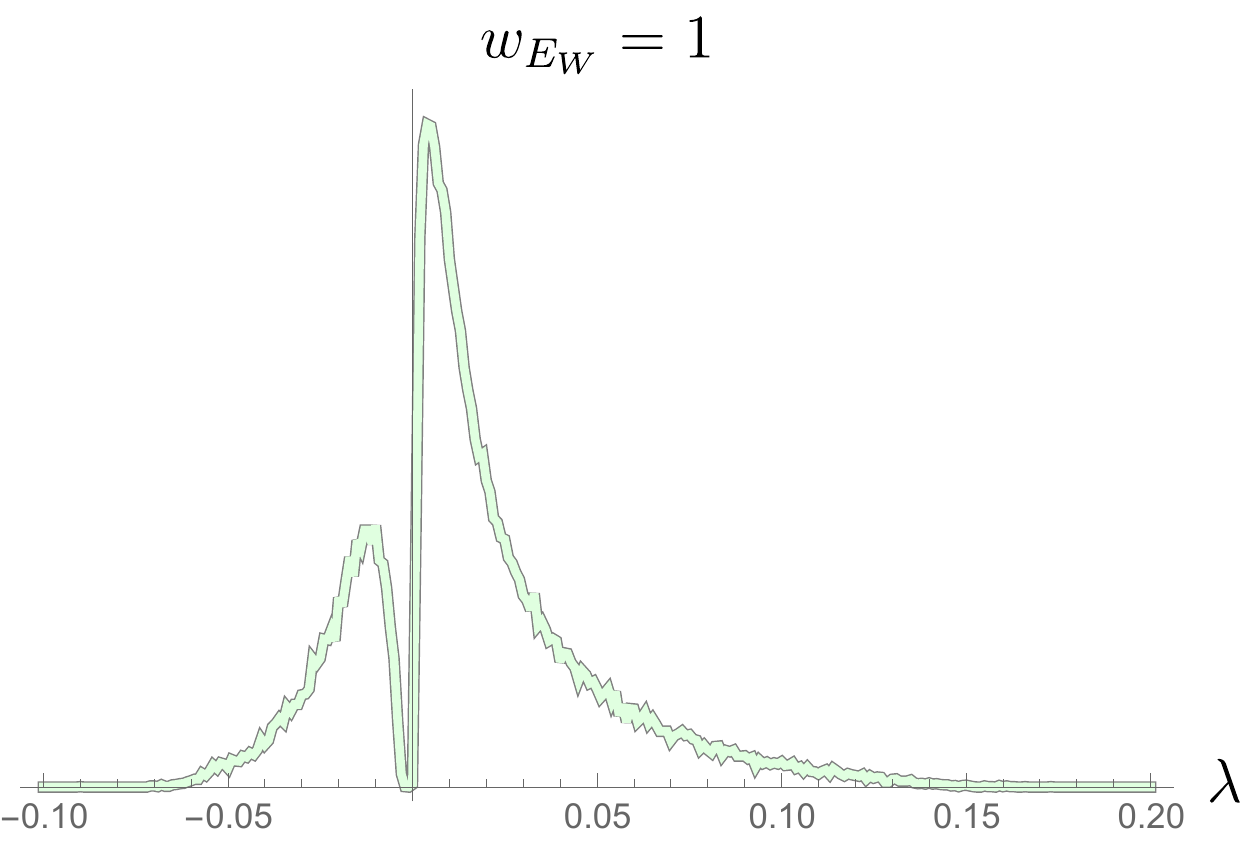}
    \includegraphics[width = .48\textwidth]{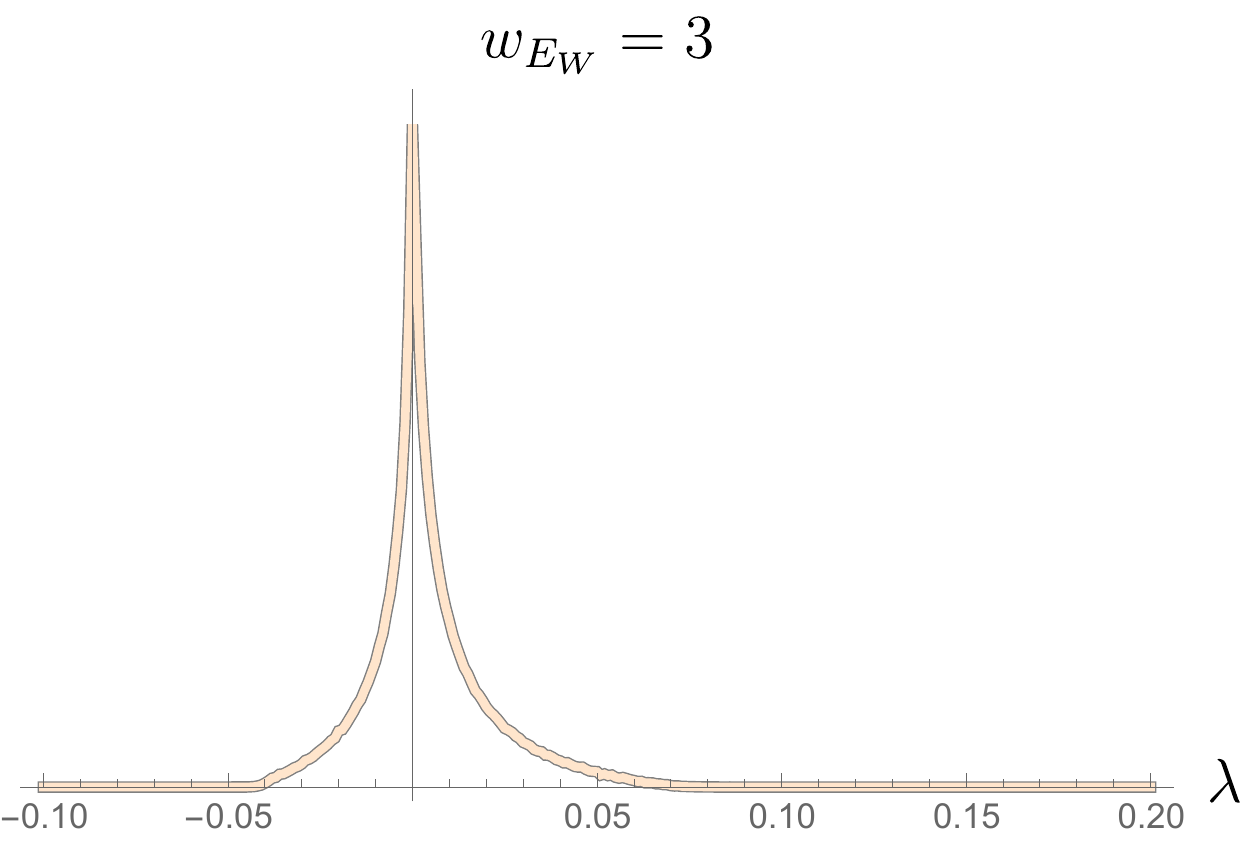}
    \includegraphics[width = .48\textwidth]{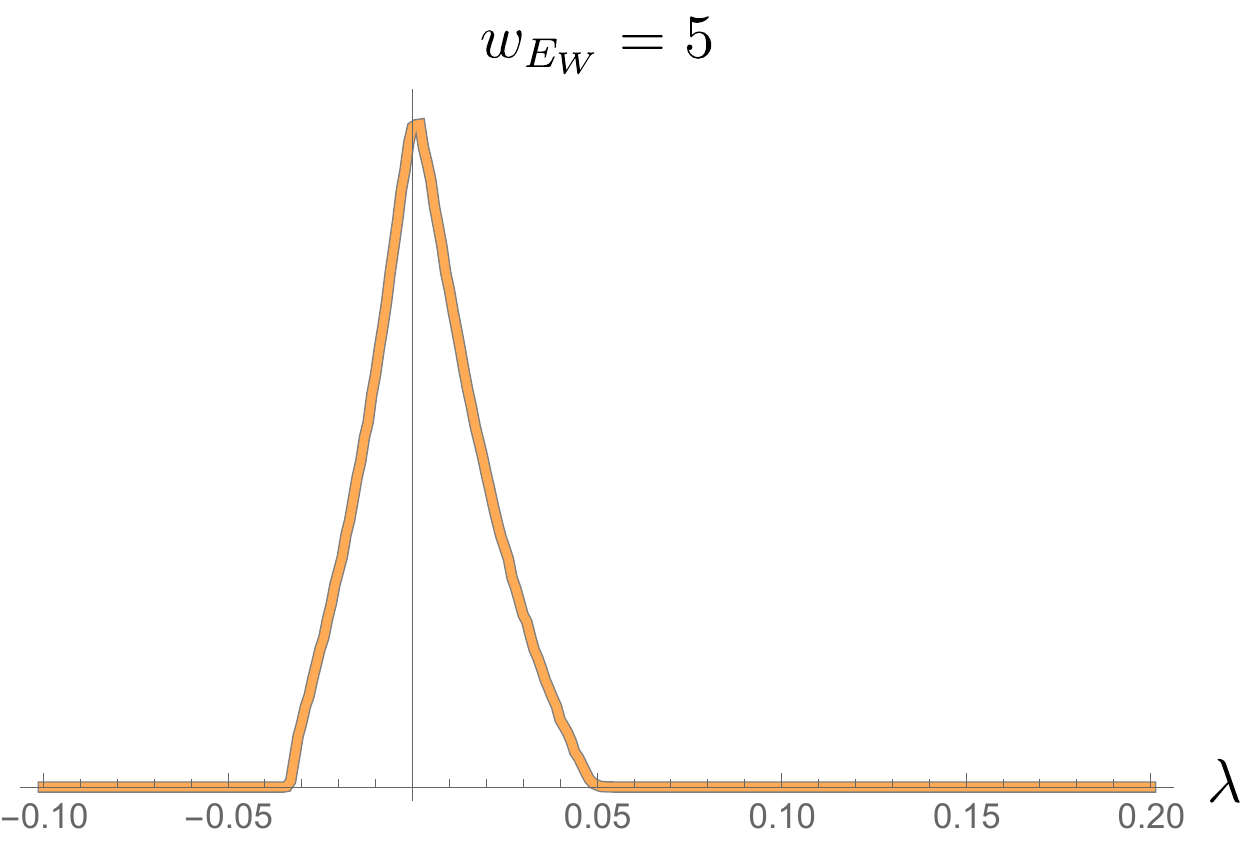}
    \includegraphics[width = .48\textwidth]{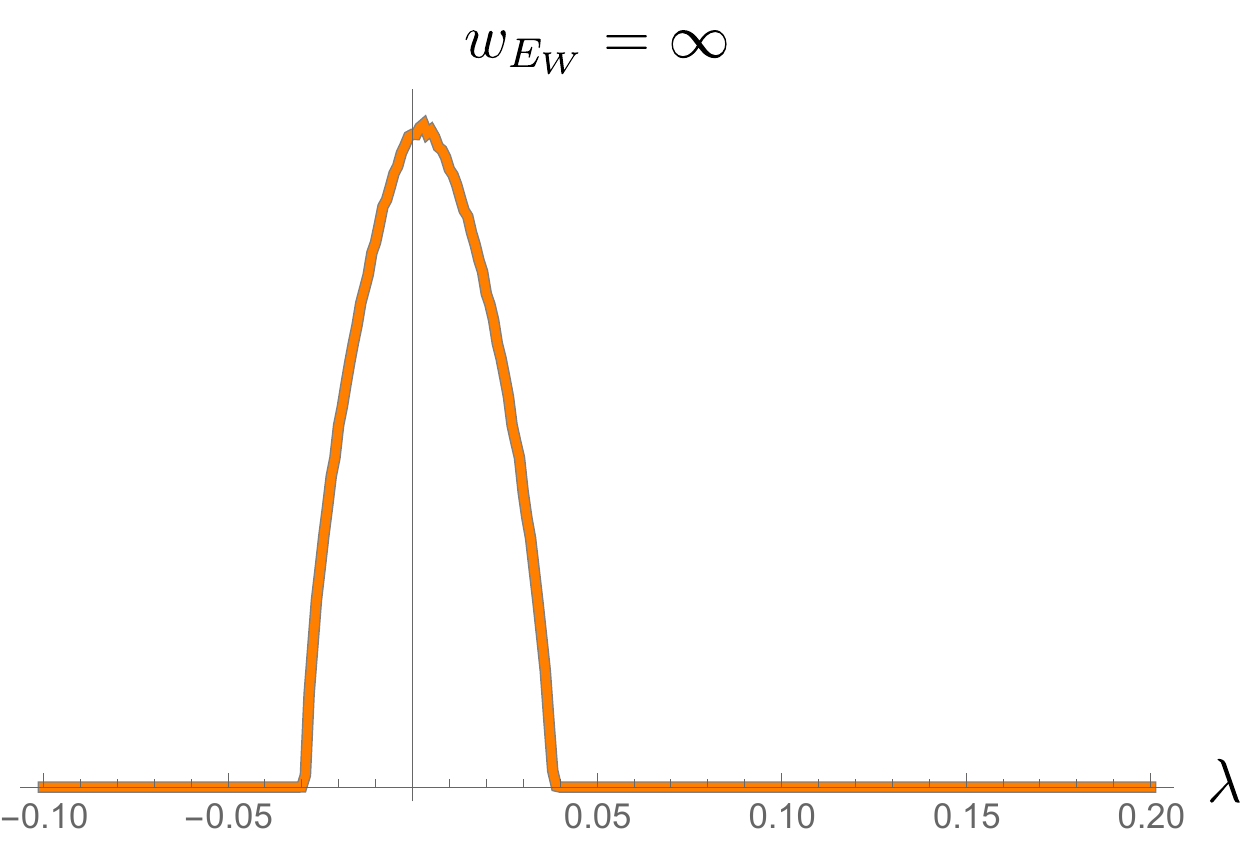}
\caption{The negativity spectrum for the two-tensor network as we vary $w_{E_W}$ and keep $w_A = w_B = 4$, $w_{C_1} = w_{C_2} = 2$, and $N= 2$. There is a topological transition between Phase 2 (upper left) with two connected components and Phase 4 (lower right) with a single connected component. The colors of the plots match the colors of the phases in Fig.~\ref{fig:2_tensor_phases_plots}. We average over $10^3$ realizations.}
    \label{two_tensor_spectrum}
\end{figure}

\begin{figure}
    \centering
    \includegraphics[width = .48\textwidth]{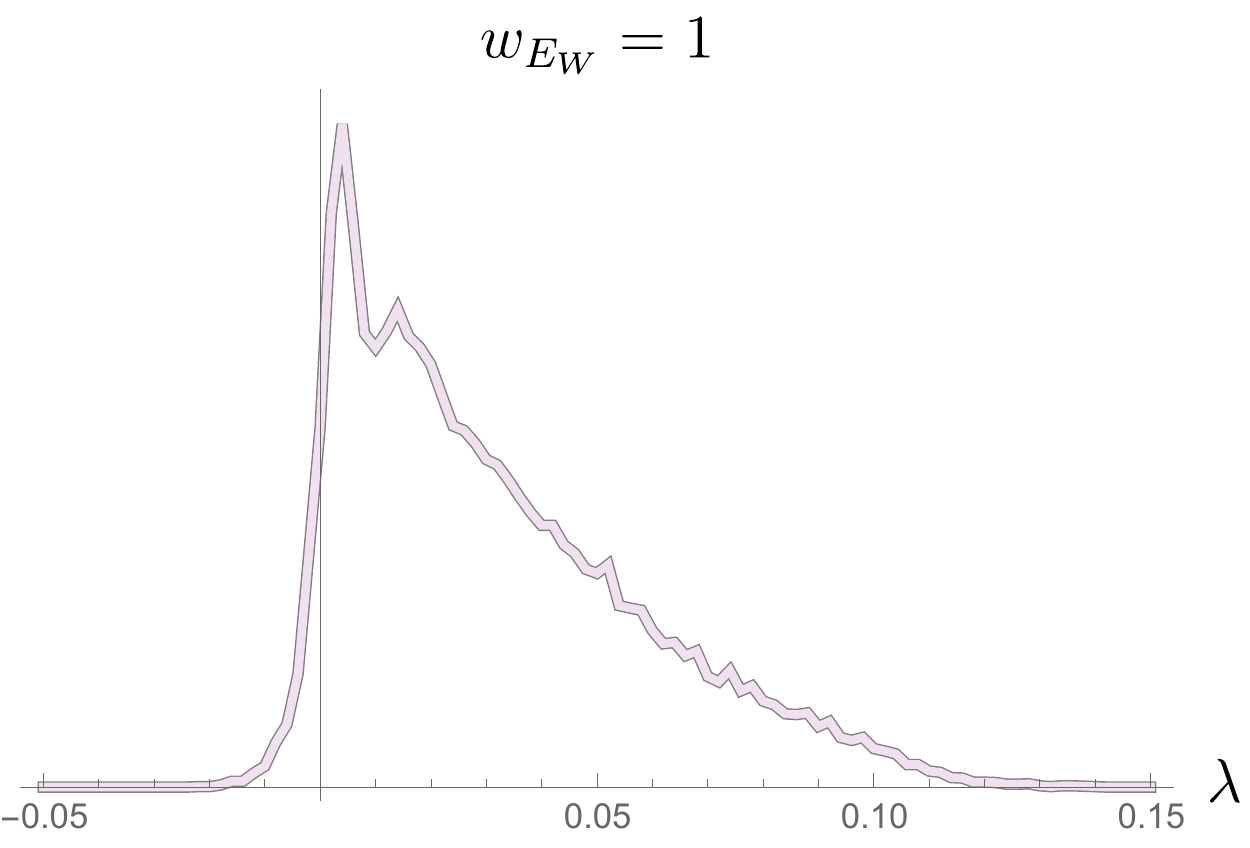}
    \includegraphics[width = .48\textwidth]{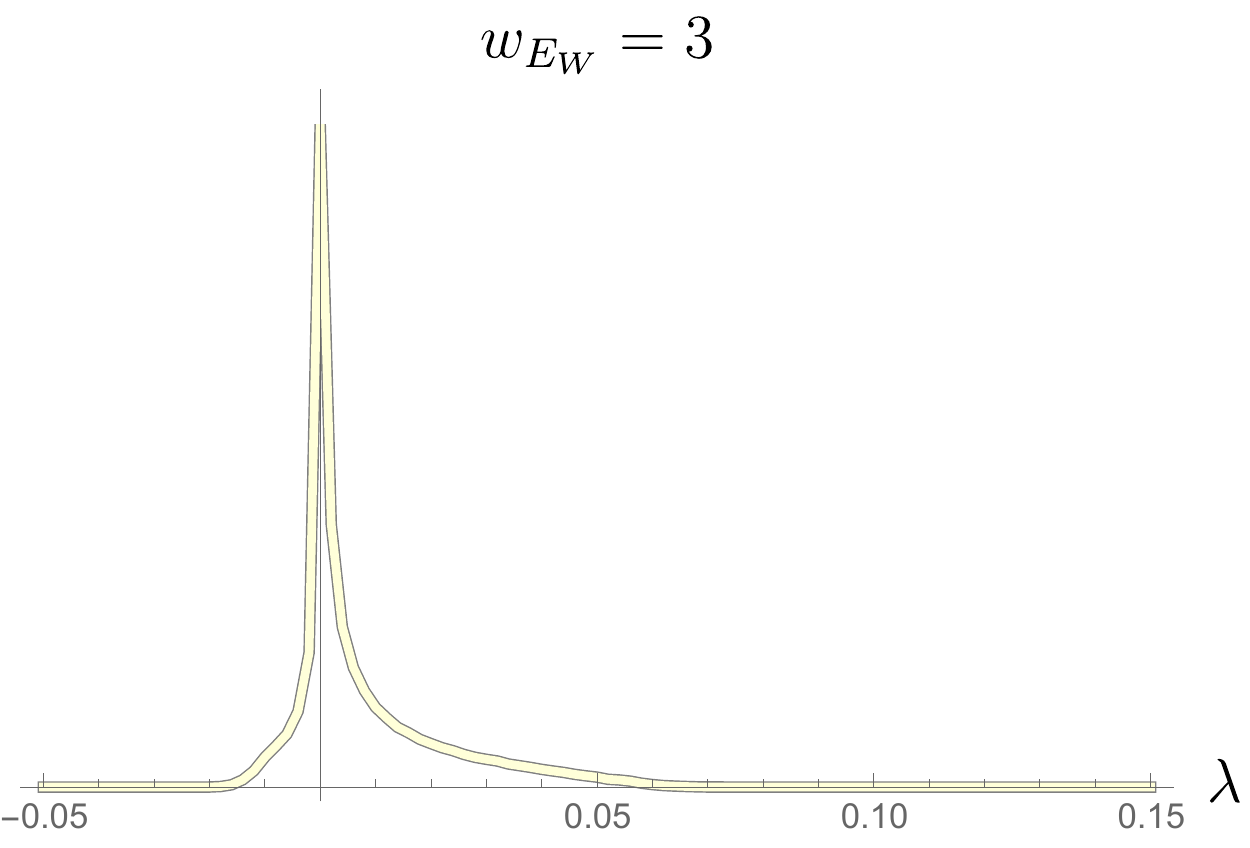}
    \includegraphics[width = .48\textwidth]{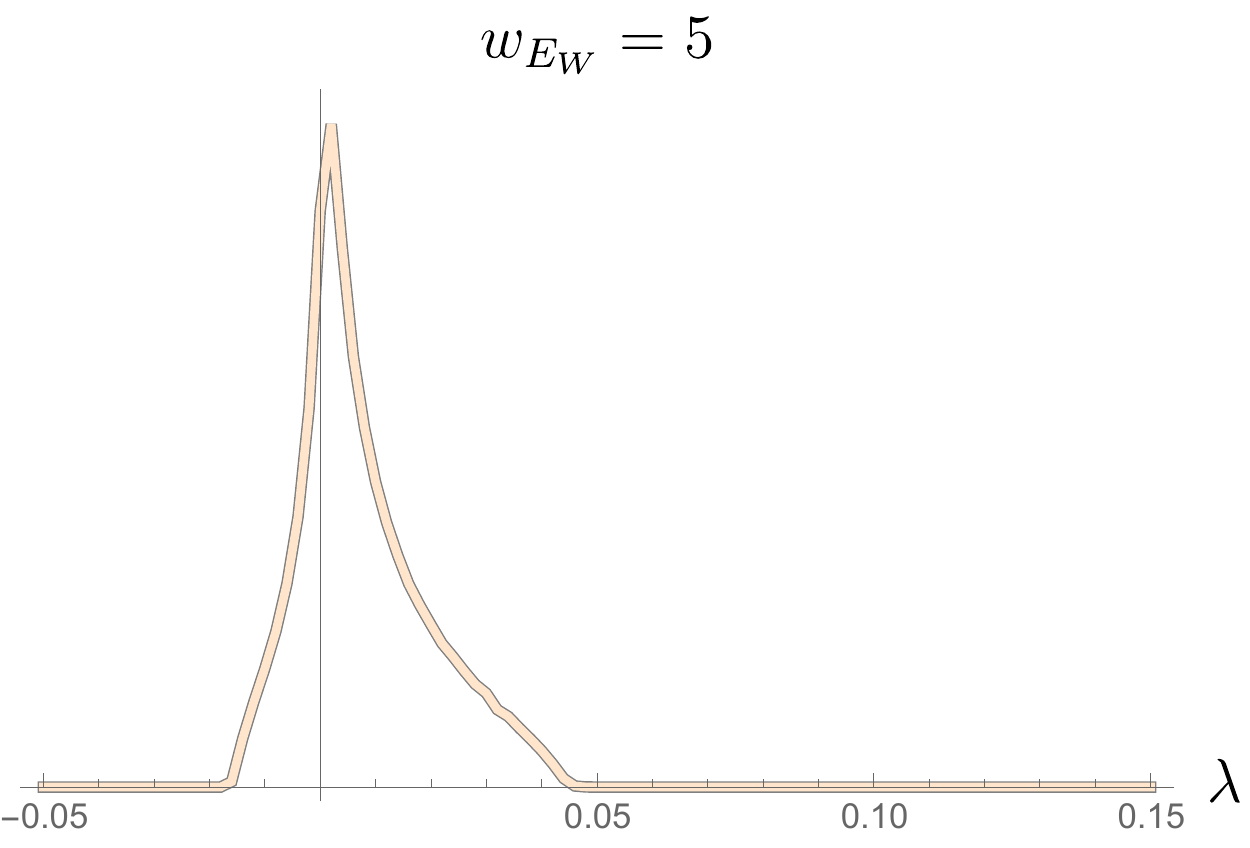}
    \includegraphics[width = .48\textwidth]{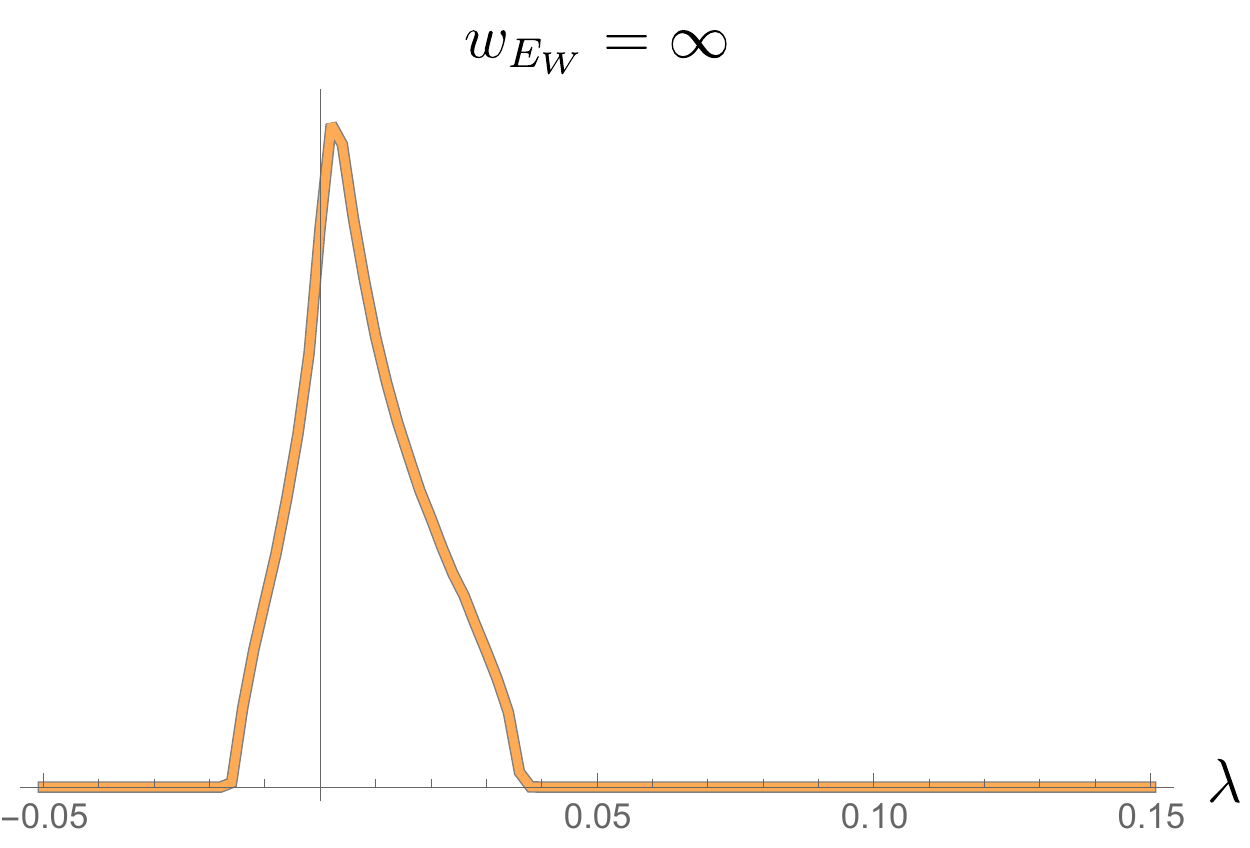}
    \caption{The negativity spectrum for the two-tensor network as we vary $w_{E_W}$ and keep $w_A = 1$,  $w_B = 6$, $w_{C_1} = w_{C_2} = 3$, and $N=2$. There is the 
    transition from vanishing negativity (Phase I in the one-tensor network) between Phase 5 (upper left) to Phase 3 to Phase 4 (lower right) where the eigenvalues gain support on the negative real axis.  The colors of the plots match the colors of the phases in Fig.~\ref{fig:2_tensor_phases_plots}. We average over $10^3$ realizations.}
    \label{two_tensor_spectrum2}
\end{figure}

We numerically study the entanglement of the two-tensor network using exact diagonalization. This strongly limits the probeable Hilbert space dimensions. In Figs.~\ref{two_tensor_spectrum} and \ref{two_tensor_spectrum2}, we show how the negativity spectrum is deformed as we tune $L_{E_W}$. When the dimension of $E_W$ approaches the total dimension of the external Hilbert space, the tensor network acts as one large random tensor. When the dimension decreases, the spectrum undergoes a variety of phase transitions due to the loss of connectivity in the graph.

\section{Negativity of pure states} \label{sec:pure_states}

We saw that in the limit of large dimensions, the two-tensor network relates to a one-tensor network. An interesting case to consider is when $\mathcal{H} _{C_1} $ and $\mathcal{H} _{C_2} $ are in fact empty, so that the state on $AB$ is pure. This case is quite rich, having various regimes with different results. 
For general dimensions, we will find that the system is governed by a highly non-trivial Schwinger-Dyson equation.


Before that, 
let us start by showing what is the negativity spectrum for a general pure state.

\subsection{Negativity spectrum of a pure state}

For any density matrix, since it is positive semi-definite, the entanglement spectrum is supported on the non-negative real line. As we saw, the negativity is defined by considering the partially-transposed density matrix and consequently is supported on the negative real line as well.

Consider a bi-partite system $AB$ that is in a global pure state. We will see that the negativity spectrum is fully determined by the entanglement spectrum, and in fact the eigenvalues of the partially transposed density matrix are simply determined by the entanglement spectrum.

A pure state can be written using the Schmidt decomposition as
\begin{equation}
|\rho \rangle  = \sum _{i=1} ^L c_i |e_i\rangle |f_i\rangle 
\end{equation}
where $L \le L_A,L_B$, and $|e_i\rangle $ and $|f_i\rangle $ are orthonormal vectors of subsystems $A$ and $B$, respectively. The corresponding density matrix is then
\begin{equation}
\rho  = \sum _{ij} c_i c_j^* |e_i\rangle |f_i\rangle \langle e_j| \langle f_j| .
\end{equation}
The reduced density matrix to system $A$ is then a diagonal matrix with values $|c_i|^2$ on the diagonal. Therefore, the entanglement spectrum is
\begin{equation}
\text{Entanglement spectrum: } \{|c_i|^2:i=1,\cdots ,L\} \text{ plus $L_A-L$ zeroes}.
\end{equation}

The partially transposed density matrix is
\begin{equation}
\rho ^{T_A} = \sum_{ij} c_i c_j^* |e_j\rangle |f_i\rangle \langle e_i|\langle f_j| .
\end{equation}
First, note that $|e_i\rangle |f_i\rangle$ 
is an eigenvector of $\rho ^{T_A}$ with eigenvalue $|c_i|^2$. This shows that the entanglement spectrum is contained in the negativity spectrum.
A general eigenvector $v = \sum _{ij} d_{ij} |e_i\rangle |f_j\rangle $ will satisfy
\begin{equation}
\rho ^{T_A} v = \sum _{ij} d_{ij} c_ic_j^* |e_j\rangle |f_i\rangle  = \sum _{ij} d_{ji} c_i^* c_j |e_i\rangle |f_j\rangle  = \lambda  \sum _{ij} d_{ij} |e_i\rangle |f_j\rangle  ,
\end{equation}
so that
\begin{equation} \label{eq:pure_neg_ev_eq}
\lambda d_{ij} =d_{ji} c_i^* c_j
\end{equation}
for all $i,j$. Now, fixing $k>l$ we claim that we have two eigenvalues for the two solutions of $\lambda ^2=|c_k|^2 |c_l|^2$. Indeed, the corresponding two vectors for the two solutions of $\lambda $
\begin{equation}
|e_k\rangle |f_l\rangle  + \frac{1}{\lambda } c_l^* c_k |e_l\rangle |f_k\rangle 
\end{equation}
are eigenvectors of $\rho ^{T_A}$ as they satisfy the equations \eqref{eq:pure_neg_ev_eq}. Note that for any $c_k=0$, the vectors $|e_k\rangle |f_l\rangle $ and $|e_l\rangle |f_k\rangle $ for all $k$ are eigenvectors with zero eigenvalues.
We have therefore found all the eigenvectors of $\rho ^{T_A}$, so the negativity spectrum is
\begin{equation}
\text{Negativity spectrum: } \{|c_i|^2:i=1,\cdots ,L\} \cup \{ \pm |c_i| \cdot |c_j|:1 \le i < j \le L \} \text{ plus } (L_AL_B-L^2) \text{ zeroes}.
\end{equation}

The negativity spectrum for pure states can thus be obtained immediately from the entanglement spectrum: any entanglement eigenvalues appear also in the negativity spectrum, and for any two distinct entanglement eigenvalues $\lambda _i,\lambda _j$ there is a pair of eigenvalues $\pm \sqrt{\lambda _i\lambda _j}$ in the negativity spectrum, and the rest are zero eigenvalues. The eigenvalues inherited from the entanglement spectrum may also vanish. This result was also noted in \cite{Ruggiero_2}.

In terms of the density of states of the entanglement spectrum $\es (\lambda )$, we therefore find that the density of the negativity spectrum $\ns(\lambda )$ is
\begin{equation} \label{eq:pure_neg_dos}
    \ns(\lambda ) = \int _{-\infty }^{\infty }d\lambda ' \left\lvert \frac{\lambda }{\lambda '} \right\rvert \es (\lambda \lambda ') \es \left( \frac{\lambda }{\lambda '}\right)  +
    \frac{1}{2} \left( \es (\lambda )-\es (-\lambda )\right) 
    + c \cdot \delta (\lambda ) 
\end{equation}
where $c$ is a constant to be determined by $\int d\lambda \, \ns(\lambda )=L_A L_B$. The first term just gives for any two eigenvalues of $\rho $ an eigenvalue in $\ns(\lambda )$ that is the square root of their product as we can see by the arguments of $\es $. The coefficient is simply a measure factor for $\es $ used with this choice of arguments. For any eigenvalue $\lambda $ of $\rho $, this term gives both a $\lambda $ and a $-\lambda $ eigenvalues in $\ns(\lambda )$ while we need only the positive part, and this is corrected by the second and third terms which are anti-symmetric in $\lambda \to -\lambda $. We do not keep track of the zero eigenvalues just as above, as those are determined according to the condition that we mentioned.

\subsection{Two-tensor pure state} \label{sec:two_tensor_pure}

The permutation interpretation allows us to find the negativity in various limits quite easily, so here we explore the various regimes.

In the pure state limit, the moments of the reduced density matrix 
$\rho$ on $AB$ are
\begin{align} \label{eq:2_tensor_pure_moment}
    \mean{\Tr \left(\rho^{T_A}\right)^n} 
    = \sum_{\tau_1,\tau_2 \in S_n} L_{A}^{C(\gamma  \circ \tau_1) -n} L_{B}^{C(\gamma ^{-1}\circ \tau_2) -n} L_{E_W}^{C(\tau_1^{-1}\circ \tau_2) -n}.
\end{align}
When $L_{A},L_{B} \gg L_{E_W}$, $\tau_1 = \gamma ^{-1}$ and $\tau_2 = \gamma $ dominate the sum and we approximately have for $n>0$
\begin{align}
    \mean{\Tr \left(\rho^{T_A}\right)^n }
    = \begin{cases}
        L_{E_W}^{1-n} & n \in 2\mathbb{Z}+1,
        \\
        L_{E_W}^{2-n} & n \in 2\mathbb{Z}.
    \end{cases}
\end{align}
This leads to a resolvent of
\begin{align}
    R(z) &= \sum_{n = 0}^{\infty} \frac{L_{E_W}^{1-n}\left(L_{E_W}(1+(-1)^n) +  (1-(-1)^n)\right)}{2z^{n+1}}
    \nonumber
    \\
    &= \frac{L_{E_W}^4 z +L_{E_W}^2}{L_{E_W}^2 z ^2-1},
\end{align}
such that the negativity spectrum is\footnote{We did not keep track of the zero eigenvalues or the zeroth moment and restored them here using $\int d\lambda \, \ns(\lambda )=L_AL_B$.}
\begin{equation}
\ns(\lambda ) = \frac{L_{E_W}^2-L_{E_W}}{2} \delta (\lambda +L_{E_W}^{-1} ) + \frac{L_{E_W}^2+L_{E_W}}{2} \delta (\lambda -L_{E_W}^{-1} ) + (L_AL_B-L_{E_W}^2)\delta(\lambda) .
\end{equation}
Subleading saddles lead to a finite width structure in the distribution at finite $N$. The negativity is
\begin{align}
    \mathcal{E} = \log (L_{E_W} -1) \simeq \log L_{E_W}.
\end{align}

Next, consider $L_{E_W} \gg L_{A}L_{B}$. In this case $\tau_1 =\tau_2$ dominates the sum and we simplify to 
\begin{align}
    \mean{\Tr \left(\rho^{T_A}\right)^n}
    = \sum_{\tau \in S_n} L_{A}^{C(\gamma \circ \tau) -n} L_{B}^{C(\gamma ^{-1}\circ \tau) -n}.
\end{align}
This is equivalent to the pure state limit of the single-tensor network so the negativity spectrum is a sum of two Marchenko–Pastur laws.

Now, consider the mixed limit $L_{A} \gg L_{B}L_{E_W}$ but $L_{B}\sim L_{E_W}$.\footnote{The systems $A$ and $B$ appear here in a symmetric way, so the case $L_B \gg L_A L_{E_W}$ is just the same with exchanging $A \leftrightarrow B$.} We have $\tau_1 = \gamma ^{-1}$
\begin{align}
    \mean{\Tr \left(\rho^{T_A}\right)^n}
    = \sum_{\tau_2 \in S_n}  L_{B}^{C(\gamma ^{-1}\circ \tau_2) -n} L_{E_W}^{C(\gamma \circ \tau_2) -n}.
\end{align}
Now, let us shift $\tau_2\rightarrow\gamma ^{-1}\circ\tau$ to get 
\begin{align}
    \mean{\Tr \left(\rho^{T_A}\right)^n}
    = \sum_{\tau \in S_n}  L_{B}^{C(\gamma ^{-2}\circ \tau) -n} L_{E_W}^{C(\tau) -n}.
\end{align}
We would like to maximize $C(\tau )+C(\tau ^{-1} \circ \gamma ^2)$. By the triangle inequality we know this is bounded by $C(\gamma ^2)+n$ and maximized for $\tau $ non-crossing in $\gamma ^2$.
For odd $n$, $C(\gamma ^{2}) = 1$, so, up to an unusual ordering of nodes, the non-crossing permutations will dominate. Moreover, it is known that the number of non-crossing permutations of $n$ elements with $k$ cycles is given by the Narayana number
\begin{equation}
N_{n,k} = \frac{1}{n} \binom{n}{k} \binom{n}{k-1} .
\end{equation}
We can therefore reorganize the sum according to the number of cycles, as
\begin{align}
    \mean{\Tr \left(\rho^{T_A}\right)^n}
&= \sum_{k = 1}^n N_{n,k}  L_{B}^{1-k} L_{E_W}^{k -n}
    \\
    &=\begin{cases}
     L_{E_W}^{1-n} \, _2F_1\left(1-n,-n;2;\frac{L_{E_W}}{L_{B}}\right), & L_{E_W} <L_B \\
     L_{B}^{1-n} \, _2F_1\left(1-n,-n;2;\frac{L_{B}}{L_{E_W}}\right), & L_{E_W} >L_B
    \end{cases}.
\end{align}
For even $n$, $\gamma ^{2}$ has two cycles, consisting of the even and odd sites. Thus, the dominant saddles will be of the form $NC_{n/2}\times NC_{n/2}$ with $NC_{n/2}$ being the non-crossing permutations of $n/2$ elements. Each noncrossing permutation acts on the even or odd copies. We can then reorganize the sum as 
\begin{align}
    \mean{\Tr \left(\rho^{T_A}\right)^n}
    &= \sum_{k_1,k_2 = 1}^{n/2} N_{n/2,k_1}N_{n/2,k_2}  L_{B}^{2-k_1 -k_2} L_{E_W}^{k_1+k_2 -n}
    \nonumber
    \\
    &=
    \begin{cases}
    L_{E_W} ^{2-n} \cdot {}_2F_1 \left( 1-\frac{n}{2} ,-\frac{n}{2} ;2;\frac{L_{E_W} }{L_B} \right) ^2, & L_{E_W} < L_B \\
        L_{B} ^{2-n} \cdot {}_2F_1 \left( 1-\frac{n}{2} ,-\frac{n}{2} ;2;\frac{L_{B} }{L_{E_W}} \right) ^2, & L_{E_W} > L_B
    \end{cases}.
\end{align}
We can take the replica limit of these even moments to find
\begin{equation}
\mathcal{E} = \begin{cases}
\log\left[ L_{E_W} \cdot {}_2F_1 \left( \frac{1}{2} ,-\frac{1}{2} ;2;\frac{L_{E_W} }{L_B} \right) ^2\right] , & L_{E_W} <L_B \\
\log\left[ L_{B} \cdot {}_2F_1 \left( \frac{1}{2} ,-\frac{1}{2} ;2;\frac{L_{B} }{L_{E_W}} \right) ^2\right] , & L_{E_W} >L_B
\end{cases}
.
\end{equation}
This smoothly interpolates between $\log L_{E_W}$ for when $L_{E_W} \ll L_{B}$ and $\log L_{B}$ when $L_{E_W} \gg L_{B}$. The resolvent is more complicated, though simple away from the phase transitions.

\subsection{Schwinger-Dyson equation} \label{sec:two_tensor_pure_gen}

Finally, we now assume the large-$N$ limit where $L_A,L_B,L_{E_W} \sim N \gg 1$ without having some dimensions that are excessively large.

Taking $\tau _i \to \gamma ^{-1} \circ \tau _i$ for $i=1,2$ in \eqref{eq:2_tensor_pure_moment} (considering it unnormalized for the moment), we get
\begin{equation} \label{eq:zero_connected_sum}
Z_n^{(PT)}=\sum _{\tau _1,\tau _2} L_A^{C(\tau _1)} L_B^{C(\gamma ^{-2} \circ \tau _2)} L_{E_W}^{C(\tau _1^{-1} \circ \tau _2)} .
\end{equation}

For any given $\tau _2$, we have $C(\tau _1)+C(\tau _1^{-1} \circ \tau _2) \le n+ C(\tau _2)$ by the triangle inequality. Therefore we obtain the maximal contribution whenever $\tau _1$ is non-crossing in cycles of $\tau _2$. Together with the term $C(\gamma ^{-2} \circ \tau _2)$ this gives $n+C(\tau _2)+C(\gamma ^{-2} \circ \tau _2) \le 2n + C(\gamma ^2)$, so again we have the dominant contribution for $\tau _2$ that is non-crossing in cycles of $\gamma ^2$.

For even $n$, $\gamma ^2=(1 ,\, 3,\, \cdots ,n-1 )(2,\, 4,\, \cdots ,n)$ consists of two cycles. Thus, we have a decoupling of the odd sites from the even sites. In the odd sites, $\tau _2$ are just all the possible non-crossing permutations, and similarly for the even sites. In each of them, $\tau _1$ is non-crossing in cycles of $\tau _2$.
For odd $n$, $\gamma ^2=(1,\, 3,\, \cdots ,n,\, 2,\, 4,\, \cdots ,n-1)$ is a single cycle.

\begin{figure}[]
\centering
\includegraphics[width=0.8\textwidth]{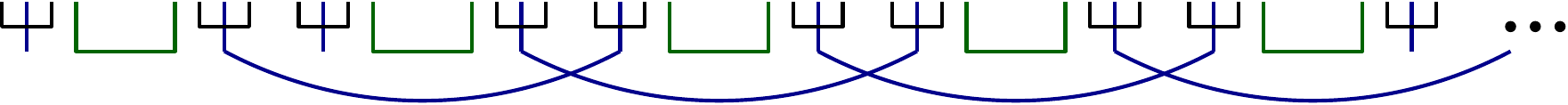}
\caption{Diagrammatic representation of the negativity for the pure two-tensor case.}
\label{fig:no_conn_diagrammatic_rep}
\end{figure}

Just as before, we can represent \eqref{eq:zero_connected_sum} diagrammatically. 	Since we do not have the subsystems $C_i$, 
we are left with the dark green, dark blue, and black lines. We get \eqref{eq:zero_connected_sum} by summing over all contractions of the diagram shown in Fig.~\ref{fig:no_conn_diagrammatic_rep}. The green lines correspond to $\tau _1$ in the new variables, and the blue lines to $\tau _2$. The ordering within the diagram has changed slightly because of the re-definition of the permutations that we did.

\begin{figure}[]
\centering
\includegraphics[width=0.8\textwidth]{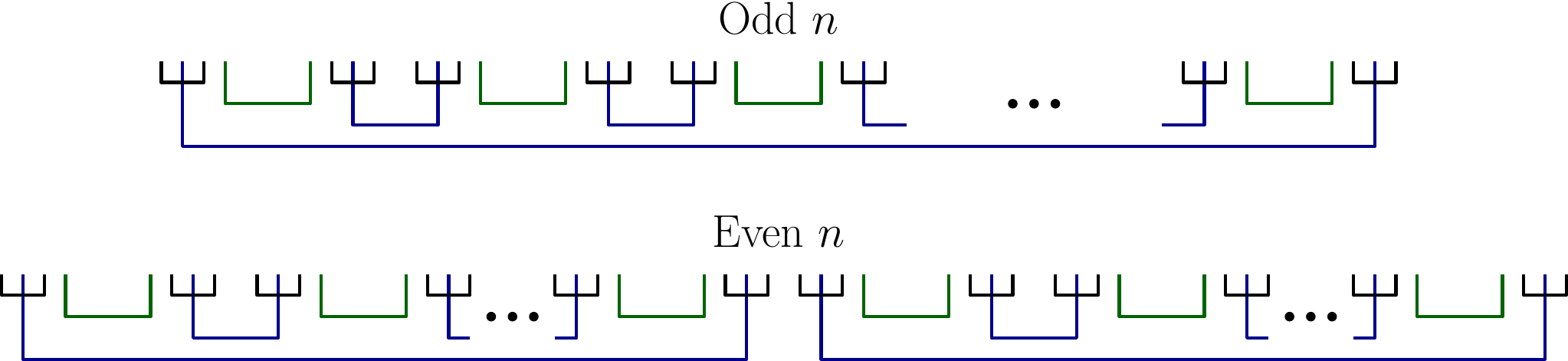}
\caption{Rearranging the negativity diagrams results in these diagrams for odd and even $n$.}
\label{fig:no_conn_diagrams_even_odd}
\end{figure}

The nodes in these diagrams can be conveniently rearranged so that each node is contracted by the dark blue line to the following node. As a result, similarly to before, for even $n$ they split into two parts, while for odd $n$ there is a single piece. The rearranged diagrams are shown in Fig.~\ref{fig:no_conn_diagrams_even_odd}. Importantly, since $\tau _2$ factorizes according to $\gamma^2$, and $\tau _1$ according to the cycles of $\tau _2$, the contractions in the two pieces of the even case factorize. That is, there are no contractions connecting the left part with the right one.

Every connected piece in Fig.~\ref{fig:no_conn_diagrams_even_odd} is what we would have gotten if we were calculating entanglement entropy instead of negativity. Diagrammatically, this is because the (say) green lines are not crossed, which is what we had to do before in order to implement the partial transpose. Instead, here, such a diagram calculates the trace of the density matrix traced over $A$, which is used for the entanglement entropy between $A$ and $B$ in the pure state on $AB$. Let us denote the $n^{th}$ moment of the density matrix by
\begin{equation}
m'_n = \mean{\tr (\rho _B)^n},\qquad \rho _B = \tr _A \rho .
\end{equation}
In fact, it is known that the negativity moments in a pure state are given simply in terms of the entanglement moments, $m'_n$, (see e.g., \cite{Calabrese:2012ew}) --- the even $n^{th}$ negativity moment is the square of the $(n/2)$'th entanglement moment, while the odd negativity and entanglement moments are the same. This is exactly what Fig.~\ref{fig:no_conn_diagrams_even_odd} shows.

However, the subtlety here is that the pure state is random, and therefore the even moments do not completely factorize to a square of entanglement moments, but rather we should average the product. What we have shown here
is that in the large-$N$ limit, we do have a full factorization because of the structure of contractions found above.

Let us denote the resolvent corresponding to $m'_n$ by
\begin{equation}
R' = \sum _{n=0} ^{\infty } \frac{m'_n}{z ^{n+1} } .
\end{equation}

\begin{figure}[]
\centering
\includegraphics[width=0.9\textwidth]{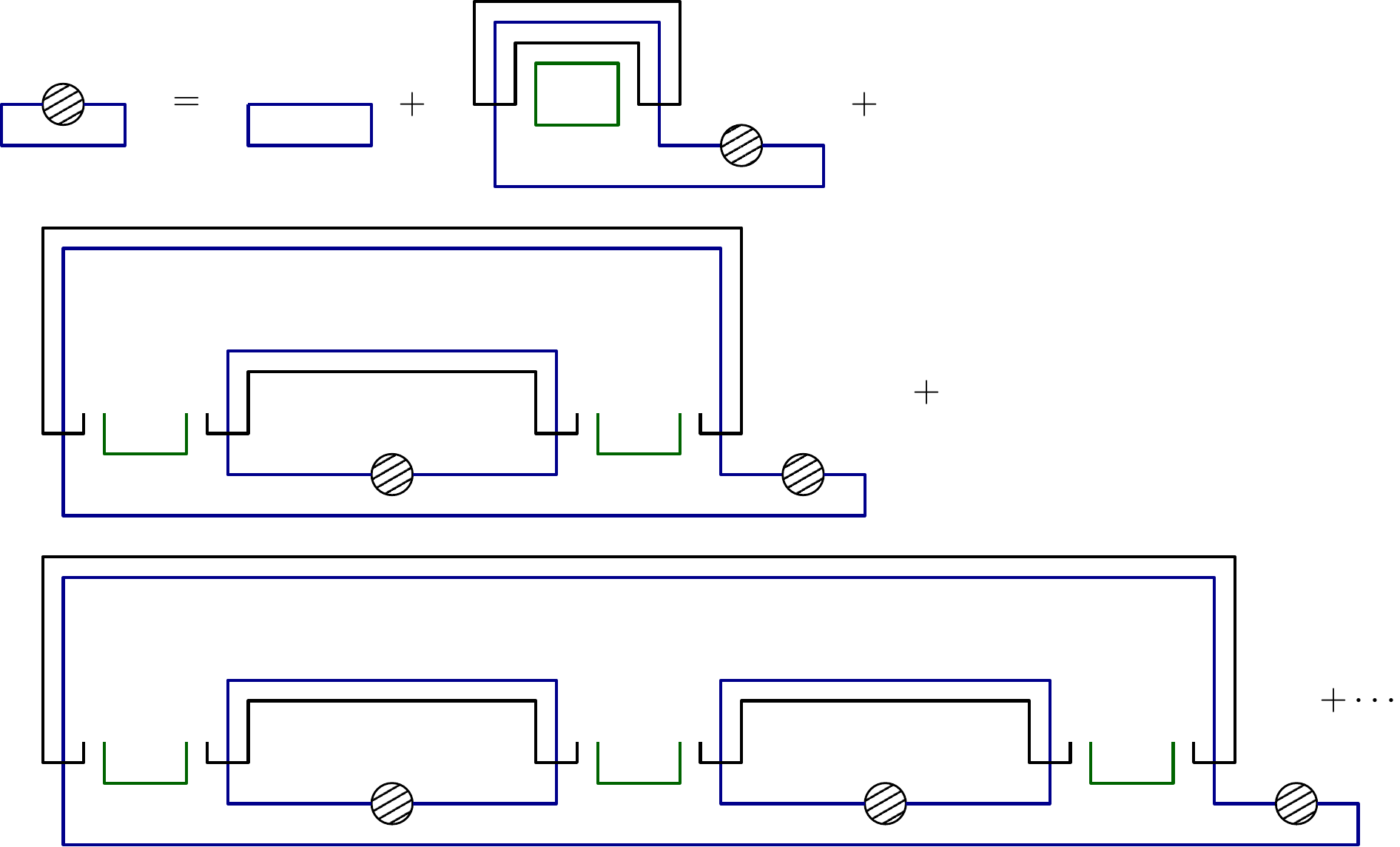}
\caption{Schwinger-Dyson equation for $R'$.}
\label{fig:no_conn_SD_eq}
\end{figure}

According to the structure of the permutations $\tau _1,\tau _2$ that we found above,
we can write a Schwinger-Dyson equation for $R'$. It is based on the drawing in Fig.~\ref{fig:no_conn_SD_eq}. We organized it according to blue contractions. However, green contractions are planar and are embedded inside cycles of the blue ones, exactly because $\tau _1$ is non-crossing in $\tau _2$. Thus, even though the green contractions are not shown, in the $(k+1)$'th term on the RHS of the Schwinger-Dyson equation, we will need to sum over the planar green contractions of $k$ nodes. We already saw how to calculate such a sum. In fact, this is the same as the sum we are after if we replace $n \to k$, use $\gamma $ instead of $\gamma ^2$ as we rearranged the diagrams so that the blue lines contract neighboring nodes, and take $\tau _2=\gamma $. Counting the blue, green, and black lines, this sum is
\begin{equation}
L_B^k \sum _{\tau \in NC_k } L_A^{C(\tau )} L_{E_W}^{C(\gamma^{-1} \circ \tau )} := L_B^k f_k(L_A,L_W),
\end{equation}
where recall that $NC_k$ are non-crossing permutations of $k$ elements.
The factor of $L_B^k$ does not enter in our SD equation, since the closed blue lines go to the trace of $R'$. We have already calculated $f_k$ using the Narayana numbers and found
\begin{equation}
\begin{split}
f_k(L_A,L_{E_W}) &= \sum _{\tau  \in NC_k} L_A^{C(\tau )} L_{E_W}^{C(\gamma^{-1} \circ \tau )} = \sum _{l=1} ^k N_{k,l} L_A^l L_{E_W}^{k+1-l} =\\
&=\begin{cases}
L_A L_{E_W} ^k \, {}_2F_1\left( 1-k,-k;2;\frac{L_A}{L_{E_W} }\right)  ,&L_A<L_{E_W} \\
L_{E_W} L_{A} ^k \, {}_2F_1\left( 1-k,-k;2;\frac{L_{E_W}}{L_A } \right) ,&L_A>L_{E_W}
\end{cases} .
\end{split}
\end{equation}

The SD equation for $R'$ is then
\begin{equation} \label{eq:pure_SD}
    z R' = L_B+ \sum_{k=1}^{\infty }\frac{(R')^k}{(L_A L_B L_{E_W})^k} f_k(L_A,L_{E_W})
\end{equation}
or more explicitly, when $L_A<L_{E_W} $
\begin{equation} \label{eq:pure_SD_2}
    z R' = L_B+ \sum _{k=1}^{\infty }\frac{(R')^k}{L_A^{k-1}L_B^k} {}_2F_1\left( 1-k,-k;2;\frac{L_A}{L_{E_W}} \right) .
\end{equation}

The resolvent of the negativity is given in terms of the entanglement moments by
\begin{equation}
R = \sum _{n=0} ^{\infty } \frac{m_n}{z ^{n+1} } = \sum _{n \text{ odd}} \frac{ m'_n}{z ^{n+1} } + \sum _{n=0} ^{\infty } \frac {m'^2_n}{z ^{2n+1} } ,
\end{equation}
and the logarithmic negativity is
\begin{equation}
\mathcal{E}= \lim _{n \text{ even } \to 1} m_n = (m'_{1/2} )^2,
\end{equation}
which is the $1/2$ R\'enyi entropy.

We have found that the negativity spectrum for a pure state is given by \eqref{eq:pure_neg_dos}.
The case that we consider now is indeed a pure state and so this formula is valid here. However, we have a random pure state. Therefore, if we are interested in the averaged negativity spectrum, we have
\begin{equation}
    \mean{\ns(\lambda )} = \int _{-\infty }^{\infty }d\lambda ' \left\lvert \frac{\lambda }{\lambda '} \right\rvert \mean{\es (\lambda \lambda ') \es \left( \frac{\lambda }{\lambda '}\right)}  +
    \frac{1}{2} \left( \mean{\es (\lambda )}-\mean{\es (-\lambda )}\right) 
    + c \cdot \delta (\lambda ) .
\end{equation}
In the first term on the RHS, we have the average of a product of density of states, while
we have studied the averaged pure state entanglement spectrum. In general, the average of the product is different from the product of averages and this would require calculating higher moments in the randomness of the tensor variables sense.

However, in large-$N$, we saw using large-$N$ diagrammatics that if we consider a product of two moments of the density matrix (without partially transposing), as in the second part of Fig.~\ref{fig:no_conn_diagrams_even_odd}, then the averaging can be done separately. This means that we can use \eqref{eq:pure_neg_dos} as is for large-$N$.

To demonstrate this consider $L_A \ll L_{E_W} $.
At leading order we can approximate the ${}_2F_1$ in \eqref{eq:pure_SD_2} by 1, giving
\begin{equation}
z R' = L_B + \frac{R' L_A}{L_A L_B - R'}.
\end{equation}
The solution is\footnote{We choose the branch such that $R' \to 0 $ as $z \to \infty $.}
\begin{equation}
R'(z ) = \frac{z L_AL_B +L_B-L_A - \sqrt{(L_A L_Bz +L_B-L_A)^2-4z L_A L_B^2}}{2z } .
\end{equation}
Therefore,
\begin{equation}
    \es(\lambda ) = \frac{L_A L_B}{2\pi \lambda } \sqrt{(L_+-\lambda )(\lambda -L_-)} ,\qquad  L_- \le \lambda  \le L_+.
\end{equation}
where we have defined
\begin{equation}
L_{\pm } = \left( \frac{1}{\sqrt{L_A}} \pm \frac{1}{\sqrt{L_B}} \right) ^2 .
\end{equation}

\begin{figure}[]
\centering
\includegraphics[width=0.5\textwidth]{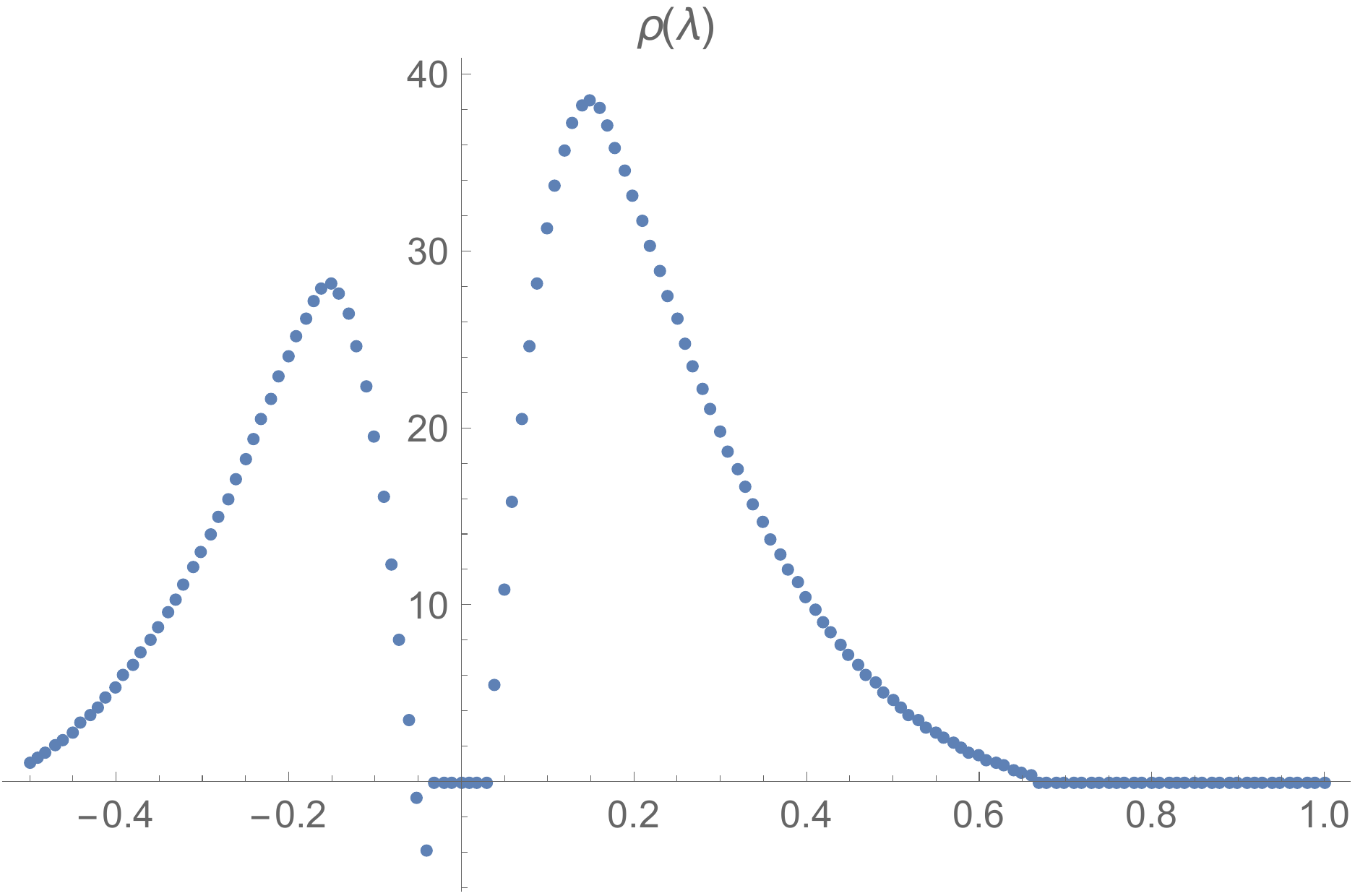}
\caption{Negativity spectrum at leading order for large $L_{E_W}$. We use here $L_A=4$ and $L_B=10$.}
\label{fig:pure_neg_leading_dW}
\end{figure}

In order to get $\ns(\lambda )$, we should just plug this in \eqref{eq:pure_neg_dos}.
We do this numerically and show the result  without the contribution going as $\delta (\lambda )$ in Fig.~\ref{fig:pure_neg_leading_dW}. Note that when finding $\ns(\lambda)$, for small $L_A,L_B$, we get significant regions with negative density of states, but as we mentioned, the current calculation is valid at large-$N$. Indeed, as we increase $L_A,L_B$ these regions disappear.

\section{Entanglement and negativity in random tensor networks} \label{sec:gen_network}

We would like to generalize the analysis to a larger class of tensor networks, such as the one shown in Fig.~\ref{fig:gen_tensor_ntw}.

Let us consider a generalized reduced density matrix that will allow us to calculate entanglement and negativity spectra. We divide the external indices of the tensor network into three sets:
\begin{itemize}
\item $T$: indices that are partially traced over, corresponding to the subsystem that we trace out,
\item $S$: no operation is performed on these indices,
\item $P$: indices that are partially transposed, allowing us to compute negativity.
\end{itemize}
Then, we define the marginal density matrix
\begin{equation}
\rho ' = \tr _T \rho ^{T_P} .
\end{equation}

\begin{figure}[]
\centering
\includegraphics[width=0.5\textwidth]{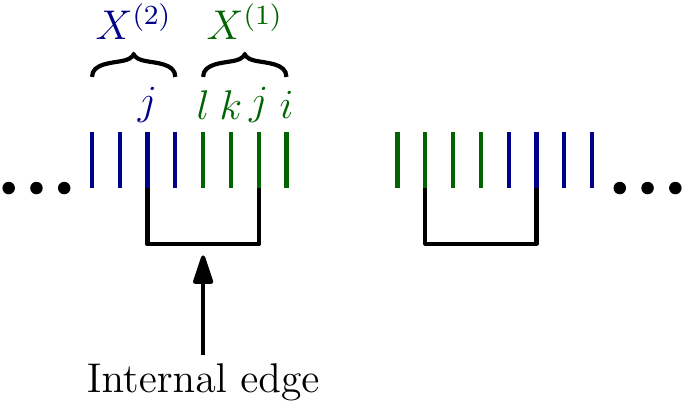}
\caption{The diagram corresponding to a single insertion of a marginal density matrix.}
\label{fig:basic_ingredient_gen}
\end{figure}

In order to calculate $\mean{  \tr \rho '^n} $, we define a diagrammatic construction generalizing what we did before. We take a basic ingredient, or node, consisting of two parts, each one made of a series of lines, one line for any index. This node represents the density matrix and generalizes what we had in Fig.~\ref{fig:density_graphically}. The lines are grouped into multi-lines, where each multi-line consists of all those lines belonging to the same $X^{(i)} $, just as before. It is convenient to represent each such tensor by a different color. This is demonstrated in Fig.~\ref{fig:basic_ingredient_gen} (compare to Fig.~ \ref{fig:gen_tensor_ntw}). The internal black connecting lines in the bottom part are just like the $E_W$ line we had in the two-tensor case. They correspond to an edge connecting two tensors.

\begin{figure}[]
\centering
\includegraphics[width=0.5\textwidth]{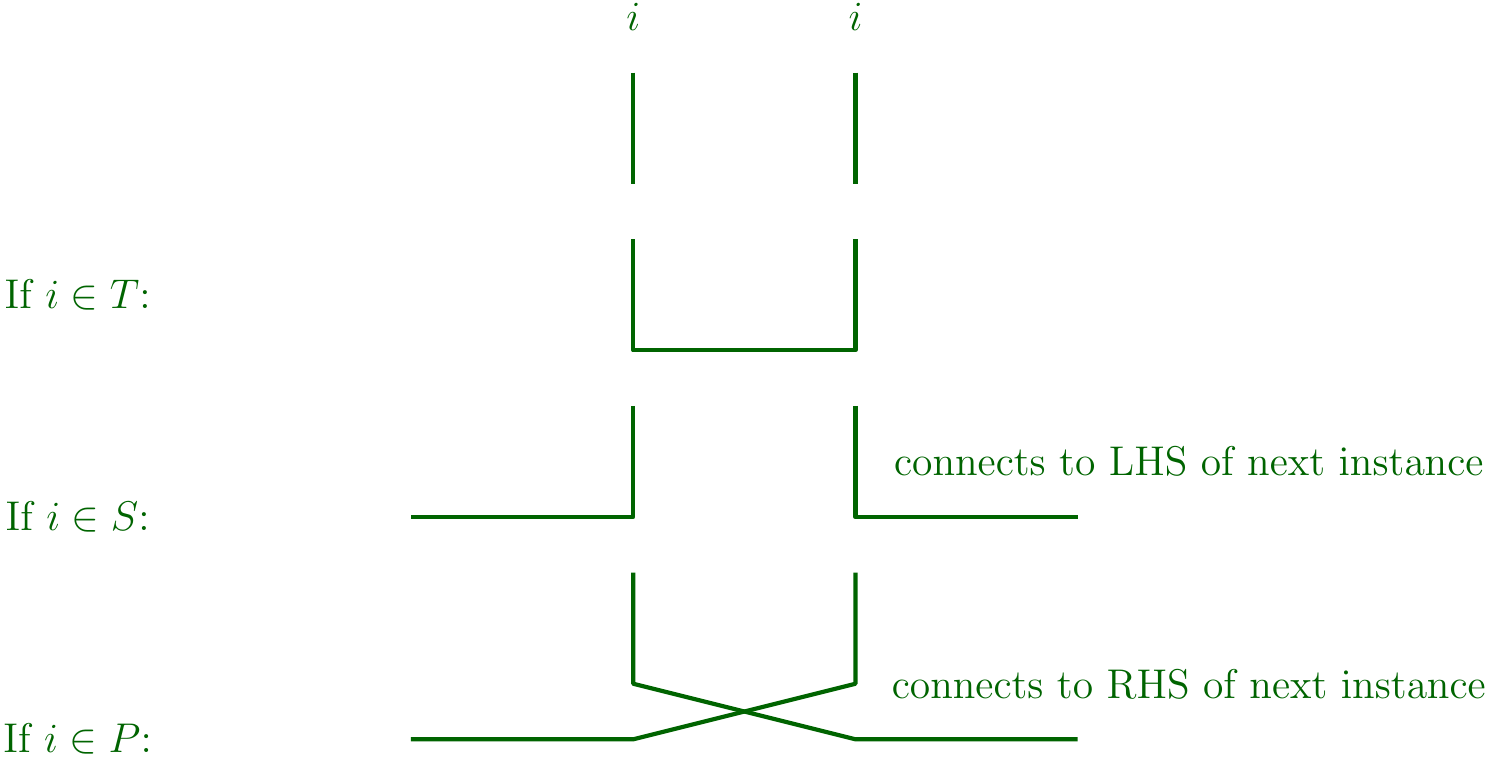}
\caption{Connecting the density matrix to its neighbors.}
\label{fig:gen_tr_contraction_types}
\end{figure}

We place $n$ copies of this basic diagram next to each other. These are connected among themselves according to the rules shown in Fig.~\ref{fig:gen_tr_contraction_types}. Any index, depending on which set it belongs to, contracts either in the same instance of the density matrix, or to the previous or the following one.

Having $n$ such copies combined with periodic boundary conditions, we should sum over all possible contractions of the multi-index sets. That is, the lines of the same color on the LHS of an instance are contracted together with lines of the same color on the RHS of some instance. Each contraction is assigned the value of the variance of the corresponding $X^{(i)} $. Every closed index loop gives the dimension of the corresponding Hilbert space.

Just as we did before, we can express this diagrammatic description of a tensor network in terms of permutations. Suppose that there are $k$ different tensors $X^{(i)} $, $i=1,\cdots ,k$, and that each index takes values in an $L_i$ dimensional space. We write $L_i=N^{w_i}d_i$ and assume that $N \gg 1$ with $w_i,d_i$ being fixed. We thus have a permutation $\beta _i$ for every $X^{(i)}$, i.e., for every tensor describing the contractions of that particular color (multi-index). The moments are then given by
\begin{equation} \label{eq:RTN_moments}
\begin{split}
\mean{ \tr \rho '^n}  = \prod _i \text{Var}\left( X^{(i)} \right)^n \cdot \sum_{\{\beta_i\}} \prod _{\text{edges}} \begin{cases}
L_i^{C(\beta _i)} & \text{ for } i \in T \\
L_i^{C(\gamma^{-1} \circ \beta _i)} & \text{ for } i \in S \\
L_i^{C(\gamma \circ \beta _i)} & \text{ for }i \in P \\
L_i^{C(\beta _{i'} ^{-1} \circ \beta _i)} & \text { for an edge connecting } i,i'
\end{cases}.
\end{split}
\end{equation}
where again $\gamma =(1,2,\cdots ,n)$ is the permutation consisting of one cycle, and note that $L_i=L_{i'} $ for an index in common to $X^{(i)} $ and $X^{(i')} $. These values are obtained from the number of closed loops of each given index.

\begin{figure}[]
\centering
\includegraphics[width=0.5\textwidth]{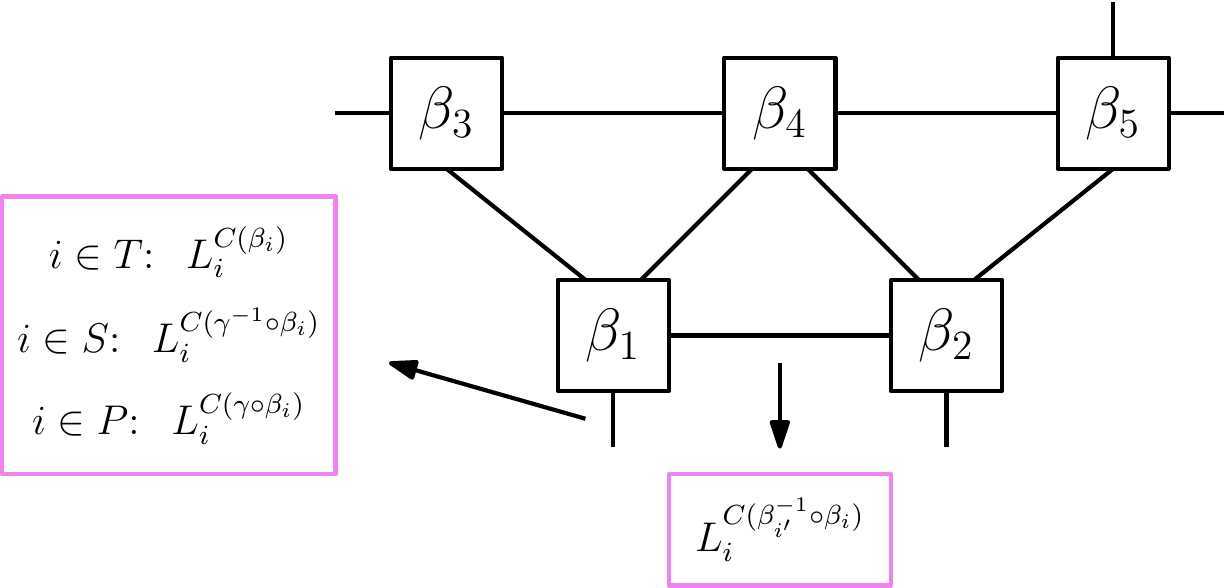}
\caption{A generic tensor network in terms of permutations.}
\label{fig:tensor_network_permutations}
\end{figure}

To summarize, we can in fact represent the permutation structure easily in terms of the same tensor network. We show this, together with the rules for evaluating the moments, in Fig.~\ref{fig:tensor_network_permutations}.

For example, the (mixed) original two-tensor network \eqref{eq:two_tensor_network}
 is given in terms of permutations as
\begin{equation} \label{eq:two_tensor_perms}
\tikz[baseline=-0.5ex]{
\path (0,0) node[circle,minimum size=1] (a) {$P$} ++(2,0) node[rectangle,draw,minimum size=2] (X) {{\large $\tau _1$}} ++(3,0) node[rectangle,draw,minimum size=2] (Y) {{\large $\tau _2$}} ++(2,0) node[circle,minimum size=1] (b) {$S$};
\node (T1) at ($(X)+(0,-1.5)$) {$T$};
\node (T2) at ($(Y)+(0,-1.5)$) {$T$};
\draw (a)--(X) node[pos=0.5,above,scale=0.9] {$j \in A $};
\draw (X)--(Y) node[pos=0.5,above,scale=0.9] {$k \in E_W $};
\draw (Y)--(b) node[pos=0.5,above,scale=0.9] {$n \in B $};
\draw (T1)--(X) node[pos=0.5,left,scale=0.9] {$i \in C_1 $};
\draw (T2)--(Y) node[pos=0.5,right,scale=0.9] {$m \in C_2 $};
} 
\end{equation}
and applying the rules of Fig.~\ref{fig:tensor_network_permutations} to this diagram gives exactly expression \eqref{eq:2_tensor_perm_partition}.

\subsection{Special case: no partial transpose (entanglement spectrum)}

We can now apply the same logic we used in Section \ref{sec:two_tensor_pure_gen} in order to get the leading, large-$N$ behavior of a random tensor network. In fact, a useful way to organize this analysis for the case of the entanglement entropy appears in \cite{2010JPhA...43A5303C}.

The appropriate language is that of flow networks and flows. The basic question motivating the problem of flow networks is having a transportation network with various routes with a limited capacity, and the goal is to find the maximal possible flow one could achieve getting from some initial point (source) to a destination point (sink). We will define the needed notions below.

A \emph{flow network} is a directed graph consisting of vertices $V$ and edges $E$. Among the vertices, there are two distinguished elements, the source $s$ and the sink $t$. There is a capacity function $w(u,v)$ from pairs of vertices into the non-negative real numbers, such that $w(u,v)>0$ for $(u,v) \in E$ and zero otherwise. A \emph{flow} is a function $f:V \times V \to \mathbb{R} $ 
such that
\begin{enumerate}
\item $f(u,v) \le w(u,v)$ (we cannot exceed the capacity),
\item $f(u,v)=-f(v,u)$ (more on this in the following),
\item $\sum _{v \in V} f(u,v)=0$ for any $u \neq s,t$ (that is, the total flow is of course zero at intermediate points). 
\end{enumerate}
For sets $X,Y$, we denote $g(X,Y) = \sum _{x \in X} \sum _{y \in Y} g(x,y)$ for any function $g$, and similarly we define $g(x,Y)$ and $g(X,y)$. A cut is a partition of $V$ into two sets $V=S \sqcup T$ such that $s \in S$ and $t \in T$.\footnote{Note that in this definition we allow each of the sets $S$ and $T$ to have disconnected components.}
The value of a flow is
\begin{equation}
|f| = \sum_{v \in V} f(s,v) = f(S,T)
\end{equation}
for any cut. This is the total flow from the source to the sink. The goal in this problem is to find the maximal possible value of a flow. There can be more than one flow saturating this value.

Given a flow network and a flow $f$, we define the residual network by having capacity $w_f(u,v)=w(u,v)-f(u,v)$. Note that the residual network is a sensible flow network by the definition of a flow, having edges $(u,v)$ for $w_f(u,v)>0$. Also note that having a positive flow from $u$ to $v$ reduces the capacity from $u$ to $v$, but increases the capacity from $v$ to $u$.

An \emph{augmenting path} is a path from the source to the sink and so consists of positive capacity edges.\footnote{A path is a finite sequence of distinct vertices such that each two adjacent vertices in the sequence are connected by an edge.} The \emph{Ford-Fulkerson method} allows us to find a maximal flow. We take the original network, look for an augmenting path, construct the residual network, and continue in this process. In each residual network we look for an augmenting path and construct another residual network. We stop when there are no augmenting paths left. Note that if $f'$ is a flow in a residual network constructed in turn using a flow $f$, then $f+f'$ is a flow in the original network with value $|f+f'|=|f|+|f'|$. An augmenting path together with a value $x>0$ that is less than or equal to the capacity for each edge along the path gives us a flow given by $f(u,v)=x$ on the edges of the path and $f(v,u)=-x$. Therefore, we get a flow by adding up the augmenting paths, and we necessarily stop at some point because $w(s,V)$ decreases at each step.

Interestingly, we are guaranteed that the resulting flow is always a maximal flow. This follows from the \emph{max-flow min-cut theorem} \cite{ford1956maximal,elias1956note,cormen2009introduction}. Moreover, this theorem tells us that the maximal flow equals the minimal cut capacity $|f| = \min_{\text{cuts}} w(S,T)$. Clearly, we cannot pass a flow that exceeds the minimal cut, and according to the theorem, we can in fact saturate this bound.

Going back to the tensor network, we can assume that we have a single edge of the same type. That is, among external edges connecting to the same tensor and having the same type $T,S$ or $P$, or internal edges connecting the same tensors. Otherwise, if we have several such edges, we can replace them by a single edge with a corresponding weight $w \in \mathbb{R} $, given by the sum of the weights of the edges we combine. The finite relative dimension of the new edge is the product $d=\prod d_i$.

Before that, we used $C(\beta )$ as counting the number of cycles in a permutation $\beta $. A complementary quantity is the minimal number of swaps needed to bring a permutation $\beta $ to the identity permutation that we denote by $|\beta |$. Indeed, for permutations of size $n$, the two are related by $|\beta |+C(\beta )=n$.
The triangle inequality claim that we have shown before becomes in this language $|\alpha \beta |+|\beta ^{-1} \gamma | \ge |\alpha \gamma |$ making it more manifest that this is a triangle inequality. As we saw, it is saturated for $\alpha \beta $ non-crossing in cycles of $\alpha \gamma $, and in such a case we will denote this by $\alpha \beta  \le \alpha \gamma $.

\begin{figure}[]
\centering
\includegraphics[width=0.8\textwidth]{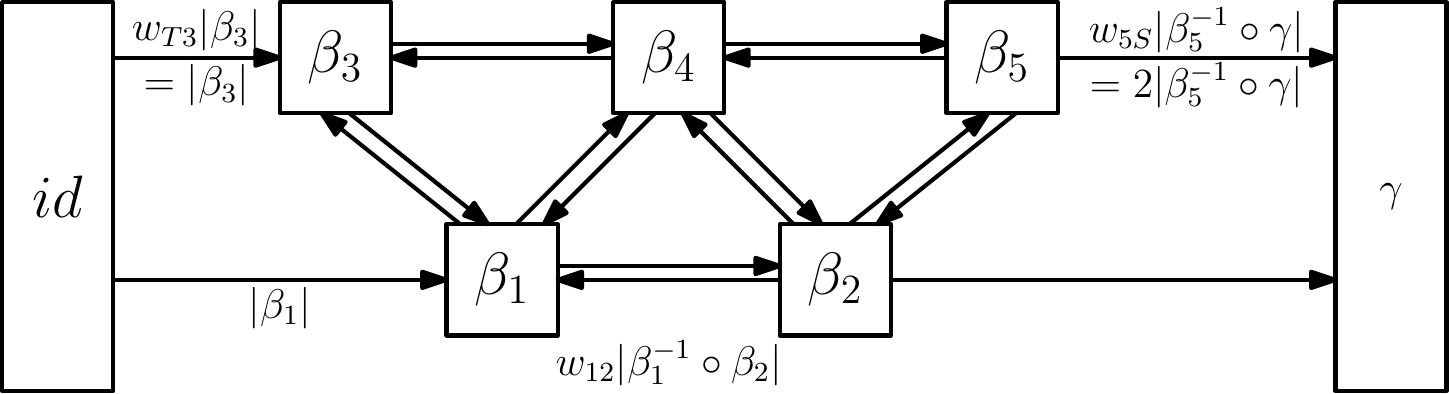}
\caption{The flow network corresponding to a tensor network. Going from a permutation $\beta $ to $\beta '$ we have a weight depending on $\beta ^{-1} \circ \beta '$.}
\label{fig:flow_tensor_ntw}
\end{figure}

Following \cite{2010JPhA...43A5303C}, we construct a flow network associated to the tensor network. Each tensor becomes a vertex in the flow network. Each vertex is associated with a permutation. We also add one vertex for every type of external edge.  For the type $T$, we assign the identity permutation, while we assign $\gamma $ for type $S$. The internal edges of the tensor network become a pair of directed edges in both directions with capacity $w_{ij} |\beta _i^{-1} \beta _j|$. The identity is taken to be the source, with every edge in the tensor network of type $T$ becoming a directed edge from the identity to $\beta _i$ with capacity $w_{Ti} |\beta _i|$. Similarly, the vertex of type $S$ becomes the sink, with capacities $w_{iS} |\beta _i^{-1} \circ \gamma |$ (see Fig.~\ref{fig:flow_tensor_ntw}). Therefore an edge $i \to j$ is assigned $w_{ij} |\beta _i^{-1} \beta _j|$.

Maximizing the moments over $N$ is the same as minimizing
\begin{equation}
F_{\beta } = \sum_i w_{Ti} |\beta _i| + \sum_{ij} w_{ij} |\beta _i^{-1} \circ \beta _j| + \sum_i w_{iS} |\beta _i^{-1} \circ \gamma  | .
\end{equation}

\begin{figure}[]
\centering
\includegraphics[width=0.9\textwidth]{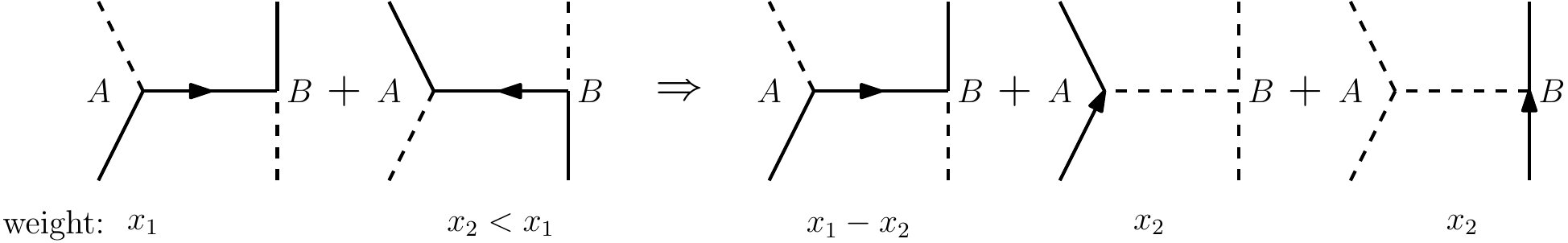}
\caption{Having two flows passing through an edge and its opposite orientation can be replaced by the flows shown in the figure, resulting in the same flow, but passing in only one of the directions of that edge.}
\label{fig:flow_reudction}
\end{figure}

We should give a remark on flows. The definition of a flow as an anti-symmetric function may appear to be confusing; one could expect that for two vertices having two edges connecting them in both directions, we could have a flow in each direction, and this flow should be limited by the capacity. This is not the case in the definition of a flow. Phrased differently, in the Ford-Fulkerson method, we are allowed to have many flows going in both directions and cancelling each other, which seems to imply physically more flow than we can. The resolution lies in the fact that we can always assume that in such a pair of edges, the flow is chosen to go only in one direction. Physically the intuition is that we just do not need to waste flow going in one direction and then going in exactly the opposite direction. More precisely, we can always replace any pair of augmenting paths shown in Fig.~\ref{fig:flow_reudction} by the other paths in that figure.\footnote{Note that the resulting paths are not necessarily valid. We may be required to remove loops from them. However, such loops will not affect the value of the resulting flows, and so we can proceed in the algorithm. Again, logically this is just the statement that we do not gain anything by running loops inside our transportation network.} Importantly, this reduction is not necessary for the Ford-Fulkerson method and the method works for any choice of augmenting paths; it may just happen that we use far too many steps than we could have used. However, we assume that this reduction has been made for the application to tensor networks below. With this assumption, we can in fact think of the edges in our graph as being undirected as long as we do not include paths that start and end on the same external node.

Having constructed the flow network associated to the tensor network, we apply the Ford-Fulkerson method where the capacities of the flow network are given by the $w_i$ defined through the Hilbert space dimensions. For each augmenting path, we apply a series of triangle inequalities
\begin{equation}
|\beta _i| + |\beta _i^{-1} \circ \beta _j| + \cdots + |\beta _k^{-1} \circ \gamma | \ge |\gamma | .
\end{equation}
Equality is obtained for $\beta $'s being non-crossing with $\beta _i \le \beta _j \le \cdots \le \beta _k$. After getting a maximal flow $f$, we have a residual network. Note that in the residual network, for each pair of edges of opposite orientation connecting the same two vertices, there will be one with smaller or equal capacity; we define the residual network to have the same capacity on both edges with value equal to the minimal value, getting symmetric edges as in the original network.\footnote{Similarly, we can drop the edge going into the source or out of the sink.} Then,
\begin{equation}
F_{\beta } \ge |f| (n-1) + F_{\beta } (\text{residual network}),
\end{equation}
where we used $|\gamma |=n-1$. As mentioned, $|f |$ is the maximal flow, so it equals the minimal cut capacity $w_{\text{min cut}} $.

As $F_{\beta } \ge 0$, we can saturate the bound, that is get $F_{\beta } =|f| (n-1)$ by precisely all $\beta _i$ satisfying that
\begin{equation} \label{eq:maximal_flow_perms}
\begin{split}
& \bullet \text{$\beta _i$ are non-crossing,} \\
& \bullet \text{$\beta _{i_1} \le \beta _{i_2} \le \cdots \le \gamma$ for the $\beta $'s along each augmenting path,} \\
& \bullet \text{$\beta _i=\text{id}$ for all vertices in the connected component of $\text{id}$ in the residual network,} \\
& \bullet \text{$\beta _i=\gamma $ for all vertices in the connected component of $\gamma $ in the residual network,} \\
& \bullet \text{All $\beta $'s are equal in the same connected component of the residual network.}
\end{split}
\end{equation}

Note that the connected components of the source and the sink are distinct since otherwise we would have a remaining augmenting path, so these conditions are consistent; they are also consistent with respect to the conditions on augmenting paths. Strictly speaking, we found that only all $\beta $'s belonging to an augmenting path or to the connected component of the source or the sink are non-crossing. The reason that all $\beta $'s are non-crossing is because any vertex is initially connected to the source or the sink (or both), so if it is not in an augmenting path, it is in the connected components of one of those in the residual network.

To summarize, we get at large-$N$
\begin{equation}
\mean{  \tr \rho '^n} =\prod _i \text{Var}\left( X^{(i)} \right)^n \cdot  N^{n \sum_i w_i - (n-1) w_{\text{min cut}} } \sum _{\beta_i \text{ satisfying } \eqref{eq:maximal_flow_perms}}  \prod _{\text{edges}} d_i^{n-|\beta _{i'}^{-1} \circ \beta _i|} .
\end{equation}
Using the variance in \eqref{eq:variance_general},\footnote{Note that there could be several different maximal flows. We should pick a single one, and they all give the same result, as we derived a bound and saturated it.}
\begin{equation}
    \mean{  \tr \rho '^n} = N^{ - (n-1) w_{\text{min cut}} } \sum _{\beta_i \text{ satisfying } \eqref{eq:maximal_flow_perms}}  \prod _{\text{edges}} d_i^{-|\beta _{i'}^{-1} \circ \beta _i|} .
\end{equation}
The $N$ dependence is fixed by the minimal cut as in the Ryu-Takayanagi (RT) prescription.

\subsection{Negativity spectrum in a tensor network}

We would like to see how to also calculate the negativity spectrum in a general state. This time, we have in addition to external states of type $T$ and $S$, the states of type $P$ that we partially transpose.
The corresponding flow network that we construct has an additional vertex $\gamma^{-1} $ that all the states in $P$ connect to.

\begin{figure}[]
\centering
\includegraphics[width=0.8\textwidth]{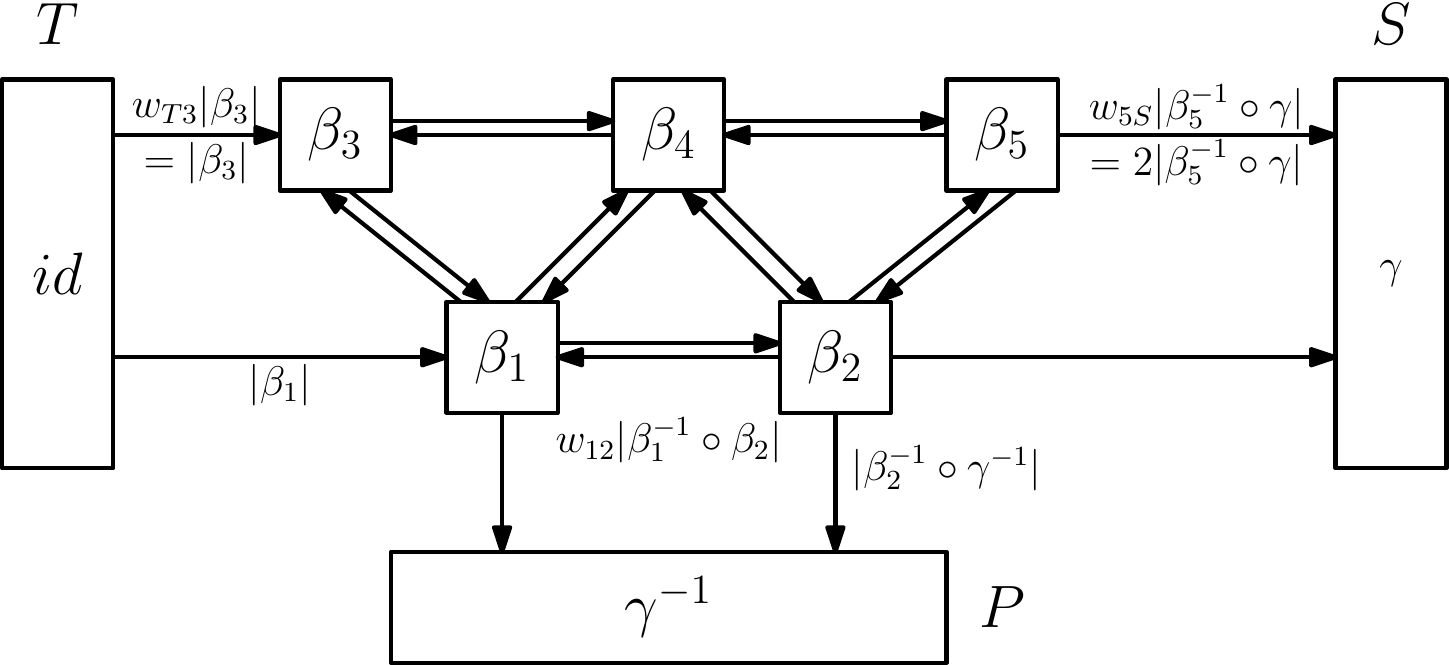}
\caption{The flow network associated to a tensor network.}
\label{fig:flow_ntw_neg}
\end{figure}

We will describe how to analyze the general case where we have a partial transpose as two steps.
Initially, let us take $\text{id}$ to be the source, and both $\gamma $ and $\gamma ^{-1} $ to be the sink. A network with several sources and/or sinks is solved similarly to a network with a single source and sink.\footnote{Formally, one could turn it into an equivalent network with a unique sink. This is done by adding a new sink vertex $s$, with an edge going from every sink $s_i$ in the network we are interested in into $s$, having an infinite capacity. We can do a similar construction when having several sources. This is not necessary though, as we can just think about $\{\gamma , \gamma ^{-1}\} $ as our sink.} This is shown in Fig.~\ref{fig:flow_ntw_neg}.

We define $F_{\beta } $ just as before as being $\sum w_{ij} |\beta _i^{-1} \circ \beta _j|$ with the external vertices being $\text{id},\gamma ,\gamma ^{-1} $. Since $|\gamma |=|\gamma ^{-1} |$, for the contribution to $F_{\beta }$ it does not matter whether an augmenting path gets to one sink or the other.

In the first step, we thus apply the Ford-Fulkerson method to this flow network. Each augmenting path corresponds to an application of the triangle inequality. Recall that as before, when having two edges connecting two vertices, we symmetrize the capacities in any residual network to the smaller value. Eventually we have no augmenting paths meaning that the source is not connected to either of the sinks. This step tells us that once again\, if $f_1$ is a maximal flow
\begin{equation}
F_{\beta } \ge |f_1| \cdot (n-1) + F_{\beta } (\text{residual network 1}),
\end{equation}
where we call the residual network at this stage ``residual network 1''.

However, in order to saturate the bound to $|f_1| \cdot (n-1)$ we would need to add to the conditions \eqref{eq:maximal_flow_perms} that $\beta _i=\gamma ^{-1}$ in the connected component of $\gamma ^{-1}$. The point is that these new conditions are not necessarily consistent and so the bound is not tight. Indeed, while the component of $\text{id}$ is different from that of $\gamma $ and $\gamma ^{-1}$, we might get that $\gamma $ and $\gamma ^{-1}$ are connected in the residual network and then we cannot impose that the permutations in this connected component equal to both $\gamma $ and $\gamma ^{-1} $.

\begin{figure}[]
\centering
\includegraphics[width=0.5\textwidth]{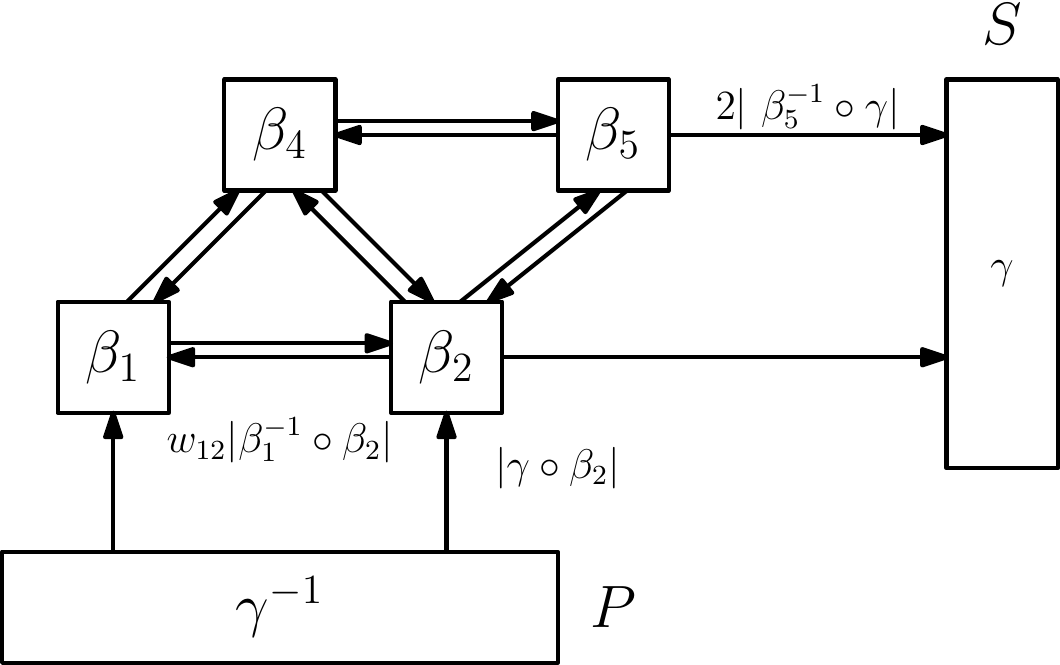}
\caption{The flow network for the second step.}
\label{fig:flow_ntw_step2}
\end{figure}

In order to resolve this, we introduce a second step. After the first step, the connected component of the source is disconnected from $\gamma ,\gamma ^{-1} $, so let us forget about it for the majority of the second step. We consider the remaining flow network, and turn $\gamma ^{-1} $ to be a source, where $\gamma $ remains the sink. We could do the other way around, with the same result. For clarity, this is shown in Fig.~\ref{fig:flow_ntw_step2}. This time we use the triangle inequality
\begin{equation}
|\gamma  \circ \beta _i| + |\beta _i^{-1} \circ \beta _j| + \cdots +|\beta _k^{-1} \circ \gamma | \ge |\gamma ^2| .
\end{equation}

Thus, we get
\begin{equation}
\begin{split}
F_{\beta }& \ge |f_1| \cdot (n-1) + F_{\beta } (\text{residual network 1}) \ge \\
& \ge |f_1| \cdot (n-1) + |f_2| \cdot |\gamma ^2| + F_{\beta } (\text{residual network 2}).
\end{split}
\end{equation}
The residual network in the final expression includes also the component of the identity, which is the same as that after the first step. Now, the conditions to saturate the bound are
\begin{equation}
\begin{split}
& \bullet \text{$\beta _{i_1} \le \beta _{i_2} \le \cdots \le \gamma \text{ or }\gamma ^{-1}$ for the $\beta $'s along each augmenting path in step 1,} \\
& \bullet \text{$\gamma  \circ \beta _{i_1} \le \gamma \circ \beta _{i_2} \le \cdots  \le \gamma ^2$ along each augmenting path in step 2,} \\
& \bullet \text{$\beta _i=\text{id}$ for all vertices in the connected component of $\text{id}$ in the final residual network,} \\
& \bullet \text{$\beta _i=\gamma ^{-1} $ for all vertices in the connected component of $\gamma ^{-1} $,} \\
& \bullet \text{$\beta _i=\gamma $ for all vertices in the connected component of $\gamma  $,} \\
& \bullet \text{All $\beta $'s are equal in the same connected component.}
\end{split}
\end{equation}
In short, we can express the last four conditions in a single statement, so that the full list of conditions can be summarized by
\begin{equation} \label{eq:flow_perms_conditios}
\begin{split}
& \bullet \text{$\beta _{i_1} \le \beta _{i_2} \le \cdots \le \gamma \text{ or }\gamma ^{-1}$ for the $\beta $'s along each augmenting path in step 1,} \\
& \bullet \text{$\gamma  \circ \beta _{i_1} \le \gamma \circ \beta _{i_2} \le \cdots  \le \gamma ^2$ along each augmenting path in step 2,} \\
& \bullet \text{All $\beta $'s are equal in the same connected component of the final residual network}\\
& \qquad \text{(including the components of $\text{id},\gamma ,\gamma ^{-1}$).}
\end{split}
\end{equation}

We can see that the only problem with consistency of \eqref{eq:flow_perms_conditios} happens when there is a vertex taking part in an augmenting path in step 1 ending on $\gamma $ but is connected to $\gamma ^{-1}$ in the final residual network (or the same with $\gamma \leftrightarrow \gamma ^{-1}$). In this case, the first and last conditions cannot be satisfied at the same time.
We will explain how this is avoided below, but for now, let us assume this does not happen.

Then, the result at large-$N$ is
\begin{empheq}{align} \label{eq:tensor_network_neg_moments}
\mean{  \tr \rho '^n} = N^{-(n-1) w^{\text{I}} _{\text{min}} - l(n) w_{\text{min}} ^{\text{II}} } \sum _{\beta_i \text{ satisfying } \eqref{eq:flow_perms_conditios}}  \prod _{\text{edges}} d_i^{-|\beta _{i'}^{-1} \circ \beta _i|},
\end{empheq}
where
\begin{equation}
l(n)=\begin{cases}
n-1 & \text{for odd } n \\
n-2 & \text{for even } n
\end{cases} .
\end{equation}
In this equation, $w^{\text{I}} _{\text{min}} $ is the minimal cut of the first step, and similarly for the second step.
Note that this description is symmetric with respect to exchanging the set $S$ with $P$, as it should be, and provides a consistency check.

\begin{figure}[]
\centering
\includegraphics[width=0.5\textwidth]{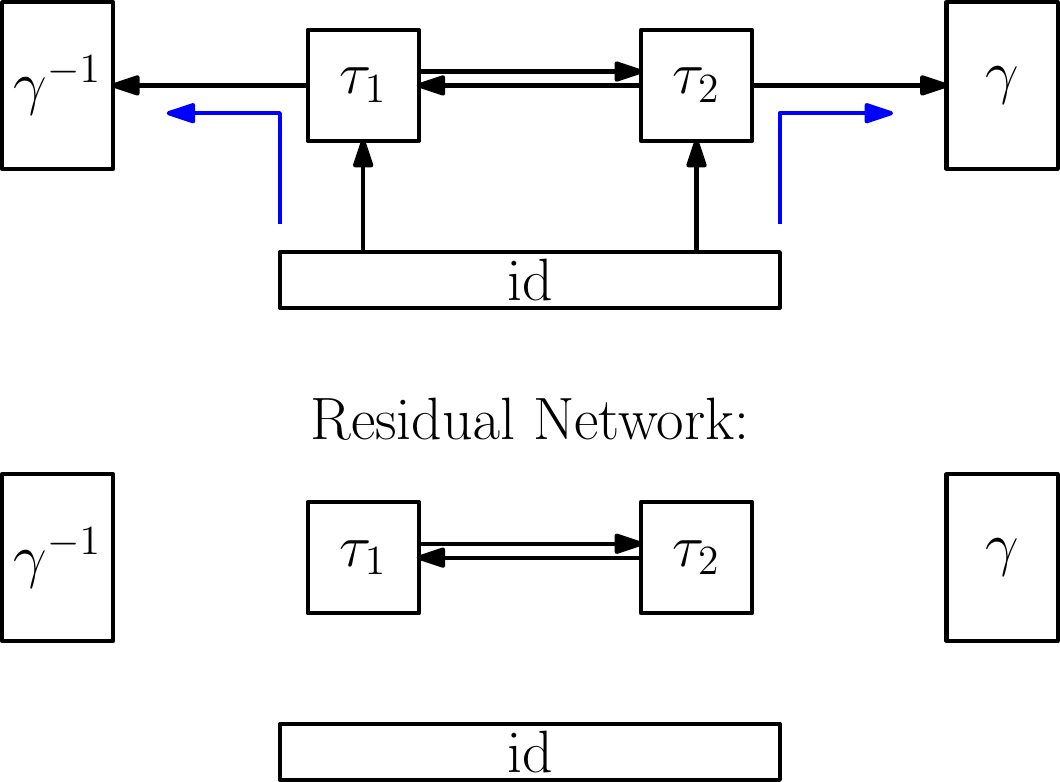}
\caption{The flow network corresponding to the two-tensor network. The flow is shown in blue. All the weights are taken to be 1 here.}
\label{fig:2_tensor_flow}
\end{figure}

Let us do some simple examples to demonstrate this. Consider the two-tensor network shown in \eqref{eq:two_tensor_perms}. Suppose for simplicity that all the weights are 1. The associated flow network is shown in Fig.~\ref{fig:2_tensor_flow}. The flow in the first step is depicted in blue. The residual network after the first step is shown in the same figure. In this case, $\gamma ,\gamma ^{-1} $ are not connected in the residual network after step 1, and therefore we do not have the second step. The conditions \eqref{eq:flow_perms_conditios} become in this case just $\tau _1=\tau _2$, as they belong to the same connected component in the residual network, with both being non-crossing in $\gamma ,\gamma ^{-1} $. This is an immediate diagrammatic way to get these constraints. Note that this case does not belong to the classification of phases we did for the two-tensor network, but rather includes a special relation among the Hilbert space dimensions. Since $\tau _1=\tau _2$, this network behaves as a one-tensor network that we analyze in Section \ref{sec:wormhole_1_tensor}.

\begin{figure}[]
\centering
\includegraphics[width=0.4\textwidth]{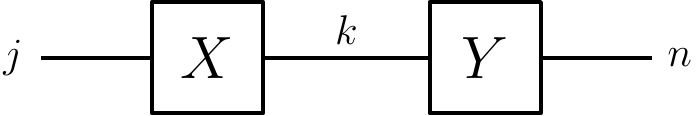}
\caption{A two-tensor network corresponding to a pure state.}
\label{fig:2_tensor_ntw_pure}
\end{figure}

In the pure case of the two-tensor network (Fig.~\ref{fig:2_tensor_ntw_pure}) the set $T$ is empty, so it is the other way around here: we do not have the first step, but do have the second step. This is what we have done in Section \ref{sec:two_tensor_pure_gen}, and we see that we get the conditions on the permutations there immediately from \eqref{eq:flow_perms_conditios}.

\begin{figure}[]
\centering
\includegraphics[width=0.4\textwidth]{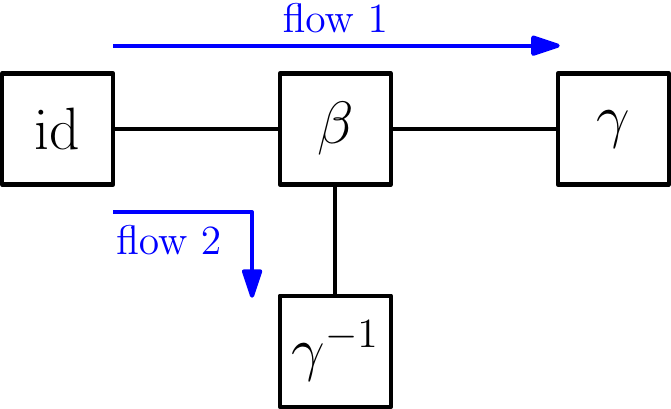}
\caption{A one-tensor network.}
\label{fig:one_tensor}
\end{figure}

Let us now return to the possible obstruction for consistency of \eqref{eq:flow_perms_conditios}. We start with the simplest example of this situation and then argue in general. This example is the one-tensor network shown in Fig.~\ref{fig:one_tensor} with unit weights. In either of the two possible choices for an augmenting path in step 1 shown in the figure we will get the clash we mentioned before.\footnote{Also, skipping step 1 and choosing the unique possible path in step 2 gives an inconsistency as well.}

\begin{figure}[]
\centering
\includegraphics[width=0.4\textwidth]{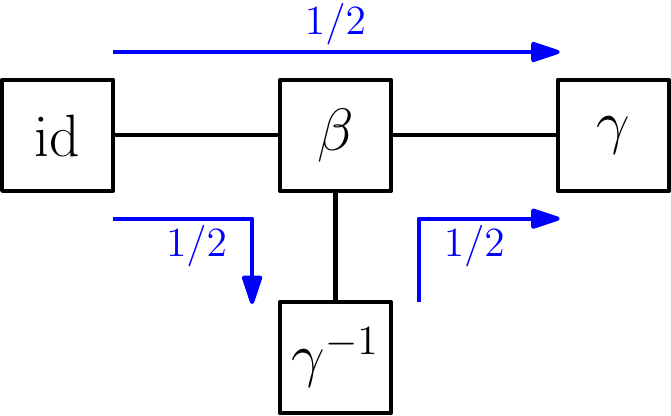}
\caption{A flow solving the one-tensor network with unit capacities.
}
\label{fig:one_tensor_flow}
\end{figure}

However, we can still solve this. Let us first find the permutations giving the maximal contribution in $N$, and then we can use the usual formula. In order to do that, we can construct the flow shown in Fig.~\ref{fig:one_tensor_flow} resulting in an empty residual network. We get that the conditions for this bound to be consistent are that (1) $\beta $ is non-crossing in $\gamma ,\gamma ^{-1} $, and (2) $\gamma  \circ \beta $ is non-crossing in $\gamma ^2$. The first condition is equivalent to $\beta $ being non-crossing with only one- and two-cycles.\footnote{This is so because higher cycles have a particular orientation that matches to either $\gamma $ or $\gamma ^{-1} $, but not both.} This condition means that
\begin{equation}
\begin{split}
& |\beta |+|\beta ^{-1} \circ \gamma |=|\gamma | ,\\
& |\beta | + |\beta ^{-1} \circ \gamma ^{-1} | = |\gamma | .
\end{split}
\end{equation}
The second condition is $|\gamma  \circ \beta |+|\beta ^{-1} \circ \gamma |=|\gamma ^2|$. Together we get that $|\gamma ^2|=2(|\gamma |-|\beta |)$ or
\begin{equation}
|\beta | = |\gamma | - \frac{|\gamma ^2|}{2} = \begin{cases}
\frac{n}{2} & n \text{ even}\\
\frac{n-1}{2} & n \text{ odd}
\end{cases},
\end{equation}
where we used that $|\gamma |=n-1$ and $|\gamma ^2|$ is $n-1$ for $n$ odd and $n-2$ for $n$ even. The conclusion is that the set of $\beta $'s satisfying these conditions are in a class denoted by $NC_2$, that are
\begin{equation} \label{eq:NC2}
NC_2=
\begin{cases}
\text{Non-crossing with only 2-cycles} & n \text{ even} \\
\text{Non-crossing with a single 1-cycle and the rest are 2-cycles} & n \text{ odd}
\end{cases} .
\end{equation}

\noindent 
For these permutations,
\begin{equation}
F_{\beta } = (n-1)+\frac{1}{2}|\gamma ^2| = \begin{cases}
\frac{3n-4}{2} & n \text{ even} \\
\frac{3n-3}{2}  & n \text{ odd}
\end{cases}.
\end{equation}

In general, the network we analyze is not a usual source-sink flow network, so a-priori there is no guarantee that all choices of flows are equivalent, as this is not a single Ford-Fulkerson process. We can think about this problem as having three external nodes, and we construct augmenting paths that go from any one of them to another one. We will think about this problem as having the goal of maximizing the total flow that we can construct, i.e., the sum of the augmenting paths irrespective of which nodes get the maximal individual flow. However, now the Ford-Fulkerson method does not necessarily work --- indeed, we saw in Fig.~\ref{fig:one_tensor} a flow that does not admit leftover augmenting paths, and yet it is not a maximal flow because the flow there was of 1 unit, while in Fig.~\ref{fig:one_tensor_flow} we found a flow of 3/2 units. Even though we are not guaranteed that the Ford-Fulkerson method works, we may still look for a maximal flow.

\begin{figure}[]
\centering
\includegraphics[width=0.5\textwidth]{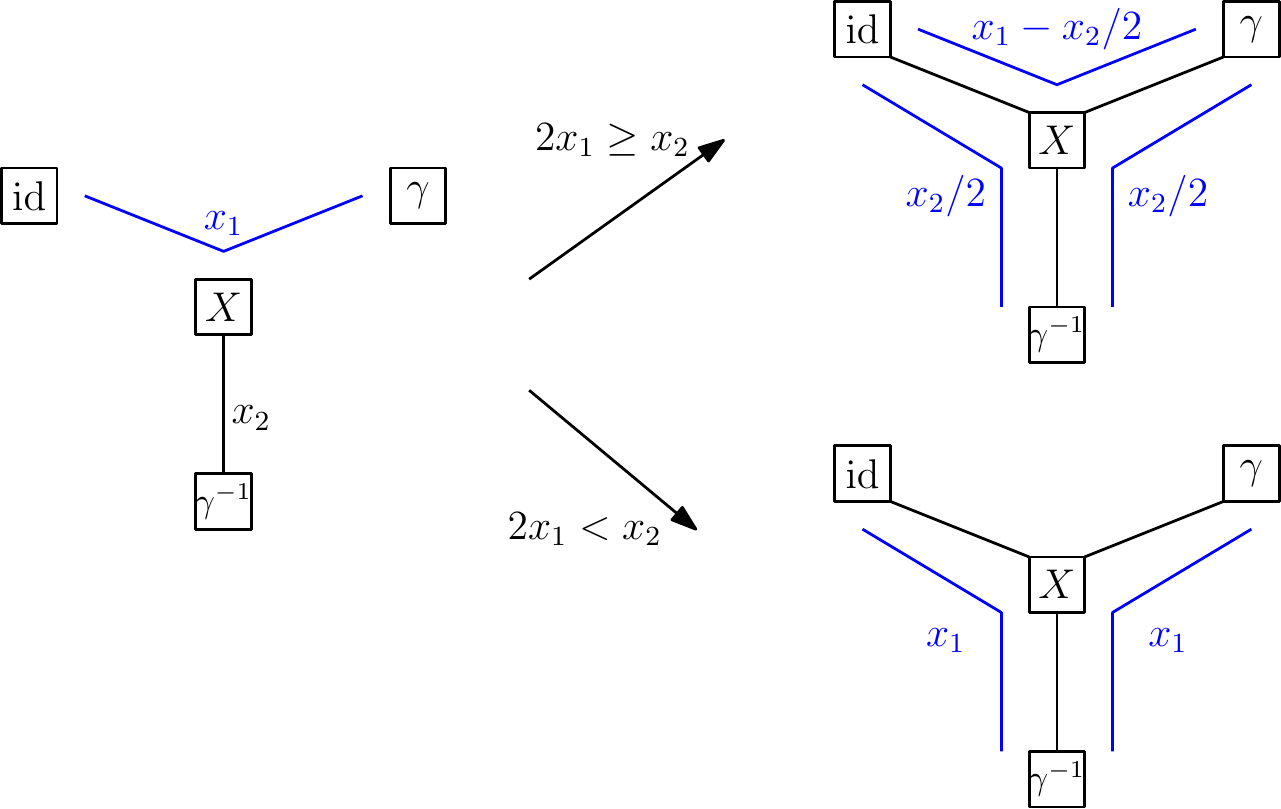}
\caption{Argument for absence of the inconsistency.}
\label{fig:clash_analysis}
\end{figure}

The claim then is that in a maximal flow the inconsistency mentioned above does not occur and the inequality is saturated, justifying \eqref{eq:tensor_network_neg_moments}. Indeed, suppose that after constructing a maximal flow, we are left with some vertex $X$ connected to $\gamma ^{-1}$ and took part in a path from id to $\gamma $. This is demonstrated on the left hand side of Fig.~\ref{fig:clash_analysis}. Note that we do not assume that this is a one-tensor network, but merely show several paths without drawing the full network. We denote by $x_1 >0$ the flow that went through the path from id to $\gamma $ passing by the node $X$, and there is a capacity of $x_2>0$ remaining in a path connecting the node to $\gamma ^{-1} $. Depending on whether $2x_1 \ge x_2$ or not, 
we show in the figure two alternative flows that we could use instead of the $x_1$ flow that was on the LHS. In either case, we increase the total flow in the network. This is a contradiction, so we cannot have this kind of inconsistency.
The same argument holds for exchanging $\gamma  \leftrightarrow \gamma ^{-1}$ and also to a vertex on an augmenting path from $\gamma ^{-1}$ to $\gamma $ that remains in the end connected to id.\footnote{We can think about the network symmetrically when constructing the maximal flow, where each of these three options should be considered. Even if we do so, the two steps above are still relevant because the augmenting paths in the first and second steps contribute differently. Still, we see that after the first step, there will not be any vertex connected to id that will take part in a path from $\gamma ^{-1}$ to $\gamma $, as we assumed.}

In fact, even though the Ford-Fulkerson method does not work in general, so that we might end up with a flow which is not maximal, we can still know whether the flow that we obtained is maximal or not. The easy way to know 
this is just by the absence of the inconsistency mentioned above: if there is an inconsistency of this form, which is a trivial condition to check, that means we did not get a maximal flow. This is a necessary and sufficient condition in fact, because then the bound is tight.
That is, if we found a flow that cannot be increased with additional augmenting paths and does not have this inconsistency, it is a maximal flow.
This provides us an alternative to the Ford-Fulkerson algorithm that we can apply in such problems.

It is natural to compare this situation to the problem of a flow network where we treat node $i$ (where $i \in \{\text{id},\gamma ,\gamma ^{-1} \}$) as the source, and the other two nodes that we will denote 
by $j,k$ as the sinks. Given a maximal flow $f$ that we constructed before to the full network, we may ask whether taking all the augmenting paths from $i$ to $j$ and from $i$ to $k$ (dropping those going from $j$ to $k$), giving us a flow $f_i$ to the problem where $i$ is the source, is a maximal flow. 
In fact, this is true. Indeed, suppose that there is an additional augmenting path from $i$ to $j$ or $k$. If it does not pass through a path going from $j$ to $k$ in $f$, we could add it to $f$, which is a contradiction. Otherwise, it means that we have a path starting at $i$, passing by some node $X$ and ending in $k$, or in short $i \to X \to k$, such that $X \to k$ 
is part of a path $j \to k$ in $f$. In this case, we cannot add $i \to X \to k$ as an additional augmenting path to $f$. However, this is just the situation that we had in Fig.~\ref{fig:clash_analysis}, where the remaining path is the one labeled by $x_2$ and $x_1$ is the flow that we had in $f$. We saw that there is a contradiction here as well.
%

Dividing the augmenting paths in $f$ into $f_{\text{id},\gamma } $, $f_{\text{id},\gamma^{-1} } $, and $f_{\gamma,\gamma^{-1} } $ in the manner mentioned above, we have shown that $f_{ij} \cup f_{ik} $ is a maximal flow in the network where $i$ is a source and $j,k$ are sinks. By the max-flow min-cut theorem, we know that this flow equals the minimal cut $w_{\text{min}} ^{(i)}$ of $i$. Therefore, the cuts of the two steps that we found before are
\begin{equation}
w_{\text{min}} ^{\text{I}} = w_{\text{min}} ^{(\text{id})} , \qquad w_{\text{min}} ^{\text{II}} = \frac{w_{\text{min}} ^{(\gamma )}+w_{\text{min}} ^{(\gamma ^{-1} )}-w_{\text{min}} ^{(\text{id})}}{2} =\frac{w_{\text{min}} ^{(\gamma )}+w_{\text{min}} ^{(\gamma ^{-1} )}-w_{\text{min}} ^{(\gamma ,\gamma ^{-1} )}}{2} .
\end{equation}
Thinking about the minimal cut as the RT surface, we see that the latter is the RT surface corresponding to the mutual information of the two subsystems $P$ and $S$.
If we are interested in the negativity $\mathcal{E} $, from \eqref{eq:tensor_network_neg_moments} we see that the coefficient of $\log (N)$ in $\mathcal{E} =w_{\text{min}} ^{\text{II}} \log (N) + \cdots $ is half the mutual information for a tensor network with any capacities, agreeing with \cite{Dong:2021clv}, but the $d_i$ dependence can be different in general. The negativity spectrum is not unique and depends on these other terms.

\subsection{Examples and new spectra}

We saw that the semi-circle and Marchenko–Pastur distributions are realized as negativity spectra of random tensor networks, but in general we can obtain other new spectra. One such spectrum will be obtained in Section \ref{sec:wormhole_1_tensor}. Here we will do another example which is the pure two-tensor network studied in Sections \ref{sec:two_tensor_pure} and \ref{sec:two_tensor_pure_gen}. We take here $w_A=w_B=w_{E_W}:= w$. While we can have any dimensions in the Schwinger-Dyson equation of Section \ref{sec:two_tensor_pure_gen}, here we take all relative dimensions to be $d_i=1$, but we can get the resolvent without further approximations.

In this case, we need a single augmenting path of magnitude $w$ in Fig.~\ref{fig:2_tensor_ntw_pure}, and there are no edges in the residual network. The solution in \eqref{eq:flow_perms_conditios} tells us that $\gamma \circ \tau _1 \le \gamma  \circ \tau _2 \le \gamma ^2$, where the permutations corresponding to the two tensors are $\tau _1$ and $\tau _2$, so defining $\beta _i=\gamma \circ \tau _i$ with $i=1,2$, this means just $\beta _1 \le \beta _2 \le \gamma ^2$, as we saw in Section \ref{sec:two_tensor_pure_gen}. It is not hard to count the number of permutations explicitly as we assume $d_i=1$. For $n $ odd, $\gamma ^2$ has a single cycle, and so we just count the number of non-crossing permutations $\beta _1,\beta _2$ such that $\beta _1 \le \beta _2$. This is known to be given by the second Fuss-Catalan number $FC_n^{(2)} = \frac{1}{2n+1} \binom{3n}{n}$. For $n$ even, $\gamma ^2$ has two cycles, and so we have two copies of the same counting with $n \to n/2$. Using \eqref{eq:tensor_network_neg_moments} we get
\begin{equation}
    m_n = \begin{cases}
    N^{(1-n)w} FC_n^{(2)} & \text{odd } n \\
    N^{(2-n)w} FC_{n/2}^{(2)} \cdot FC_{n/2}^{(2)} & \text{even } n
    \end{cases} .
\end{equation}
The resolvent is therefore
\begin{equation} \label{eq:2_tensor_pure_phase_trans}
    R(z) = \frac{1}{z^2} \cdot {}_4F_3\left[ \begin{matrix}
    \frac{2}{3} \, \frac{5}{6} \, \frac{7}{6} \, \frac{4}{3} \\ \, \frac{5}{4} \, \frac{3}{2} \, \frac{7}{4} \,
    \end{matrix} ; \frac{729}{16N^{2w}z^2} \right] + \frac{N^{2w}}{z} \cdot {}_4F_3 \left[ \begin{matrix}
    \frac{1}{3} \, \frac{1}{3} \, \frac{2}{3} \, \frac{2}{3} \\ \, 1 \, \frac{3}{2} \, \frac{3}{2} \,
    \end{matrix} ; \frac{729}{16N^{2w}z^2} \right] .
\end{equation}

\begin{figure}[]
\centering
\includegraphics[width=0.6\textwidth]{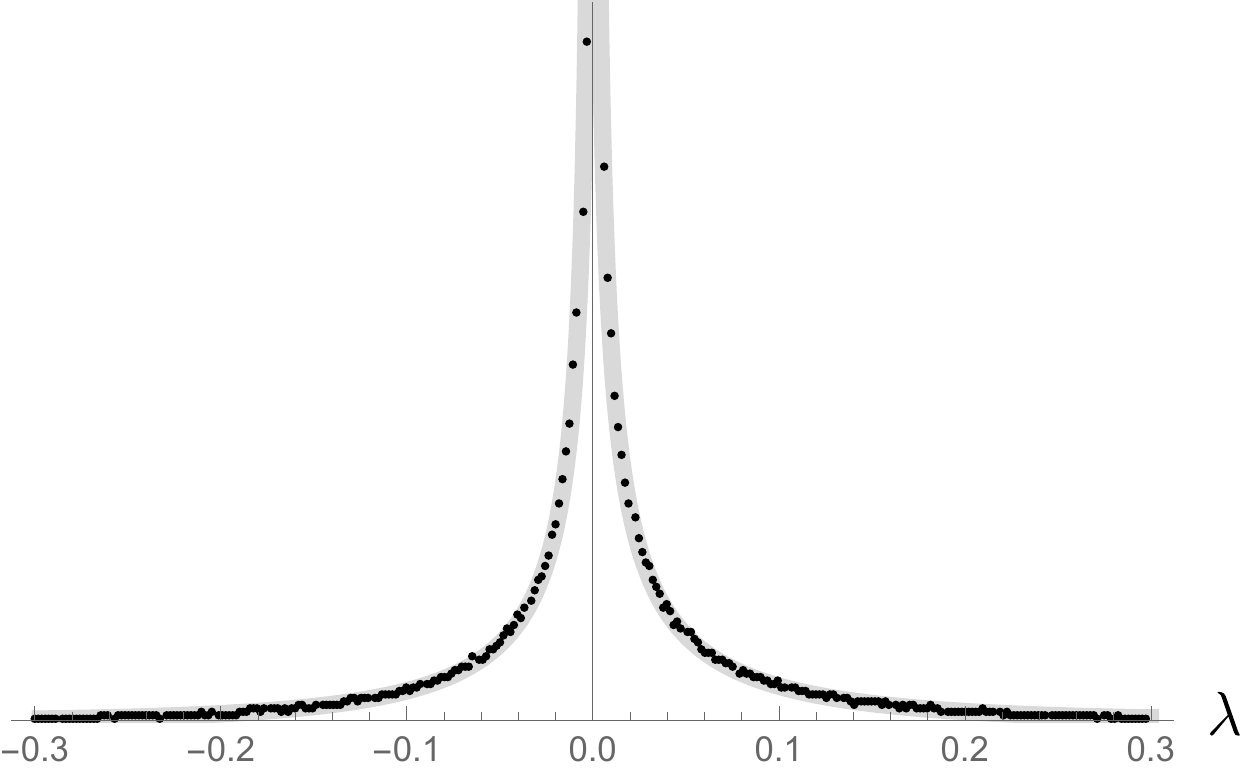}
\caption{A plot of the spectrum of the pure two-tensor network for $N^{2w}=256$. 
We compare it to a simulation of this tensor network, averaging over $10^3$ disorder realizations.
}
\label{fig:2_tensor_phase_trans_spectrum}
\end{figure}

It is not hard to find numerically the spectrum given the resolvent \eqref{eq:2_tensor_pure_phase_trans} using \eqref{eq:neg_spectrum_from_resolvent}. We show this in Fig.~\ref{fig:2_tensor_phase_trans_spectrum}. This spectrum is bounded, but differs from the previous cases.

Let us give another example of a more complicated network, and apply the formalism of the previous subsection. 
We consider the cycle graphs of \cite{2010JPhA...43A5303C}. If we take a three-cycle graph
\begin{align}
\begin{tikzpicture}
    \node[draw, shape=rectangle] (id) at (-1.5,0) {id};
    \node[draw, shape=rectangle] (gamma) at (3.5,0) {$\gamma$};
    \node[draw, shape=rectangle] (gammainv) at (1,-3) {$\gamma^{-1}$};
    \node[draw, shape=rectangle] (T1) at (0,0) {$T_1$};
    \node[draw, shape=rectangle] (T2) at (2,0) {$T_2$};
    \node[draw, shape=rectangle] (T3) at (1,-1.5) {$T_3$};
    \draw [thick] (T1) -- (T2) node[pos=0.5, above] {0.5};
    \draw [thick] (T1) -- (T3) node[pos=0.5, left] {0.5};
    \draw [thick] (T2) -- (T3) node[pos=0.5, right] {1};
    \draw [thick] (T1) -- (id) node[pos=0.5, above] {1};
    \draw [thick] (T2) -- (gamma) node[pos=0.5, above] {1};
    \draw [thick] (T3) -- (gammainv) node[pos=0.5,right] {1};
    \end{tikzpicture}
\end{align}
with the capacities indicated in the figure, we need to maximize
\begin{align}
    \frac{1}{2} \left( 2C(\beta_1) + C(\beta_1^{-1}\circ \beta_2)+ C(\beta_1^{-1}\circ \beta_3)+ 2C(\beta_2^{-1}\circ \beta_3)+2C(\gamma^{-1}\circ \beta_2)+2C(\gamma\circ \beta_3) \right) .
\end{align}
From the flow network with augmenting paths $\text{id} \to \beta _1 \to \beta _2 \to \gamma $, $\text{id} \to \beta _1 \to \beta _3 \to \gamma ^{-1}$ and $\gamma ^{-1} \to \beta _3 \to \beta _2 \to \gamma $, we find the following conditions on the permutations
\begin{align}
    \beta_1 \leq \beta_2 = \beta_3,
\end{align}
with $\beta_2 = \beta_3$ the same permutations of type $NC_2$ as we saw in the last subsection (see \eqref{eq:NC2}). The number of non-crossing permutations of size $2n$ consisting of only two-cycles is given by the $n^{th}$ Catalan number, so 
there are
\begin{align}
    \begin{cases}
    C_{n/2}, & n \in 2\mathbb{Z}
    \\
    n C_{(n-1)/2}, & n \in 2\mathbb{Z}+1
    \end{cases}
\end{align}
ways to choose $\beta_2 = \beta_3$, and for each specific choice, there are
\begin{align}
    \begin{cases}
    2^{n/2}, & n \in 2\mathbb{Z}
    \\
    2^{(n-1)/2}, & n \in 2\mathbb{Z}+1
    \end{cases}
\end{align}
ways to choose $\beta_1$. Therefore, the total degeneracy is 
\begin{align}
    \begin{cases}
    2^{n/2} C_{n/2}, & n \in 2\mathbb{Z}
    \\
    n2^{(n-1)/2} C_{(n-1)/2}, & n \in 2\mathbb{Z}+1
    \end{cases} .
\end{align}
Choosing all $d_i=1$, we get the negativity moments
\begin{align}
    m_n=
    \begin{cases}
        2^{n/2} C_{n/2}N^{2-\frac{3n}{2}}, & n \in 2\mathbb{Z}
        \\
        n2^{(n-1)/2} C_{(n-1)/2}N^{ 3(1-n)/2}, & n \in 2\mathbb{Z}+1
    \end{cases}.
\end{align}
Let us now compute the resolvent
\begin{equation}
    \begin{split}
    &\sum_{n \in 2\mathbb{Z}}^{\infty} \frac{2^{n/2} C_{n/2}N^{2-\frac{3n}{2}}}{z^{n+1}} =-\frac{1}{4} z N^5
   \left(\sqrt{1-\frac{8}{z ^2N^3}}-1\right)
   \\
    &\sum_{n \in 2\mathbb{Z}+1}^{\infty} \frac{n2^{(n-1)/2} C_{(n-1)/2}N^{ 3(1-n)/2}}{z^{n+1}} =\frac{2N^3}{z ^2N^3 \left(\sqrt{1-\frac{8}{z ^2
  N^3}}+1\right)-8} .
    \end{split}
\end{equation}
This is the same spectrum as we find in phase III of the one-tensor network, i.e., a semi-circle law, except that the radius of the semi-circle is larger by a factor of $\sqrt{2}$.

The logarithmic negativity may be computed from the even moments as
\begin{align}
    \mathcal{E} = \frac{1}{2}\log \left[N\right] + \log \left[\frac{8}{3\pi}\right] +\frac{1}{2}\log\left[2 \right].
\end{align}
We have separated this out into three terms to emphasize that this is the same as the single-tensor network except there is an additional half qubit of information.


\section{Micro-canonical JT gravity as a one-tensor network} \label{sec:west_coast_1_tensor}

The simplest random tensor network that we have mentioned, which is the one-tensor network of Section \ref{sec:1_tensor}, has an immediate relation to a holographic model that we recall in this section, with practically the same diagrams used in both contexts. 

Recently, the quantum extremal surface prescription with an island for the entanglement entropy of the Hawking radiation of a black hole was explained using Euclidean replica wormholes \cite{Almheiri:2019qdq,2019arXiv191111977P}. In this section, we will study mixed state entanglement in a setting of a black hole with its radiation, using the model of \cite{2019arXiv191111977P} that is based on Jackiw-Teitelboim gravity.\footnote{We note that different details and perspectives on negativity in this model are considered in Refs.~\cite{Dong_unpublished, Neg_equ_pure_state_paper}.}

The authors of \cite{2019arXiv191111977P} consider JT gravity with an end-of-the-world (EOW) brane similarly to \cite{Kourkoulou:2017zaj}. The EOW brane resides behind the horizon in Lorentzian signature, and can be thought of as the trajectories of the Hawking pairs of the radiation. It has a large number of orthogonal internal states to account for the radiation particles. In addition, there is an auxiliary quantum system $R$ corresponding to the radiation that came from the black hole. For our purposes, we will imagine that we can divide the space where we have the radiation into two parts such that one has $k_1$ possible orthonormal states $|i\rangle _{R_1} $, while the other has $k_2$ states $|a\rangle _{R_2} $. Correspondingly, the states of the black hole with the EOW brane in state $i,a$ are denoted by $|\psi_{ia} \rangle _B$. The matrix elements $\langle \psi_{ia} |\psi_{jb} \rangle $ are given by gravity amplitudes having a single asymptotic boundary, where on one endpoint an EOW brane of type $i,a$ should end, and on the other an EOW brane of type $j,b$ ends. Allowing gravitational wormhole configurations, when considering a single such matrix element, two different states appear to be orthogonal, while the absolute value squared of this matrix element is non-zero. This can be interpreted as if the gravitational theory implements a disorder average, consistent with having vanishing average value with a non-vanishing variance of the matrix element. This non-orthogonality is the origin of the Page curve-like behavior in this model. It is also the origin of the distinguishability of different black hole microstates \cite{2021arXiv210205053K,2021arXiv210800011K}.

As the radiation is entangled with the interior particles, we consider the state
\begin{equation} \label{eq:west_coast_state}
|\psi\rangle  = \frac{1}{\sqrt{k}} \sum _{i,a} |\psi_{i,a} \rangle _B |i\rangle _{R_1} |a\rangle _{R_2} ,
\end{equation}
where $k := k_1k_2$.

\begin{figure}[]
\centering
\includegraphics[width=0.6\textwidth]{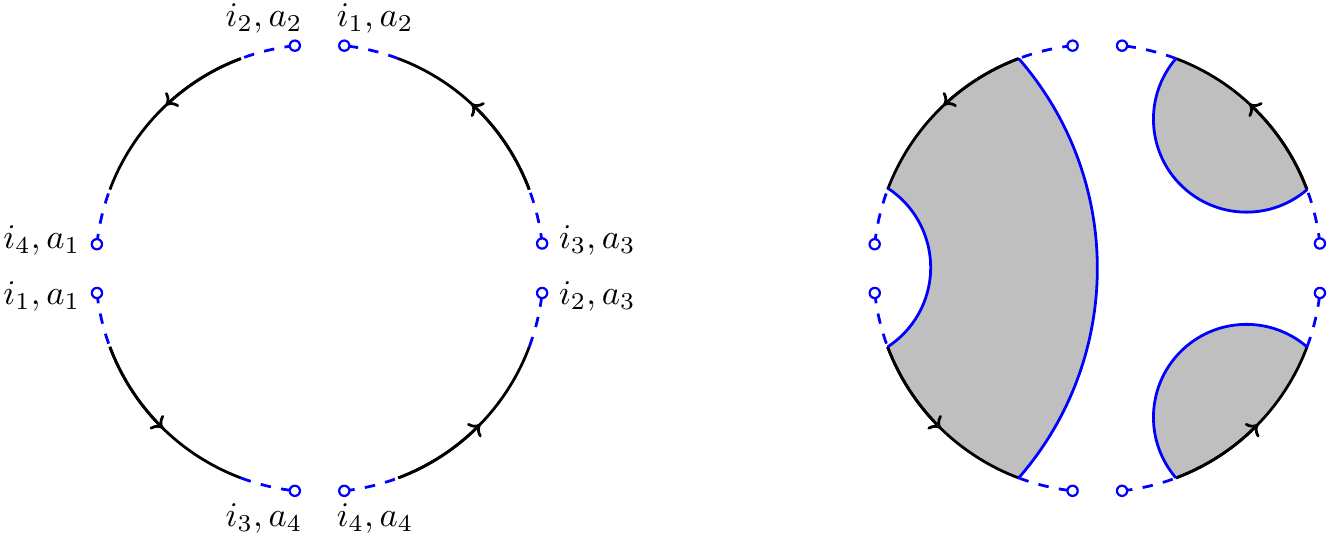}
\caption{Replica trick for negativity after tracing out part of the radiation.}
\label{fig:west_coast_tr_rad}
\end{figure}

\begin{figure}[]
\centering
\includegraphics[width=0.6\textwidth]{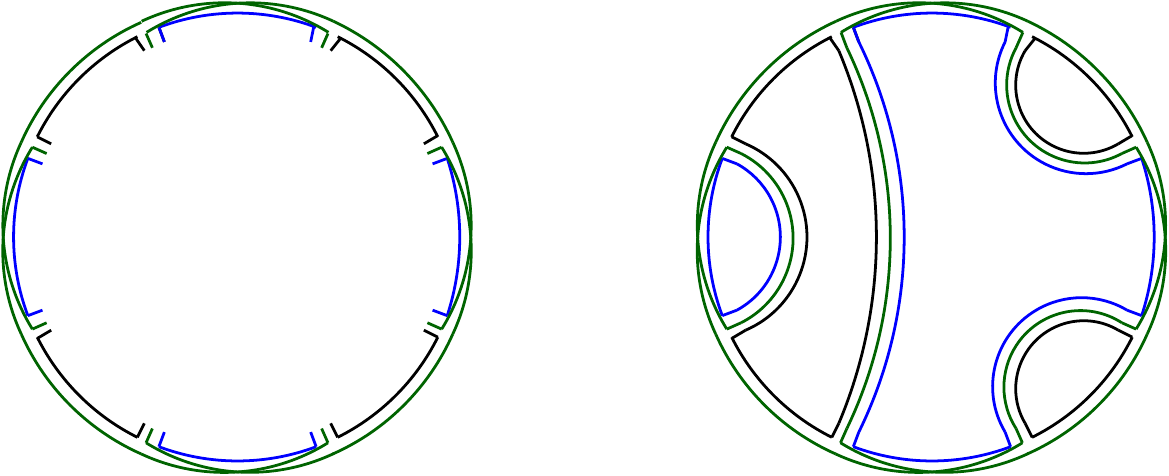}
\caption{Equivalent one-tensor network diagrammatic representation of Fig.~\ref{fig:west_coast_tr_rad}.}
\label{fig:west_coast_tr_rad_1_tensor}
\end{figure}

First, suppose that we do not have access to part of the radiation corresponding to system $R_2$, and would like to measure the entanglement between subsystem $R_1$ of the radiation, and the black hole. In this case, we should trace out $R_2$ and remain with a density matrix, omitting the subscripts in the states describing the corresponding system
\begin{equation}
\rho =\tr _{R_2} |\psi\rangle \langle \psi| = \frac{1}{k} \sum_{i,a,j} |\psi_{i,a} \rangle |i\rangle \langle \psi_{j,a} |\langle j| .
\end{equation}
For the negativity, we consider the partially transposed density matrix $\rho ^{T_{R_1} } $, having moments
\begin{equation} \label{eq:west_coast_tr_rad_moment}
\begin{split}
& \tr \left(\rho ^{T_{R_1} } \right) ^n = \frac{1}{k^n} \tr \left[ |\psi_{i_1,a_1} \rangle |j_1\rangle \langle \psi_{j_1,a_1} |\langle i_1| \cdot |\psi_{i_2,a_2} \rangle |j_2\rangle \langle \psi_{j_2,a_2} |\langle i_2| \cdots \right] =\\
& = \frac{1}{k^n } \langle \psi_{j_1,a_1} |\psi_{i_2,a_2} \rangle \langle \psi_{j_2,a_2} |\psi_{i_3,a_3} \rangle  \langle \psi_{j_3,a_3} |\psi_{i_4,a_4} \rangle \cdots \langle \psi_{j_n,a_n} |\psi_{i_1,a_1} \rangle \delta _{i_1,j_2} \delta _{i_2,j_3} \cdots \delta _{i_n,j_1} = \\
&= \frac{1}{k^n} \langle \psi_{i_n,a_1} |\psi_{i_2,a_2} \rangle \langle \psi_{i_1,a_2} |\psi_{i_3,a_3} \rangle \langle \psi_{i_2,a_3} |\psi_{i_4,a_4} \rangle \cdots \langle \psi_{i_{n-1} ,a_n} |\psi_{i_1,a_1} \rangle
\end{split}
\end{equation}
with an implicit summation over the indices. The boundary condition for the gravitational theory that this expression gives is shown on the LHS of Fig.~\ref{fig:west_coast_tr_rad} with the explicit appearance of the indices, and an example for a contributing configuration including a wormhole is shown on the RHS for $n=4$. This index structure can conveniently be represented using the diagrammatic notation that we used before, as shown in figure
\ref{fig:west_coast_tr_rad_1_tensor}. In this representation, the black lines correspond to the asymptotic boundary of JT gravity, the green line corresponds to the $R_1$ part, and the blue line to the $R_2$ part. The diagram implements precisely the index constraints, so that now we do not need to write the indices explicitly. Any bulk configuration is given by contracting a triple line on the left of the blue segments on the boundary with a triple line on the right of a blue segment. This gives the possible geometries, excluding handles that we can neglect when the entropy parameter $S_0$ that multiplies the Euler term in JT gravity is large. These are precisely the rules we had for the one-tensor model. In the JT language, this multiline corresponds to the worldline of the EOW particle.

The value that we should assign in JT gravity to each configuration contributing to the $n^{th}$ moment is a factor of $Z_m$ for every wormhole having $m$ boundaries, and normalize by $\frac{1}{(k Z_1)^n} $ where the $k$ dependence comes from the normalization of the state in \eqref{eq:west_coast_tr_rad_moment} and the $Z_1$ dependence normalizes the bulk integral. $Z_m$ here is the JT gravity partition function with $m$ asymptotic boundaries in the presence of EOW branes. Every closed loop of the $R_1$ system gives a factor of $k_1$ and similarly for $R_2$. As explained in \cite{2019arXiv191111977P}, in the micro-canonical ensemble with entropy $\Smicro $, $Z_m$ is given by $e^{\Smicro}$ times an $m^{th}$ power law which simply cancels with the normalization $1/Z_1^n$, so that we can effectively replace $Z_m \rightarrow e^{\Smicro} $. Therefore, in the micro-canonical ensemble, where we assume that $e^{\Smicro} $ and $k_1,k_2$ are large such that we can neglect handles, the model is the same as a one-tensor network where we consider the negativity between two systems of Hilbert space dimensions $e^{\Smicro} $ and $k_1$, after tracing out a system of size $k_2$. The diagrammatic rules are indeed the same as in Section \ref{sec:1_tensor}.

\begin{figure}[]
\centering
\includegraphics[width=0.5\textwidth]{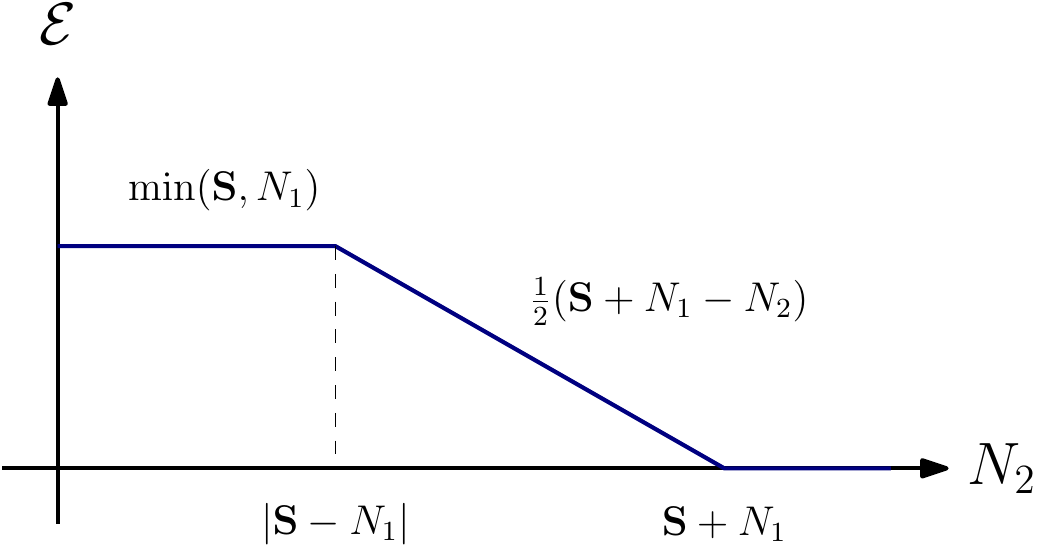}
\caption{Entanglement as measured by negativity between a fixed amount of radiation that we have access to, and the black hole, as a function of an increasing amount of escaping radiation.}
\label{fig:west_coast_fixed_rad_amount}
\end{figure}

Suppose that while the black hole evaporates, which is described by increasing $k$, we only have access to a fixed amount of radiation, and 
ask how the entanglement between this radiation and the black hole behaves. This is described by keeping $k_1$ and $e^{\Smicro} $ fixed while increasing $k_2$. It corresponds to moving along a vertical line in the phase diagram of Fig.~\ref{fig:one_tensor_phase_diagram}. Let us denote by $N_1=\log k_1$ and similarly for $N_2$, representing the number of Hawking particles. At first when $N_2$ is very small, we are in phase II where the negativity is $\min\left( \Smicro, \log k_1\right) $. This is the regime where we have approximately a pure state, reproducing the result of \cite{2019arXiv191111977P} and corresponding to the Page curve. As we increase $N_2$, after passing $|\Smicro - N_1|$, we move to phase III where the negativity starts decreasing and is $\frac{1}{2} \left(\Smicro + N_1-N_2\right) $.\footnote{There is an $O(1)$ contribution of $\log \frac{8}{3\pi } $ in the negativity in this regime, but we are interested in large dimensions.} Eventually, for $N_2 > \Smicro+N_1$ the negativity remains zero. In the case that $\Smicro=N_1$, we do not have the first step (which is phase II). This behavior as a function of $N_2$ is shown in Fig.~\ref{fig:west_coast_fixed_rad_amount}.
Note that even though we still have some portion of the radiation for $N_2>\Smicro+N_1$, the negativity vanishes.

\begin{figure}[]
\centering
\includegraphics[width=0.6\textwidth]{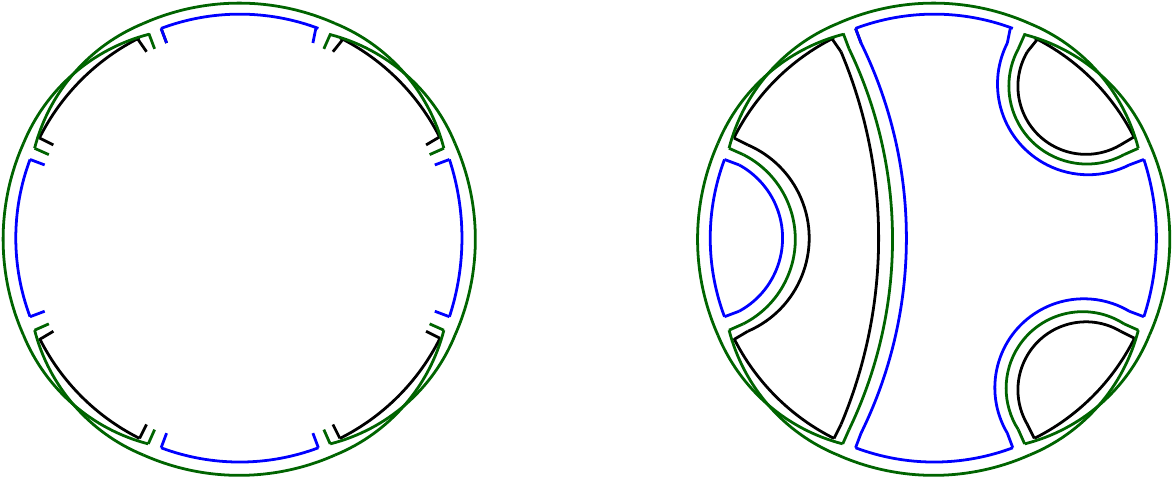}
\caption{one-tensor network diagrams for the negativity between the two parts of the radiation.}
\label{fig:west_coast_tr_BH}
\end{figure}

Alternatively, we can also trace out the black hole and consider instead the entanglement between the two parts of the radiation. Similarly to before, the moments are now
\begin{equation}
\begin{split}
& \tr \left[ \left( \tr _{B} |\psi\rangle \langle \psi|\right) ^{T_{R_2} } \right] ^n = \frac{1}{k^n} \langle \psi_{i_2,a_1} |\psi_{i_1,a_2} \rangle \langle \psi_{i_3,a_2} |\psi_{i_2,a_3} \rangle  \langle \psi_{i_4,a_3} |\psi_{i_3,a_4} \rangle \cdots \langle \psi_{i_1,a_n} |\psi_{i_n,a_1} \rangle .
\end{split}
\end{equation}
This can be represented in the diagrammatic notation as before by the diagrams shown in Fig.~\ref{fig:west_coast_tr_BH}. The difference from before is that the basic element corresponding to the density matrix was built on top of the blue segments before, while now it is on the black segments. This is of course precisely consistent with what system we trace out if we compare to the one-tensor network notation. The rules for evaluating diagrams are again as in the one-tensor case, where now we consider the negativity between spaces of dimensions $k_1$ and $k_2$, after tracing out a system of dimension $e^{\Smicro} $.

We again get a description of the different cases of the negativity if we think about fixing the radiation sizes $N_1$ and $N_2$, while changing the size of the traced out system, that is changing $\Smicro$ starting from large to small. It also makes sense to consider this physically. In this case we again move vertically in the phase diagram of Fig.~\ref{fig:one_tensor_phase_diagram}, now from bottom to top. If we consider the state \eqref{eq:west_coast_state} naively, as if the different $B$ states are orthonormal, then after reducing to the radiation subsystem, we have a density matrix corresponding to a statistical mixture, but quantum mechanically an unentangled state. Indeed, for large $\Smicro$ the negativity vanishes. Note that naively, before the correction giving the Page curve, the black hole system has (exponent of) $N_1+N_2$ states which keeps increasing. Indeed, there is a transition that begins precisely once we get to $\Smicro=N_1+N_2$, where the negativity is no longer vanishing. We can understand it as follows. Once there are $|\psi_{i,a}\rangle $ states that are identical, we start generating parts of the density matrix that correspond to an entangled state $\sum _{i,a} |i\rangle |a\rangle $ where the sum is not over the full range but rather part of it. This leads to a linear growth. However, once we reach $\Smicro=|N_1-N_2|$, we get to a plateau where the negativity remains constant and does not grow anymore, as it cannot exceed $\min(N_1,N_2)$. Note that in the diagrams of Fig.~\ref{fig:west_coast_tr_BH} (differently from those of Fig.~\ref{fig:west_coast_tr_rad_1_tensor}), the permutation structure corresponds directly to the wormhole structure in JT gravity in the sense that no transformation is needed, since the density matrix sits on the black segments. At the beginning, when $\Smicro$ was very large, we were dominated by the identity permutation with no wormholes at all. In the plateau region phase on the contrary, we are dominated by a 1-cycle permutation, so that the wormholes span over all the copies of the system.

\section{Negativity in holographic fixed-area states} \label{sec:holography}

\subsection{Fixed-area states}

The information theoretic aspects of random tensor networks appear in holography as holographic fixed-area states \cite{Akers:2018fow,Dong:2018seb}. Let us review this starting with entanglement entropy.

\begin{figure}[]
\centering
\includegraphics[width=0.6\textwidth]{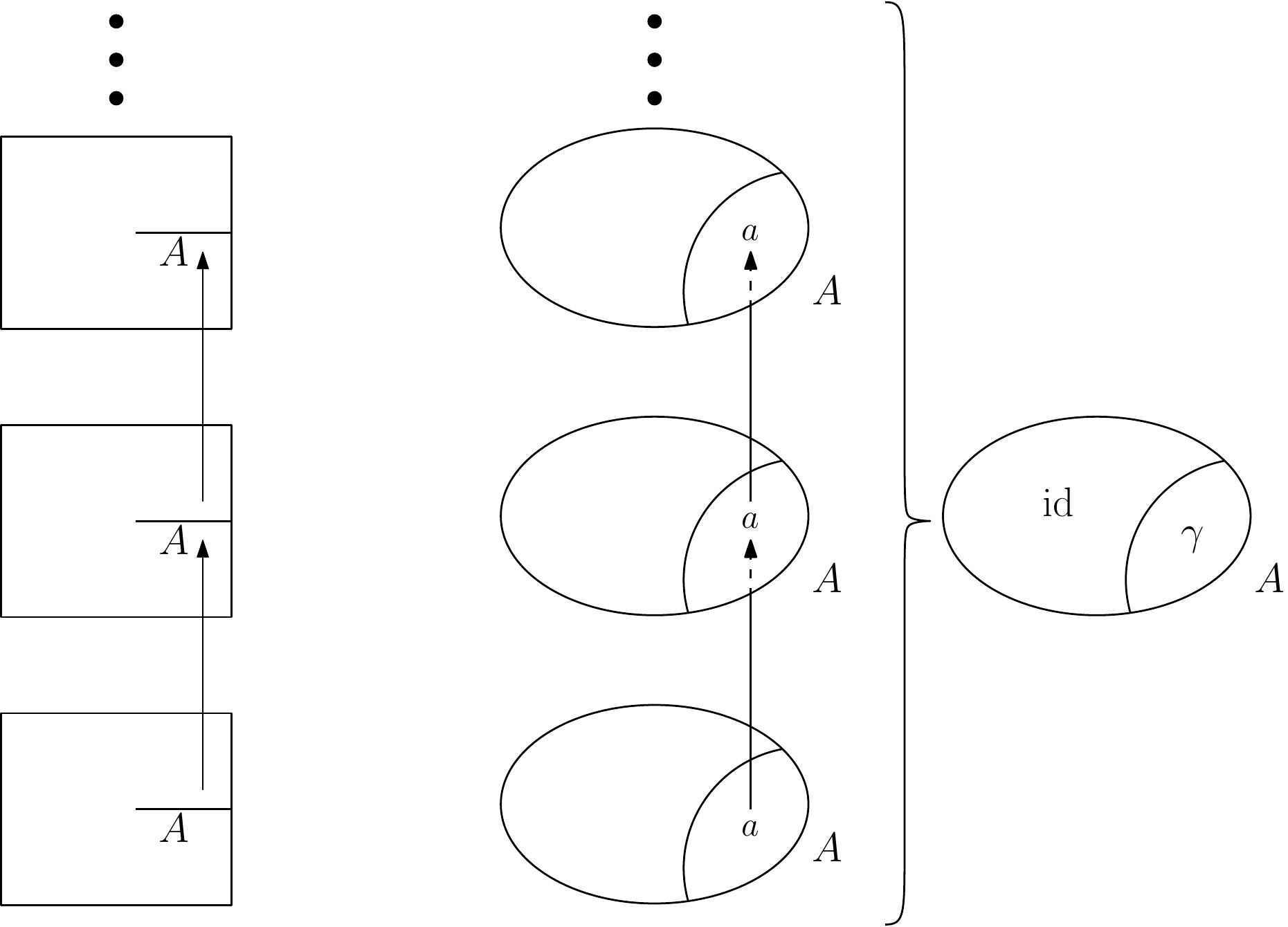}
\caption{Branched surfaces for the R\'enyi entropies. On the LHS we show the gluing on the field theory side. On the RHS, a spatial slice of the bulk is shown.}
\label{fig:replica_EE}
\end{figure}

In quantum field theory we can calculate the entanglement entropy by analytic continuation from the R\'enyi entropies using the replica trick. Given a subsystem $A$ and its complement $B$, the entanglement between $A$ and $B$ is studied using the reduced density matrix $\rho _A$ on $A$. For states prepared using the path integral we calculate the moments $\tr \rho _A^n$ by taking $n$ copies of the system, cutting each one open along $A$, and gluing the copies cyclically, i.e., using the permutation $\gamma $. In system $B$ we have no cuts, so we can think about it as gluing the $n$ copies using the identity permutation along $B$. This is shown on the left hand side of Fig.~\ref{fig:replica_EE}.

In holographic states we can calculate the $n^{th}$ moment by considering the bulk dual of this branched covering space. The simplest candidate for the bulk topology is to take the bulk dual of a single copy of the system, and consider a subregion $a$ that intersects the boundary at $A$. We denote the boundary of $a$ by $\partial a=\gamma _A \cup A$. Taking $n$ copies of this system branched over $a$, and glued cyclically using $\gamma$ along $a$ we get an admissible topology for the corresponding bulk geometry (see Fig.~\ref{fig:replica_EE}).

Semi-classically, we can approximate the path integral of the bulk partition function by $Z=e^{-I[g_n]} $ where $I$ is the action evaluated on the solution $g_n$ for the geometry. For simplicity, let us consider Einstein-Hilbert as the bulk effective theory. Then, the $n^{th}$ R\'enyi entropy is approximated by
\begin{equation} \label{eq:Renyi_bulk_approx}
S_n = \frac{I[g_n]-nI[g_1]}{n-1} .
\end{equation}
In general, the geometry $g_n$ differs from $g_1$, even locally. This is due to backreaction from coupling the replicas. In the $n \to 1$ limit that gives the entanglement entropy, the surface $\gamma _A$ is extremal in $g_1$, giving a replica trick explanation \cite{Lewkowycz:2013nqa} of the Ryu-Takayanagi formula \cite{Ryu:2006bv, Ryu:2006ef}.

We can now consider a state where we fix the area of the extremal surface homologous to $A$ in the bulk to some value $\mathcal{A} $ \cite{Akers:2018fow,Dong:2018seb}. That is, we include in the path integral a delta function setting it to a fixed value. If we enforce this using a Lagrange multiplier, it enters the action only locally along the RT surface. This means that for $n=1$, away from the RT surface, the equations of motion are unchanged compared to the situation where we do not fix the area. At the RT surface, there is a cosmic brane with a conical deficit angle $\phi $ set such that the area in the classical solution agrees with the fixed area. Now, for general integer $n$, we can glue $n$ copies of this solution along the entanglement wedge $a$ with the same geometry locally and still get a solution. This is because the equations of motion are satisfied away from $\gamma _A$ and the area agrees with the fixed area. The total (uniform) deficit angle around $\gamma _A$ is now $n\phi $. This is the reason for the simplicity of fixed-area states. The Einstein-Hilbert action (without a cosmic brane term that vanishes for the fixed area) evaluated with this solution has a linear piece in $n$ coming from the geometry away from the RT surface. Because of the deficit angle, we get a piece proportional to $(n\phi -2\pi )$ from the conical defect at the RT surface times the fixed area. The terms linear in $n$ do not contribute to the R\'enyi entropies, as they cancel in \eqref{eq:Renyi_bulk_approx}. Therefore, we get that the R\'enyi entropies are $\frac{\mathcal{A} }{4G_N} $ where $G_N$ is Newton's constant. That is, we get the Bekenstein-Hawking entropy with the fixed area and a flat entanglement spectrum as the entropies are independent of $n$.

\begin{figure}[]
\centering
\includegraphics[width=0.6\textwidth]{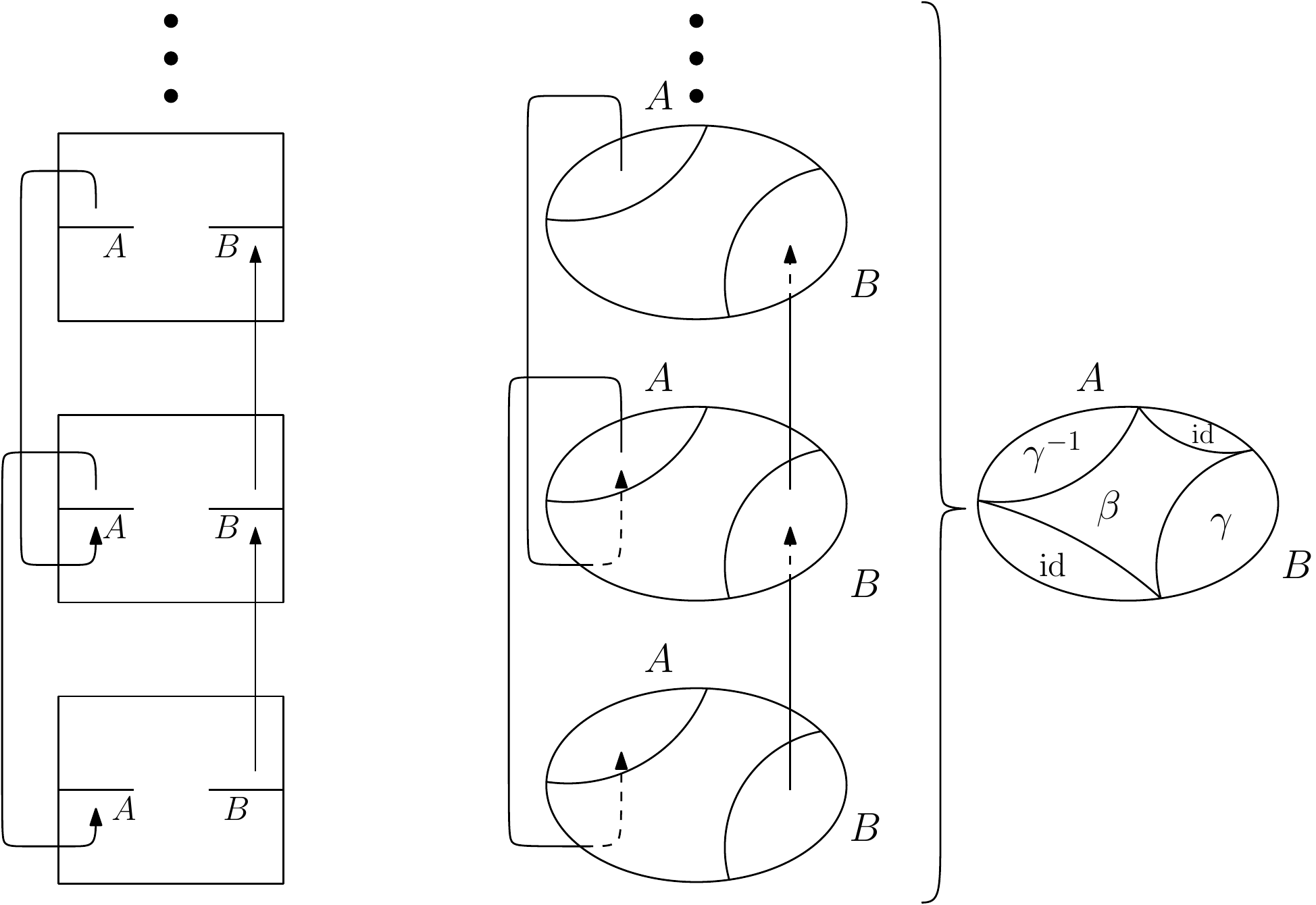}
\caption{Branched surfaces for the negativity moments.}
\label{fig:Renyi_negativity}
\end{figure}

Let us now move on to negativity which was previously considered in \cite{Dong:2021clv}. Now we partition the quantum field theory into three subsystems $A$ ,$B$, and $C$. We trace out system $C$ and do a partial transpose over system $A$. The moments of the partially transposed density matrix $\tr _C \rho ^{T_A} $ in this case are obtained by taking $n$ copies, cutting them open along $A$ and $B$, gluing along $B$ with $\gamma $, while gluing along $A$ with $\gamma ^{-1} $, implementing the partial transpose. Along $C$, we glue with the identity as we did before for the traced system. This is shown on the LHS of Fig.~\ref{fig:Renyi_negativity}.

We can construct a bulk topology by extending $A$, $B$, and $C$ into the bulk as in \cite{Dong:2021clv}. In the configuration shown in Fig.~\ref{fig:Renyi_negativity}, the gluing in the bulk is done in correspondence with the boundary permutations, while in the remaining region we can allow a-priori any permutation $\beta  $.

Again, it is simplest to see what happens when we fix the areas of the RT surfaces of all these regions. As before, the contribution away from the entangling surfaces is linear in $n$ and does not contribute to the negativity moments as in \eqref{eq:Renyi_bulk_approx}. Now, consider an entangling surface such that on one side of it, we glue according to a permutation $\sigma _1$ and on the other side $\sigma _2$. As we encircle the entangling surface, we move between the $n$ copies according to $\sigma _1 \circ \sigma _2^{-1} $. For any cycle of size $n_i$ of this permutation, we get a contribution proportional to $(n_i \phi -2\pi ) \mathcal{A} $, with $\mathcal{A} $ being the fixed area, to $I[g_n]$, as we mentioned before. Since $\sum _in_i=n$, the $n_i$ terms again cancel in the moments. We remain with $2\pi  \mathcal{A} $ times the number of cycles. Therefore, each such entangling surface contributes to the $n^{th}$ moment simply as $\frac{\mathcal{A}}{4G_N}  C(\sigma _1 \circ \sigma _2^{-1} )$. We get that the moments are
\begin{equation}
m_n=\frac{Z_n^{(PT)}}{Z_1^n} 
\end{equation}
with
\begin{equation} \label{eq:fixed_area_negativity_moments}
Z_n^{(PT)} = \exp \left[  \sum _i \frac{\mathcal{A} _i}{4G_N}  C(\sigma ^{(i)} _1 \circ \sigma ^{(i)} _2)\right] ,
\end{equation}
where the sum goes over the fixed areas in the bulk.
For the example of Fig.~\ref{fig:Renyi_negativity}, we have
\begin{equation} \label{eq:fixed_area_1_tensor}
Z_n^{(PT)} = \exp \left[ \frac{A[\gamma _A]}{4G_N} C(\gamma \circ \beta  )+\frac{A[\gamma _B]}{4G_N} C(\gamma ^{-1} \circ \beta  )+\frac{A[\gamma _C]}{4G_N} C(\beta  )\right] ,
\end{equation}
where $A[\gamma_B]$ is the fixed area of the RT surface of $B$ and so on.

\begin{figure}[]
\centering
\includegraphics[width=0.3\textwidth]{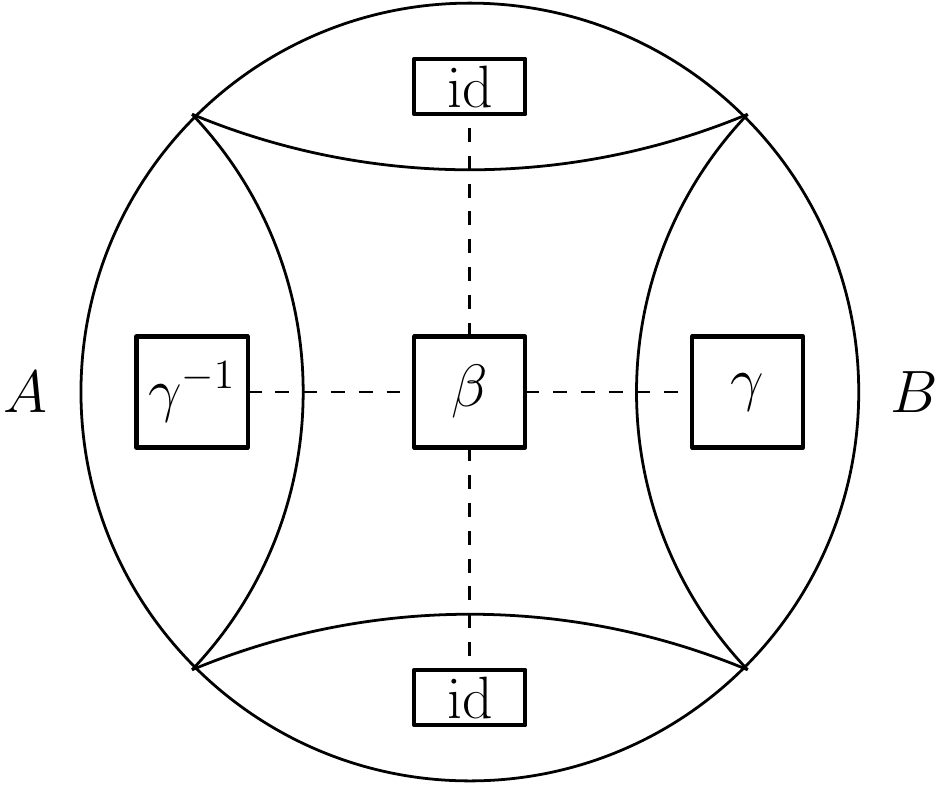}
\caption{Matching a fixed-area state with the corresponding tensor network.}
\label{fig:fixed_area_1_tensor}
\end{figure}

Equation \eqref{eq:fixed_area_negativity_moments} is exactly the same as the moments of a random tensor network \eqref{eq:RTN_moments}. For any internal region in the bulk, we place a random tensor. The bulk regions touching boundary subsystems that we partially transpose have a network node of $\gamma ^{-1} $. Those that are traced out have an id node and those corresponding to subsystems that we do not trace or partially transpose have a $\gamma $ node. The negativity moments here and in \eqref{eq:RTN_moments} then agree once we identify the Hilbert space dimension $L$ of a tensor network edge connecting two tensors with the exponential of the fixed area crossed by this edge $\exp\left[ \mathcal{A} /(4G_N)\right] $. Fig.~\ref{fig:Renyi_negativity} with its moments \eqref{eq:fixed_area_1_tensor} is the simplest tensor network which is the one-tensor network, as shown explicitly in Fig.~\ref{fig:fixed_area_1_tensor}.

\subsection{Two-tensor network holographically} \label{sec:holography_2_tensor}

\begin{figure}[]
\centering
\includegraphics[width=0.4\textwidth]{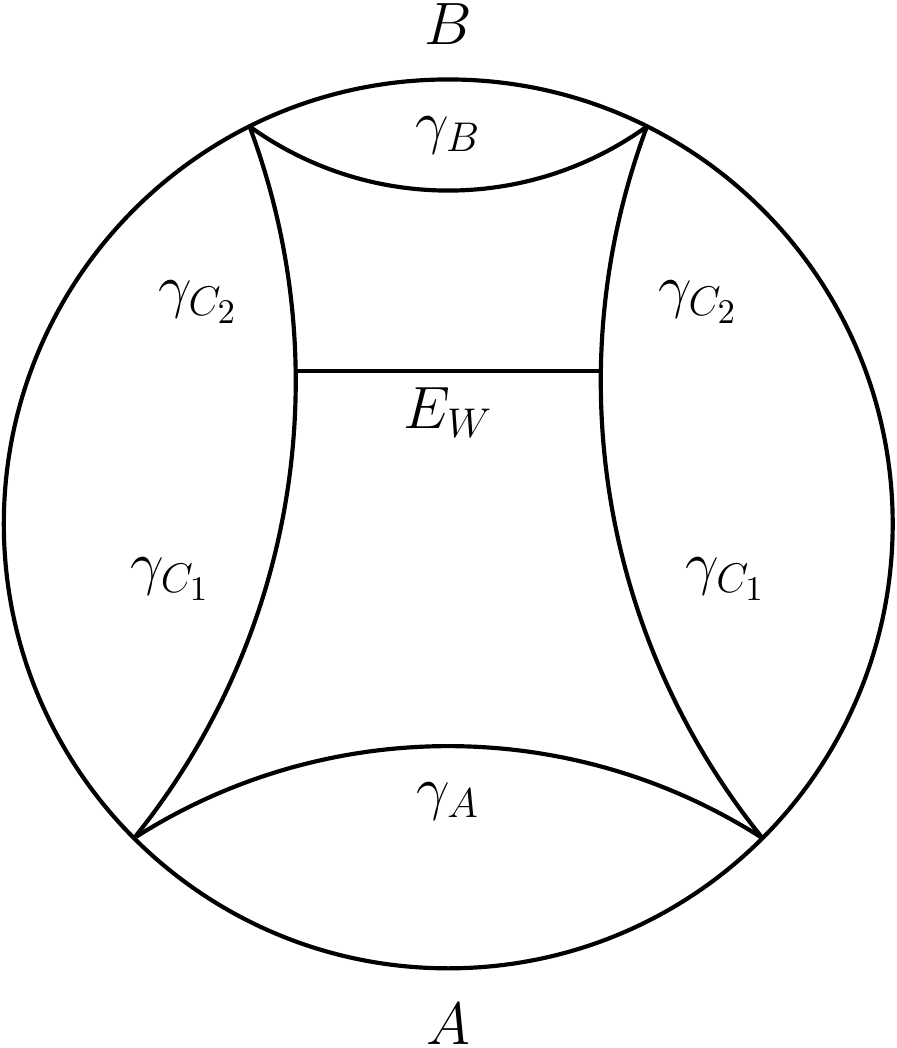}
\caption{The holographic setting described by the two-tensor system.}
\label{fig:two_intervals_wedge}
\end{figure}

We may also have higher tensor networks corresponding to states with more fixed areas. Let us mention the holographic setting that is described by the two-tensor network of Section \ref{sec:2_tensor}. It is shown in Fig.~ \ref{fig:two_intervals_wedge}. We have three subsystems $A$, $B$, and $C$ of the boundary theory, just as before. In the bulk, the RT surface of $A$ is denoted by $\gamma _A$, and similarly for $B$ and $C$. The connected RT surface for the $AB$ system is shown as well, and it is divided in two by the entanglement wedge cross section, the extremal surface with two ends on $\gamma_{AB}$. The entanglement wedge cross section is denoted by $E_W$ and the piece of the connected RT surface that connects to $A$ is $\gamma _{C_1} $, while the remaining piece is $\gamma _{C_2} $.

In the general phase structure of the two-tensor network shown in Table \ref{tab:2_tensor_phases}, this holographic setting can be achieved in phases 1 and 4, while the rest of the phases are necessarily non-holographic. 
This is because in the holographic setting we have $A[\gamma _{C_1}]+A[E_W] > A[\gamma _A]$ and $A[\gamma _{conn}] +A[\gamma _B] > A[\gamma _A]$\ as $\gamma _A$, $\gamma _B$ are RT surfaces, where $\gamma _{conn} =\gamma _{C_1} \cup \gamma _{C_2} $ and similarly for switching $A$ and $B$. In both holographic phases, we saw that the two-tensor network essentially behaves as a one-tensor network since at leading order the two permutations are equal.
One regime is when $L_{C_1} L_{C_2} \gg L_A L_B$, which belongs to phase 1 in Table \ref{tab:2_tensor_phases}. It corresponds in holography to the disconnected regime where the dominant RT surface for the entanglement entropy of $AB$ is $\gamma _A \cup \gamma _B$. By matching to the corresponding one-tensor network dimensions, we see in Fig.~\ref{fig:one_tensor_phase_diagram} that we land on region I in the phase diagram where the negativity vanishes. The other regime is when $L_{C_1} L_{C_2} \ll L_A L_B$, belonging to phase 4, which corresponds to the connected regime in holography where the dominant RT surface for $AB$ is $\gamma _{conn} =\gamma _{C_1} \cup \gamma _{C_2} $. Since in holography we always have $A[\gamma _{conn}] +A[\gamma _B] > A[\gamma _A]$ and $A[\gamma _{conn}]+A[\gamma _A]>A[\gamma _B]$, that means that we are in region III of the phase diagram in Fig.~\ref{fig:one_tensor_phase_diagram}. The negativity spectrum in this region is a semi-circle as we saw, and as was also found in \cite{Dong:2021clv}. Note that region II with the Marchenko–Pastur distribution is not realized holographically in the setting of Fig.~\ref{fig:two_intervals_wedge}. It turns out that region II can be realized holographically in multiboundary wormholes.

\subsection{Wormholes} \label{sec:wormhole_1_tensor}

\begin{figure}[]
\centering
\includegraphics[width=0.5\textwidth]{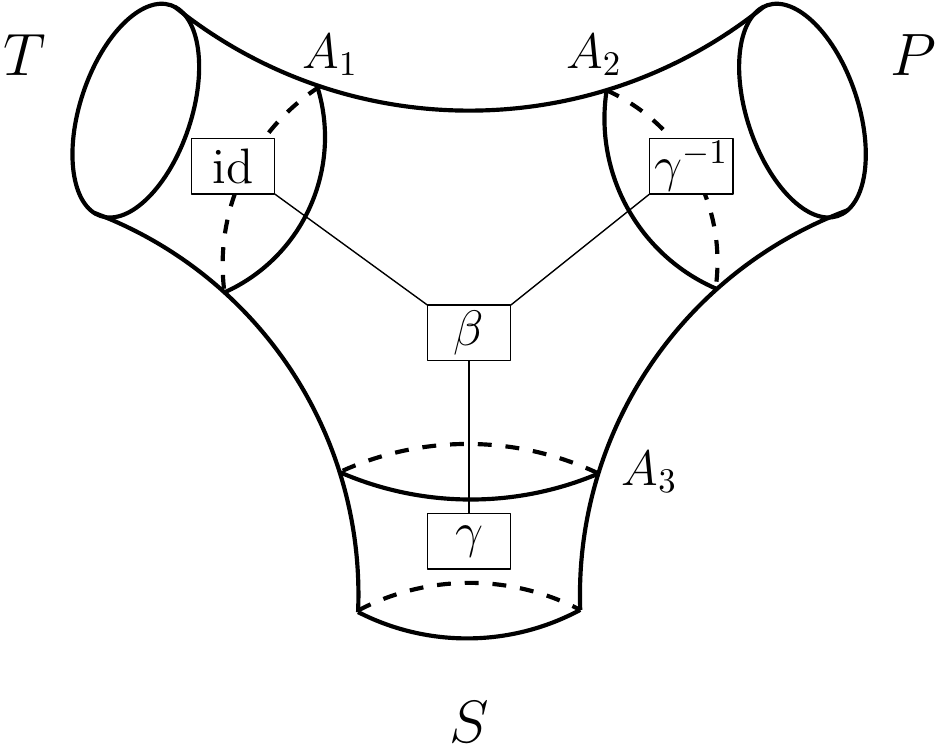}
\caption{A three-boundary wormhole configuration.}
\label{fig:three_wormhole}
\end{figure}

Let us consider a three-boundary wormhole configuration with three fixed areas $A_1$ ,$A_2$, and $A_3$ that corresponds to the one-tensor networks (see Fig.~\ref{fig:three_wormhole}). Note that while either of (say) $A_1$ or $A_2 + A_3$ could be an RT surface for the same boundary region, they differ topologically.

As a first case, consider the situation where $A_1=A_2+A_3$. Let us denote by $N^{w_i}$ the value $e^{A_i/(4G_N)} $
for the minimal area from now on, so that here $w_1=w_2+w_3$. In this case the flow is $\text{id} \to \beta  \to \gamma ^{-1} $ and $\text{id} \to \beta  \to \gamma $. This means that $\beta $ is a non-crossing configuration made of one- and two-cycles only. Therefore the moments are
\begin{equation}
\mean{  \tr \rho '^n}  = N^{-w_1(n-1)} |NC_{1,2} |
\end{equation}
where $|NC_{1,2} |$ is the number of non-crossing partitions of a cycle made of $n$ elements consisting of parts of size 1 or 2 only.

The number of non-crossing partitions of a cycle of size $n$ with $s_k$ parts of size $k$ is given by \cite{kreweras1972partitions}
\begin{equation}
v(s) = \frac{(n)_{h-1} }{s_1! s_2! \cdots } = \frac{n!}{(n-h+1)! s_1! s_2! \cdots }
\end{equation}
where
\begin{equation}
h=s_1+s_2+\cdots 
\end{equation}
is the total number of parts in the partition.

Let us denote by $k=s_2$ the number of two-cycles. Then there are $n-2k$ one-cycles, and so the total number of partitions we are interested in is
\begin{equation}
|NC_{1,2} |=\sum _{k=0} ^{\lfloor n/2 \rfloor} \frac{n!}{(k+1)!k!(n-2k)!} ={}_2F_1\left( \frac{1-n}{2} ,-\frac{n}{2} ,2;4\right) .
\end{equation}
Denoting this number by $a_n$, we note that it satisfies the recursion relation
\begin{equation}
(n+3)a_{n+2} -(3+2n)a_{n+1} -3na_n=0,\qquad a_0=a_1=1.
\end{equation}

$|NC_{1,2}|$ are the Motzkin numbers $M_n$ as those are defined precisely as the number of possibilities to draw non-crossing chords on a circle with $n$ nodes, not necessarily covering all the nodes so that we think about the leftover nodes as being 1-cycles. We find
\begin{equation}
    N^{w_1 n} \mean{ \tr \rho '^n } =N^{w_1} M_n .
\end{equation}
The generating function of the Motzkin numbers is
\begin{equation}
    \sum _{n=0}^{\infty } M_n x^n = \frac{1-x-\sqrt{1-2x-3x^2}}{2x^2}.
\end{equation}
The resolvent of $\rho '$ is therefore
\begin{equation}
    R(z )=
    \frac{N^{2w_1}}{2}\left( N^{w_1}z -1-N^{w_1}\sqrt{\left( z -\frac{3}{N^{w_1}}\right) \left( z +\frac{1}{N^{w_1}}\right) } \right) ,
\end{equation}
so
\begin{equation}
    \ns (\lambda ) = \frac{N^{3w_1}}{2\pi } \sqrt{\left( \frac{3}{N^{w_1}}-\lambda \right) \left( \lambda +\frac{1}{N^{w_1}}\right) } \cdot \mathds{1}_{\left[ -\frac{1}{N^{w_1}},\frac{3}{N^{w_1}}\right] }.
\end{equation}
This is a semi-circle distribution, non-symmetric with respect to the origin, having support in the negative part as well.

Another interesting regime is when $A_3=A_1+A_2$ (equivalently $w_3=w_1+w_2$). The flow is then $\text{id}\to \beta  \to \gamma $ and $\gamma ^{-1} \to \beta  \to \gamma $. This gives conditions
\begin{equation}
    \begin{split}
        & |\beta |+|\beta ^{-1} \circ \gamma | = |\gamma |,\\
        & |\gamma \circ \beta |+|\beta ^{-1} \circ \gamma | = |\gamma ^2|,
    \end{split}
\end{equation}
or alternatively the contributing permutations are such that for odd $n$
\begin{equation}
    \begin{split}
        & \beta \text{ is NC, i.e., } |\beta |+|\beta ^{-1} \circ \gamma | = |\gamma |,\\
        & C(\beta )=C(\gamma  \circ \beta )
    \end{split}
\end{equation}
(where NC stands for non-crossing) while for even $n$
\begin{equation}
    \begin{split}
        & \beta \text{ is NC, i.e., } |\beta |+|\beta ^{-1} \circ \gamma | = |\gamma |,\\
        & C(\beta )+1=C(\gamma  \circ \beta ).
    \end{split}
\end{equation}

We find the number of such permutations is $\frac{1}{n+1}\binom{3n/2}{n/2}$ for even $n$, while it is $\binom{(3n-1)/2}{(n-1)/2}$ for odd $n$ by checking many values explicitly.
\begin{figure}[]
\centering
\includegraphics[width=0.4\textwidth]{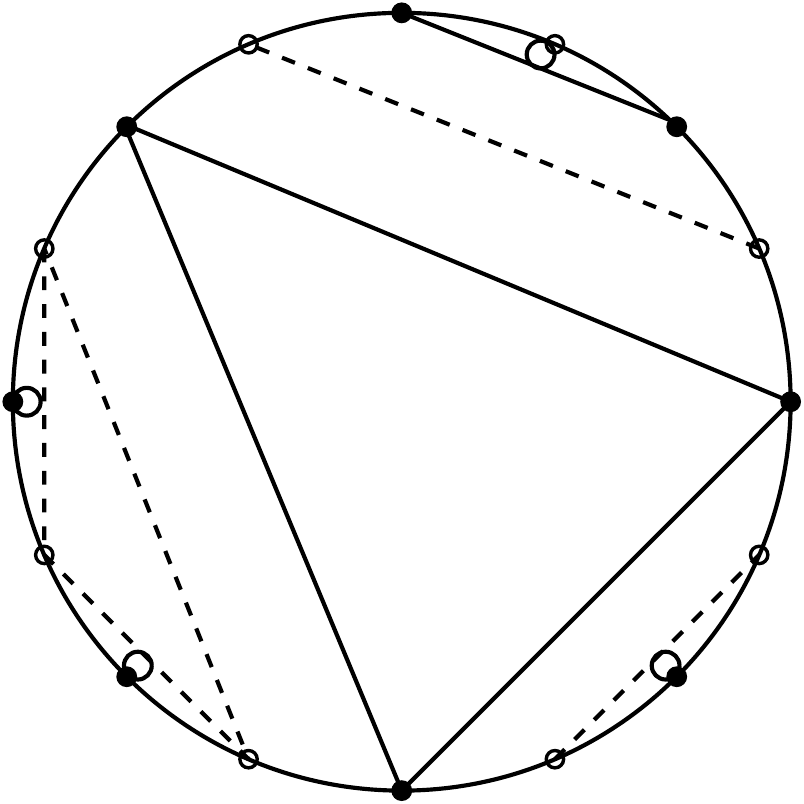}
\caption{Dual non-crossing partition to another non-crossing partition, also known as the Kreweras complement. One partition is shown using solid lines, while the other corresponds to dashed lines.}
\label{fig:dual_non_crossing}
\end{figure}
We will give a compact proof of this for even $n$.
As a first step, we notice that the two equations above are in fact equivalent to the condition that the permutation $\gamma \circ \beta ^{-1}$ is non-crossing in $\gamma $ and $\gamma ^2$. As a second step, we can count the number of such permutations as follows. Since for even $n$ the permutation $\gamma ^2$ has two cycles consisting of the even and odd sites, what we are after is the number of non-crossing permutations where each cycle is entirely contained either in the odd sub-lattice or the even sub-lattice. In order to count this, we denote by $\tau $ the corresponding partition of the odd sub-lattice. For every such non-crossing partition, there corresponds a unique non-crossing partition $\bar \tau $ of the even sub-lattice \cite{kreweras1972partitions}, where $\bar \tau $ is the maximal non-crossing partition such that the union of $\tau ,\bar \tau $ is a non-crossing partition; see Fig.~\ref{fig:dual_non_crossing} for an example. The even cycles of the partitions we are after must be precisely the non-crossing sub-partitions of $\bar \tau $. We thus count the number of $\tau ,\sigma \in S_{n/2} $ such that $\tau $ is non-crossing and $\sigma \le \bar \tau $ is non-crossing. This is the same as the number of non-crossing $\sigma \le \bar \tau $, which by definition is the second Fuss-Catalan number $FC^{(2)}_{n/2} = \frac{1}{n+1}\binom{3n/2}{n/2}$.

\begin{figure}[]
\centering
\includegraphics[width=0.5\textwidth]{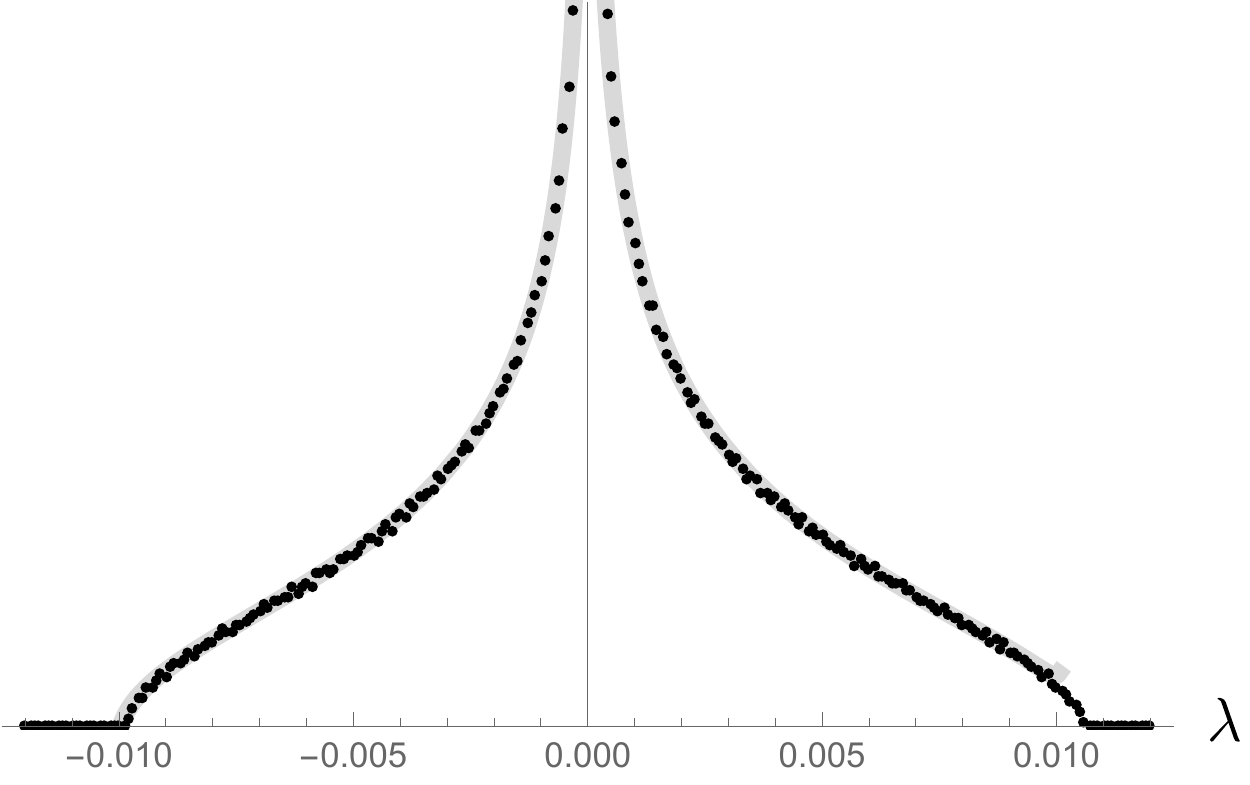}
\caption{Negativity spectrum at the phase transition when $w_3 = 2w_1 = 2w_2 = 4$ and $N= 2$. The average of $10$ realizations (black dots) is compared to \eqref{one_tensor_transition} (gray line).}
\label{fig:1_tensor_PT_simulation}
\end{figure}

The moments are thus
\begin{equation}
    \mean{ \tr \rho '^n} = \begin{cases}
    N^{(1-n)w_3+w_2} \cdot \frac{1}{n+1} \binom{3n/2}{n/2} & \text{even } n\\
    N^{(1-n)w_3} \binom{(3n-1)/2}{(n-1)/2} & \text{odd } n
    \end{cases}.
\end{equation}
From this, we get that the resolvent is
\begin{equation}
    R(\lambda ) = \frac{2N^{2w_3+w_2}}{\sqrt{3}} \sin \left[ \frac{1}{3} \arcsin \frac{3\sqrt{3}}{2N^{w_3}\lambda }\right]  + \frac{N^{w_3}}{\sqrt{3}} \cdot \frac{\sin \left[ \frac{2}{3}\arcsin\frac{3\sqrt{3}}{2N^{w_3} \lambda } \right] }{\lambda \sqrt{1 - \frac{27}{4N^{2w_3}\lambda ^2}}} .
\end{equation}
Let us denote
\begin{equation}
    x = \frac{3\sqrt{3}}{2N^{w_3} \lambda }.
\end{equation}
In order to get the density of states, we should look for the imaginary part using \eqref{eq:neg_spectrum_from_resolvent}. Both sine terms have an imaginary part for $|x|>1$. It can be verified that they do not have any imaginary parts outside of this range including delta function like terms. However, the square-root in the second term also has an imaginary part in this range. Therefore, for the second term we actually need the real part of the sine. Thus, there is no contribution to the density of states from the second term as well for $|x| \le 1$. We then obtain that the density of states is
\begin{equation}
\label{one_tensor_transition}
\ns (\lambda ) = \begin{cases}
\frac{N^{2w_3+w_2}}{2\pi } \left[ \left( \sqrt{x^2-1}+x\right) ^{1/3} -\left( \sqrt{x^2-1}+x\right) ^{-1/3} \right]   \\
\qquad+ \frac{N^{2w_3}}{6\sqrt{3}\pi } \frac{\left( \sqrt{x^2-1}+x\right) ^{2/3} +\left(  \sqrt{x^2-1}+x\right) ^{-2/3} }{\sqrt{1-x^{-2} }} & 0 \le \lambda \le \frac{3 \sqrt{3}}{2N^{w_3}}  \\
\frac{N^{2w_3+w_2}}{2\pi } \left[ \left( \sqrt{x^2-1}-x\right) ^{1/3} -\left( \sqrt{x^2-1}-x\right) ^{-1/3} \right]  \\
\qquad - \frac{N^{2w_3}}{6 \sqrt{3}\pi } \frac{\left( \sqrt{x^2-1}-x\right) ^{2/3} +\left( \sqrt{x^2-1}-x\right) ^{-2/3} }{\sqrt{1-x^{-2} }} & -\frac{3 \sqrt{3}}{2N^{w_3}} \le \lambda \le 0
\end{cases}.
\end{equation}
This is the closed form result at large-$N$. A plot of this is shown in Fig.~\ref{fig:1_tensor_PT_simulation}, 
where this formula is compared to a simulation of the one-tensor network. The spectrum in \eqref{one_tensor_transition} indeed satisfies $\int d\lambda \, \ns (\lambda )=N^{w_2+w_3}$ and $\int d\lambda \, \lambda  \ns (\lambda )=1$.


In terms of Fig.~\ref{fig:one_tensor_phase_diagram} (and Ref.~\cite{Shapourian:2020mkc}) this gives the negativity spectrum for the regime corresponding to the phase transition between phase II (called the maximally entangled phase in \cite{Shapourian:2020mkc}) with Marchenko–Pastur distribution and phase III (called the entanglement saturation phase) with a semi-circle distribution.

\subsection{Quantum corrections}

\begin{figure}
    \centering
    \includegraphics[height = 5cm]{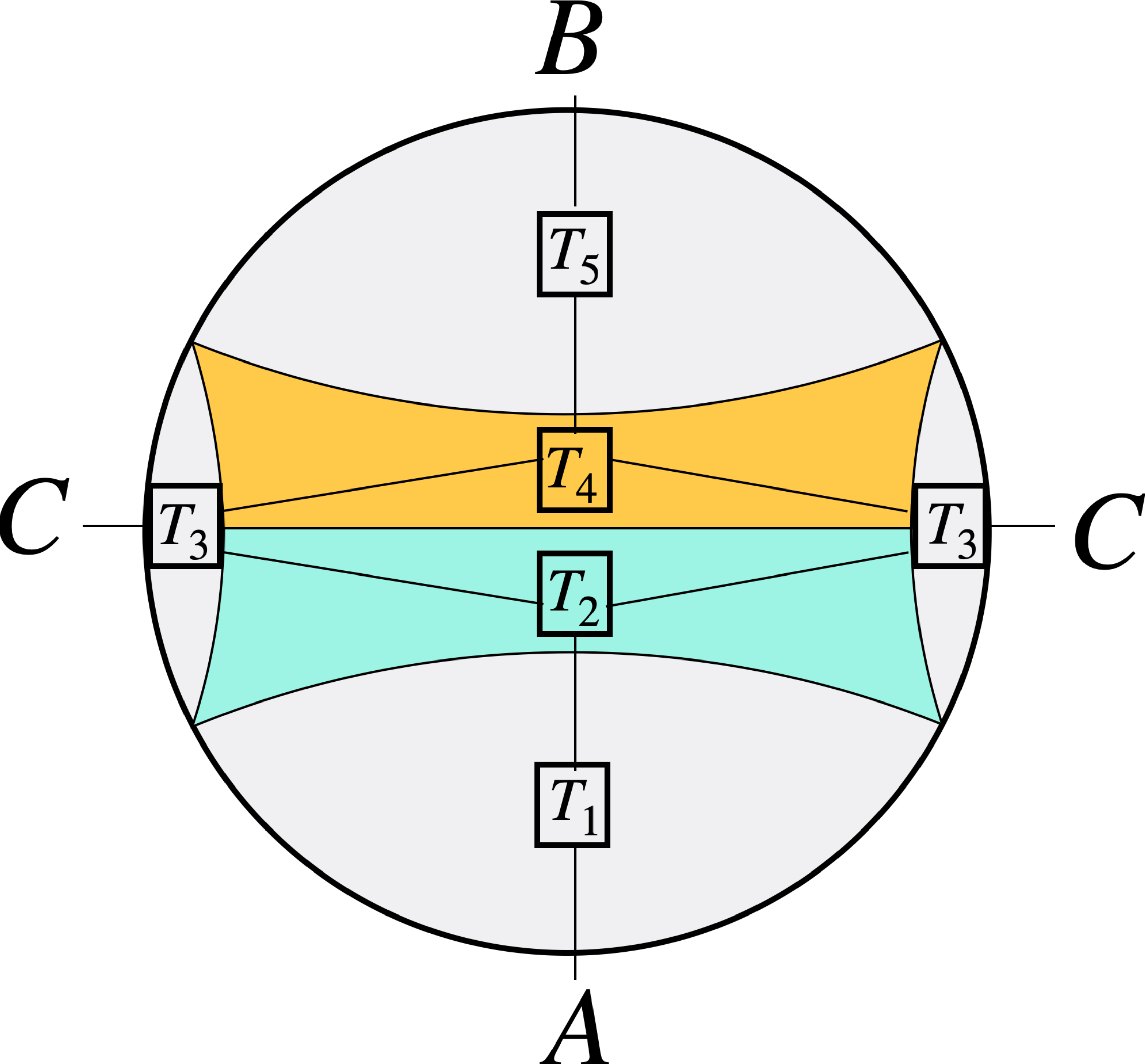}
    \caption{The bulk geometry for the negativity moments with the random tensor network overlaid.}
    \label{tensor_flow_corr_fig}
\end{figure}

So far, we have effectively been studying ``pure gravity'' without quantum fields propagating on the space. We now generalize these results to when there is a code subspace of states for each fixed geometry. First consider the case of disjoint regions, $A$ and $B$, where there are no bulk quantum fields. In this case, the flow network is (Fig.~\ref{tensor_flow_corr_fig})
\begin{align}
\begin{tikzpicture}
    \node[draw, shape=rectangle] (T3) at (0,-1.5) {$T_3$};
    \node[draw, shape=rectangle] (T2) at (3,0) {$T_2$};
    \node[draw, shape=rectangle] (T1) at (6,0) {$T_1$};
    \node[draw, shape=rectangle] (T4) at (3,-3) {$T_4$};
    \node[draw, shape=rectangle] (T5) at (6,-3) {$T_5$};
    \node[draw, shape=rectangle] (so) at (-3,-1.5) {source};
    \node[draw, shape=rectangle] (si1) at (9,0) {sink 1};
    \node[draw, shape=rectangle] (si2) at (9,-3) {sink 2};
    \draw [thick] (T3)  -- node[above] {$w_{32}$}(T2);
    \draw [thick] (T2) --  node[above] {$w_{21}$}(T1);
    \draw [thick] (T3) --  node[above] {$w_{C}$}(so);
    \draw [thick] (T1) --  node[above] {$w_{A}$}(si1);
    \draw [thick] (T5) --  node[above] {$w_{B}$}(si2);
    \draw [thick] (T3) --  node[left] {$w_{34}$}(T4);
    \draw [thick] (T5) --  node[above] {$w_{45}$}(T4);
    \draw [thick] (T2) --  node[right] {$w_{24}$}(T4);
    \end{tikzpicture}\hspace{.5cm}.
\end{align}
There is a hierarchy of scales $w_A,w_B,w_C \gg w_{21}, w_{45}, w_{32},w_{34} \gg w_{24}$. In order to have nontrivial correlation between $A$ and $B$, we take $w_{32}+w_{34} < w_{21}+w_{45}$. For simplicity, we take the more symmetric case where $w_{32}< w_{21}$ and $w_{34} < w_{45}$ individually. In this case, the following augmenting paths disconnect the source from the sinks
\begin{align}
\begin{tikzpicture}
    \node[draw, shape=rectangle] (T3) at (0,-1.5) {$T_3$};
    \node[draw, shape=rectangle] (T2) at (3,0) {$T_2$};
    \node[draw, shape=rectangle] (T1) at (6,0) {$T_1$};
    \node[draw, shape=rectangle] (T4) at (3,-3) {$T_4$};
    \node[draw, shape=rectangle] (T5) at (6,-3) {$T_5$};
    \node[draw, shape=rectangle] (so) at (-3,-1.5) {source};
    \node[draw, shape=rectangle] (si1) at (9,0) {sink 1};
    \node[draw, shape=rectangle] (si2) at (9,-3) {sink 2};
    \draw [ultra thick,-latex,preaction={draw,red,-,double=red,double distance=2\pgflinewidth,}] (T3)  -- node[above] {$w_{32}$}(T2);
    \draw [ultra thick,-latex,preaction={draw,red,-,double=red,double distance=2\pgflinewidth,}] (T2) --  node[above] {$w_{21}$}(T1);
    \draw [ultra thick,-latex,preaction={draw,orange,-,double=orange,double distance=2\pgflinewidth,}] (so) --  node[above] {$w_{C}$}(T3);
    \draw [ultra thick,-latex,preaction={draw,red,-,double=red,double distance=2\pgflinewidth,}] (T1) --  node[above] {$w_{A}$}(si1);
    \draw [ultra thick,-latex,preaction={draw,yellow,-,double=yellow,double distance=2\pgflinewidth,}] (T5) --  node[above] {$w_{B}$}(si2);
    \draw [ultra thick,-latex,preaction={draw,yellow,-,double=yellow,double distance=2\pgflinewidth,}] (T3) --  node[left] {$w_{34}$}(T4);
    \draw [ultra thick,-latex,preaction={draw,yellow,-,double=yellow,double distance=2\pgflinewidth,}] (T4) --  node[above] {$w_{45}$}(T5);
    \draw [thick] (T2) --  node[right] {$w_{24}$}(T4);
    \end{tikzpicture}\hspace{.5cm},
\end{align}
leading to residual network
\begin{align}
\begin{tikzpicture}
    \node[draw, shape=rectangle] (T3) at (0,-1.5) {$T_3$};
    \node[draw, shape=rectangle] (T2) at (3,0) {$T_2$};
    \node[draw, shape=rectangle] (T1) at (6,0) {$T_1$};
    \node[draw, shape=rectangle] (T4) at (3,-3) {$T_4$};
    \node[draw, shape=rectangle] (T5) at (6,-3) {$T_5$};
    \node[draw, shape=rectangle] (so) at (-3,-1.5) {source};
    \node[draw, shape=rectangle] (si1) at (9,0) {sink 1};
    \node[draw, shape=rectangle] (si2) at (9,-3) {sink 2};
    \draw [thick] (T2) --  node[above] {\tiny$w_{21}-w_{32}$}(T1);
    \draw [thick] (T3) --  node[above] {\tiny$w_{C}-w_{32}-w_{34}$}(so);
    \draw [thick] (T1) --  node[above] {\tiny$w_{A}-w_{32}$}(si1);
    \draw [thick] (T5) --  node[above] {\tiny$w_{B}-w_{34}$}(si2);
    \draw [thick] (T5) --  node[above] {\tiny$w_{45}-w_{34}$}(T4);
    \draw [thick] (T2) --  node[right] {\tiny$w_{24}$}(T4);
    \end{tikzpicture}\hspace{.5cm}.
\end{align}
In the next step, we must maximize the flow from the first sink to the second sink
\begin{align}
\begin{tikzpicture}
    \node[draw, shape=rectangle] (T3) at (0,-1.5) {$T_3$};
    \node[draw, shape=rectangle] (T2) at (3,0) {$T_2$};
    \node[draw, shape=rectangle] (T1) at (6,0) {$T_1$};
    \node[draw, shape=rectangle] (T4) at (3,-3) {$T_4$};
    \node[draw, shape=rectangle] (T5) at (6,-3) {$T_5$};
    \node[draw, shape=rectangle] (so) at (-3,-1.5) {source};
    \node[draw, shape=rectangle] (si1) at (9,0) {sink 1};
    \node[draw, shape=rectangle] (si2) at (9,-3) {sink 2};
    \draw [ultra thick,-latex,preaction={draw,green,-,double=green,double distance=2\pgflinewidth,}] (T1) --  node[above] {\tiny$w_{21}-w_{32}$}(T2);
    \draw [thick] (T3) --  node[above] {\tiny$w_{C}-w_{32}-w_{34}$}(so);
    \draw [ultra thick,-latex,preaction={draw,green,-,double=green,double distance=2\pgflinewidth,}] (si1) --  node[above] {\tiny$w_{A}-w_{32}$}(T1);
    \draw [ultra thick,-latex,preaction={draw,green,-,double=green,double distance=2\pgflinewidth,}] (T5) --  node[above] {\tiny$w_{B}-w_{34}$}(si2);
    \draw [ultra thick,-latex,preaction={draw,green,-,double=green,double distance=2\pgflinewidth,}] (T4) --  node[above] {\tiny$w_{45}-w_{34}$}(T5);
    \draw [ultra thick,-latex,preaction={draw,green,-,double=green,double distance=2\pgflinewidth,}] (T2) --  node[right] {\tiny$w_{24}$}(T4);
    \end{tikzpicture}\hspace{.5cm}.
\end{align}
If $w_{24} < w_{21}-w_{32}, w_{45}-w_{34}$ (not relevant for holography), the residual network is 
\begin{align}
\begin{tikzpicture}
    \node[draw, shape=rectangle] (T3) at (0,-1.5) {$T_3$};
    \node[draw, shape=rectangle] (T2) at (3,0) {$T_2$};
    \node[draw, shape=rectangle] (T1) at (6,0) {$T_1$};
    \node[draw, shape=rectangle] (T4) at (3,-3) {$T_4$};
    \node[draw, shape=rectangle] (T5) at (6,-3) {$T_5$};
    \node[draw, shape=rectangle] (so) at (-3,-1.5) {source};
    \node[draw, shape=rectangle] (si1) at (9,0) {sink 1};
    \node[draw, shape=rectangle] (si2) at (9,-3) {sink 2};
    \draw [thick] (T2) --  node[above] {\tiny$w_{21}-w_{32}-w_{24}$}(T1);
    \draw [thick] (T3) --  node[above] {\tiny$w_{C}-w_{32}-w_{34}$}(so);
    \draw [thick] (T1) --  node[above] {\tiny${w_{A}-w_{32}-w_{24}}$}(si1);
    \draw [thick] (T5) --  node[above] {\tiny${w_{B}-w_{34}-w_{24}}$}(si2);
    \draw [thick] (T5) --  node[above] {\tiny$w_{45}-w_{34}-w_{24}$}(T4);
    \end{tikzpicture}\hspace{.5cm}.
\end{align}
Let us instead choose $ w_{21}-w_{32} =  w_{45}-w_{34}< w_{24}$, in which case the residual network is
\begin{align}
\begin{tikzpicture}
    \node[draw, shape=rectangle] (T3) at (0,-1.5) {$T_3$};
    \node[draw, shape=rectangle] (T2) at (3,0) {$T_2$};
    \node[draw, shape=rectangle] (T1) at (6,0) {$T_1$};
    \node[draw, shape=rectangle] (T4) at (3,-3) {$T_4$};
    \node[draw, shape=rectangle] (T5) at (6,-3) {$T_5$};
    \node[draw, shape=rectangle] (so) at (-3,-1.5) {source};
    \node[draw, shape=rectangle] (si1) at (9,0) {sink 1};
    \node[draw, shape=rectangle] (si2) at (9,-3) {sink 2};
    \draw [thick] (T3) --  node[above] {\tiny$w_{C}-w_{32}-w_{34}$}(so);
    \draw [thick] (T1) --  node[above] {\tiny$w_{A}-w_{21}$}(si1);
    \draw [thick] (T5) --  node[above] {\tiny$w_{B}-w_{45}$}(si2);
    \draw [thick] (T2) --  node[right] {\tiny$w_{24}+w_{32}-w_{21}$}(T4);
    \end{tikzpicture}.
\end{align}
The dominant permutations are therefore $\tau_2 = \tau_4 \in NC_2$, $\tau_3 = \text{id}$, $\tau_1 = \gamma$, $\tau_5 = \gamma^{-1}$.

\begin{figure}
    \centering
    \includegraphics[height = 5cm]{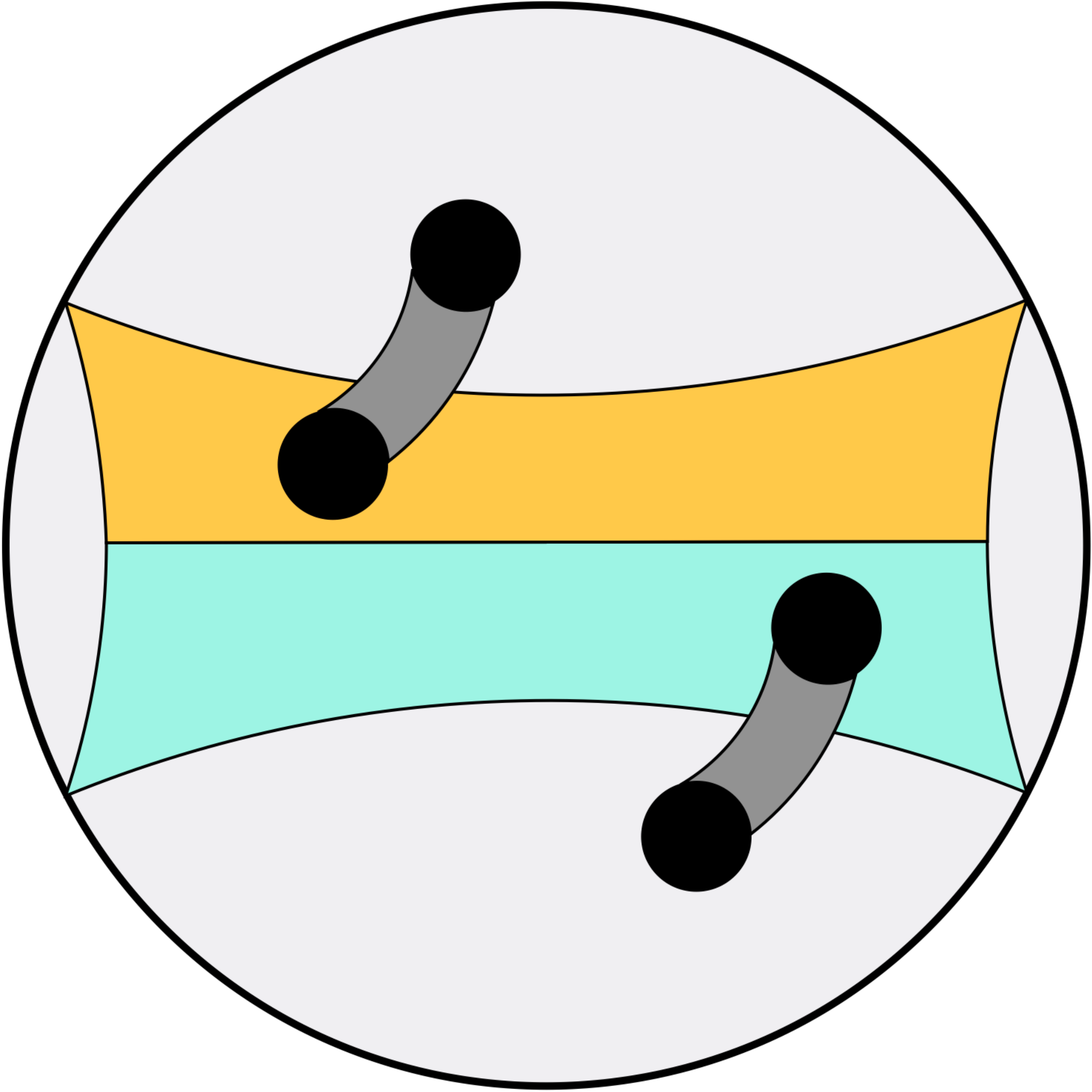}
    \caption{A time slice of AdS is shown with large bulk matter. We consider highly entangled black holes. The entanglement can geometrically be thought of as a wormhole \'a la ER = EPR \cite{2013ForPh..61..781M}.}
    \label{exotic_EW}
\end{figure}

Consider the exotic single-sided geometry involving two wormholes (Fig.~\ref{exotic_EW}). The flow network is
\begin{align}
\begin{tikzpicture}[scale =0.75]
    \node[draw, shape=rectangle] (T3) at (0,-1.5) {$T_3$};
    \node[draw, shape=rectangle] (T2) at (3,0) {$T_2$};
    \node[draw, shape=rectangle] (T1) at (6,0) {$T_1$};
    \node[draw, shape=rectangle] (T4) at (3,-3) {$T_4$};
    \node[draw, shape=rectangle] (T5) at (6,-3) {$T_5$};
    \node[draw, shape=rectangle] (so) at (-3,-1.5) {source};
    \node[draw, shape=rectangle] (si1) at (9,0) {sink 1};
    \node[draw, shape=rectangle] (si2) at (9,-3) {sink 2};
    \node[draw, shape=rectangle] (T6) at (4.5,2) {$T_6$};
    \node[draw, shape=rectangle] (T7) at (4.5,-5) {$T_7$};
    \draw [thick] (T3)  -- node[above] {$w_{32}$}(T2);
    \draw [thick] (T2) --  node[above] {$w_{21}$}(T1);
    \draw [thick] (T3) --  node[above] {$w_{C}$}(so);
    \draw [thick] (T1) --  node[above] {$w_{A}$}(si1);
    \draw [thick] (T5) --  node[above] {$w_{B}$}(si2);
    \draw [thick] (T3) --  node[left] {$w_{34}$}(T4);
    \draw [thick] (T5) --  node[above] {$w_{45}$}(T4);
    \draw [thick] (T2) --  node[right] {$w_{24}$}(T4);
    \draw [thick] (T1) --  node[right] {$w_{16}$}(T6);
    \draw [thick] (T2) --  node[left] {$w_{26}$}(T6);
    \draw [thick] (T7) --  node[right] {$w_{75}$}(T5);
    \draw [thick] (T4) --  node[left] {$w_{74}$}(T7);
    \end{tikzpicture}\hspace{.5cm}.
\end{align}
After the same first step, we have residual network
\begin{align}
\begin{tikzpicture}[scale =0.75]
    \node[draw, shape=rectangle] (T3) at (0,-1.5) {$T_3$};
    \node[draw, shape=rectangle] (T2) at (3,0) {$T_2$};
    \node[draw, shape=rectangle] (T1) at (6,0) {$T_1$};
    \node[draw, shape=rectangle] (T4) at (3,-3) {$T_4$};
    \node[draw, shape=rectangle] (T5) at (6,-3) {$T_5$};
    \node[draw, shape=rectangle] (so) at (-3,-1.5) {source};
    \node[draw, shape=rectangle] (si1) at (9,0) {sink 1};
    \node[draw, shape=rectangle] (si2) at (9,-3) {sink 2};
    \node[draw, shape=rectangle] (T6) at (4.5,2) {$T_6$};
    \node[draw, shape=rectangle] (T7) at (4.5,-5) {$T_7$};
    \draw [thick] (T2) --  node[above] {\tiny$w_{21}-w_{32}$}(T1);
    \draw [thick] (T3) --  node[above] {\tiny$w_{C}-w_{32}-w_{34}$}(so);
    \draw [thick] (T1) --  node[above] {\tiny$w_{A}-w_{32}$}(si1);
    \draw [thick] (T5) --  node[above] {\tiny$w_{B}-w_{34}$}(si2);
    \draw [thick] (T5) --  node[above] {\tiny$w_{45}-w_{34}$}(T4);
    \draw [thick] (T2) --  node[right] {\tiny$w_{24}$}(T4);
    \draw [thick] (T7) --  node[right] {$w_{75}$}(T5);
    \draw [thick] (T4) --  node[left] {$w_{74}$}(T7);
    \draw [thick] (T1) --  node[right] {$w_{16}$}(T6);
    \draw [thick] (T2) --  node[left] {$w_{26}$}(T6);
    \end{tikzpicture}\hspace{.5cm}.
\end{align}
We again take an augmenting path through the network
\begin{align}
\begin{tikzpicture}[scale =0.75]
    \node[draw, shape=rectangle] (T3) at (0,-1.5) {$T_3$};
    \node[draw, shape=rectangle] (T2) at (3,0) {$T_2$};
    \node[draw, shape=rectangle] (T1) at (6,0) {$T_1$};
    \node[draw, shape=rectangle] (T4) at (3,-3) {$T_4$};
    \node[draw, shape=rectangle] (T5) at (6,-3) {$T_5$};
    \node[draw, shape=rectangle] (so) at (-3,-1.5) {source};
    \node[draw, shape=rectangle] (si1) at (9,0) {sink 1};
    \node[draw, shape=rectangle] (si2) at (9,-3) {sink 2};
    \node[draw, shape=rectangle] (T6) at (4.5,2) {$T_6$};
    \node[draw, shape=rectangle] (T7) at (4.5,-5) {$T_7$};
    \draw [ultra thick,-latex,preaction={draw,green,-,double=green,double distance=2\pgflinewidth,}] (T1) --  node[above] {\tiny$w_{21}-w_{32}$}(T2);
    \draw [thick] (T3) --  node[above] {\tiny$w_{C}-w_{32}-w_{34}$}(so);
    \draw [ultra thick,-latex,preaction={draw,green,-,double=green,double distance=2\pgflinewidth,}] (si1) --  node[above] {\tiny$w_{A}-w_{32}$}(T1);
    \draw [ultra thick,-latex,preaction={draw,green,-,double=green,double distance=2\pgflinewidth,}] (T5) --  node[above] {\tiny$w_{B}-w_{34}$}(si2);
    \draw [ultra thick,-latex,preaction={draw,green,-,double=green,double distance=2\pgflinewidth,}] (T4) --  node[above] {\tiny$w_{45}-w_{34}$}(T5);
    \draw [ultra thick,-latex,preaction={draw,green,-,double=green,double distance=2\pgflinewidth,}] (T2) --  node[right] {\tiny$w_{24}$}(T4);
    \draw [thick] (T7) --  node[right] {$w_{75}$}(T5);
    \draw [thick] (T4) --  node[left] {$w_{74}$}(T7);
    \draw [thick] (T1) --  node[right] {$w_{16}$}(T6);
    \draw [thick] (T2) --  node[left] {$w_{26}$}(T6);
    \end{tikzpicture}\hspace{.5cm},
\end{align}
leading to
\begin{align}
\begin{tikzpicture}[scale =0.75]
    \node[draw, shape=rectangle] (T3) at (0,-1.5) {$T_3$};
    \node[draw, shape=rectangle] (T2) at (3,0) {$T_2$};
    \node[draw, shape=rectangle] (T1) at (6,0) {$T_1$};
    \node[draw, shape=rectangle] (T4) at (3,-3) {$T_4$};
    \node[draw, shape=rectangle] (T5) at (6,-3) {$T_5$};
    \node[draw, shape=rectangle] (so) at (-3,-1.5) {source};
    \node[draw, shape=rectangle] (si1) at (9,0) {sink 1};
    \node[draw, shape=rectangle] (si2) at (9,-3) {sink 2};
    \node[draw, shape=rectangle] (T6) at (4.5,2) {$T_6$};
    \node[draw, shape=rectangle] (T7) at (4.5,-5) {$T_7$};
    \draw [thick] (T3) --  node[above] {\tiny$w_{C}-w_{32}-w_{34}$}(so);
    \draw [thick] (T1) --  node[above] {\tiny$w_{A}-w_{21}$}(si1);
    \draw [thick] (T5) --  node[above] {\tiny$w_{B}-w_{45}$}(si2);
    \draw [thick] (T2) --  node[right] {\tiny$w_{24}+w_{32}-w_{21}$}(T4);
        \draw [thick] (T7) --  node[right] {$w_{75}$}(T5);
    \draw [thick] (T4) --  node[left] {$w_{74}$}(T7);
    \draw [thick] (T1) --  node[right] {$w_{16}$}(T6);
    \draw [thick] (T2) --  node[left] {$w_{26}$}(T6);
    \end{tikzpicture}\hspace{.5cm}.
\end{align}
Because of the quantum corrections, there is a remaining residual path
\begin{align}
\begin{tikzpicture}[scale =0.75]
    \node[draw, shape=rectangle] (T3) at (0,-1.5) {$T_3$};
    \node[draw, shape=rectangle] (T2) at (3,0) {$T_2$};
    \node[draw, shape=rectangle] (T1) at (6,0) {$T_1$};
    \node[draw, shape=rectangle] (T4) at (3,-3) {$T_4$};
    \node[draw, shape=rectangle] (T5) at (6,-3) {$T_5$};
    \node[draw, shape=rectangle] (so) at (-3,-1.5) {source};
    \node[draw, shape=rectangle] (si1) at (9,0) {sink 1};
    \node[draw, shape=rectangle] (si2) at (9,-3) {sink 2};
    \node[draw, shape=rectangle] (T6) at (4.5,2) {$T_6$};
    \node[draw, shape=rectangle] (T7) at (4.5,-5) {$T_7$};
    \draw [thick] (T3) --  node[above] {\tiny$w_{C}-w_{32}-w_{34}$}(so);
    \draw [ultra thick,-latex,preaction={draw,green,-,double=green,double distance=2\pgflinewidth,}] (si1) --  node[above] {\tiny$w_{A}-w_{21}$}(T1);
    \draw [ultra thick,-latex,preaction={draw,green,-,double=green,double distance=2\pgflinewidth,}] (T5) --  node[above] {\tiny$w_{B}-w_{45}$}(si2);
    \draw [ultra thick,-latex,preaction={draw,green,-,double=green,double distance=2\pgflinewidth,}] (T2) --  node[right] {\tiny$w_{24}+w_{32}-w_{21}$}(T4);
    \draw [ultra thick,-latex,preaction={draw,green,-,double=green,double distance=2\pgflinewidth,}] (T7) --  node[right] {$w_{75}$}(T5);
    \draw [ultra thick,-latex,preaction={draw,green,-,double=green,double distance=2\pgflinewidth,}] (T4) --  node[left] {$w_{74}$}(T7);
    \draw [ultra thick,-latex,preaction={draw,green,-,double=green,double distance=2\pgflinewidth,}] (T1) --  node[right] {$w_{16}$}(T6);
    \draw [ultra thick,-latex,preaction={draw,green,-,double=green,double distance=2\pgflinewidth,}] (T6) --  node[left] {$w_{26}$}(T2);
    \end{tikzpicture}\hspace{.5cm}.
\end{align}
Depending on the relative sizes, we can have different residual networks. For simplicity, we take $w_{24}+w_{32}-w_{21} := w_W$ to be the smallest flow weight.
\begin{align}
\begin{tikzpicture}[scale =0.75]
    \node[draw, shape=rectangle] (T3) at (0,-1.5) {$T_3$};
    \node[draw, shape=rectangle] (T2) at (3,0) {$T_2$};
    \node[draw, shape=rectangle] (T1) at (6,0) {$T_1$};
    \node[draw, shape=rectangle] (T4) at (3,-3) {$T_4$};
    \node[draw, shape=rectangle] (T5) at (6,-3) {$T_5$};
    \node[draw, shape=rectangle] (so) at (-3,-1.5) {source};
    \node[draw, shape=rectangle] (si1) at (9,0) {sink 1};
    \node[draw, shape=rectangle] (si2) at (9,-3) {sink 2};
    \node[draw, shape=rectangle] (T6) at (4.5,2) {$T_6$};
    \node[draw, shape=rectangle] (T7) at (4.5,-5) {$T_7$};
    \draw [thick] (T3) --  node[above] {\tiny$w_{C}-w_{32}-w_{34}$}(so);
    \draw [thick] (T1) --  node[above] {\tiny$w_{A}-w_{21}-w_{W}$}(si1);
    \draw [thick] (T5) --  node[above] {\tiny$w_{B}-w_{45}-w_W$}(si2);
    \draw [thick] (T7) --  node[right] {$w_{75}-w_W$}(T5);
    \draw [thick] (T4) --  node[left] {$w_{74}-w_W$}(T7);
    \draw [thick] (T1) --  node[right] {$w_{16}-w_W$}(T6);
    \draw [thick] (T2) --  node[left] {$w_{26}-w_W$}(T6);
    \end{tikzpicture}\hspace{.5cm}.
\end{align}
With this residual network, there is only a single configuration of dominant permutations, namely $T_3 = \text{id}$, $T_2 = T_6=T_1 = \gamma$, and $T_4 = T_7= T_5 = \gamma^{-1}$. This resembles the entanglement wedge cross section.

We can also consider bulk matter in mixed states. A simple way to implement this geometrically is by having a black hole entangled with another asymptotically AdS region, but tracing over this other space, leaving a mixed state. In the tensor network, this amounts to adding a tensor that connects to one of the bulk tensors and to the source. For example, if we place a mixed state black hole in the entanglement wedge of $A$, the flow network is given by
\begin{align}
\begin{tikzpicture}[scale = .75]
    \node[draw, shape=rectangle] (T3) at (0,-1.5) {$T_3$};
    \node[draw, shape=rectangle] (T2) at (3,0) {$T_2$};
    \node[draw, shape=rectangle] (T1) at (6,0) {$T_1$};
    \node[draw, shape=rectangle] (T4) at (3,-3) {$T_4$};
    \node[draw, shape=rectangle] (T5) at (6,-3) {$T_5$};
    \node[draw, shape=rectangle] (T6) at (6,3) {$T_6$};
    \node[draw, shape=rectangle] (so) at (-3,-1.5) {source};
    \node[draw, shape=rectangle] (si1) at (9,0) {sink 1};
    \node[draw, shape=rectangle] (si2) at (9,-3) {sink 2};
    \draw [thick] (T3)  -- node[above] {$w_{32}$}(T2);
    \draw [thick] (T2) --  node[above] {$w_{21}$}(T1);
    \draw [thick] (T3) --  node[above] {$w_{C}$}(so);
    \draw [thick] (T1) --  node[above] {$w_{A}$}(si1);
    \draw [thick] (T6) --  node[right] {$w_{16}$}(T1);
    \draw [thick] (T6) --  node[above] {$w_{BH}$}(so);
    \draw [thick] (T5) --  node[above] {$w_{B}$}(si2);
    \draw [thick] (T3) --  node[left] {$w_{34}$}(T4);
    \draw [thick] (T5) --  node[above] {$w_{45}$}(T4);
    \draw [thick] (T2) --  node[right] {$w_{24}$}(T4);
    \end{tikzpicture}\hspace{.5cm}.
\end{align}
Using the same tools, this may also be straightforwardly solved.

\subsection{Bit threads}

 The bit thread formulation of holographic entanglement \cite{2017CMaPh.352..407F} is highly reminiscent of the solution of entanglement in random tensor networks using flow networks. Bit threads are an alternative perspective on the Ryu-Takayanagi formula. Rather than identifying the minimal extremal surface in the bulk corresponding to a boundary region $A$, one is instructed to find the maximal flux of a divergenceless vector field, $v^{\mu}$
\begin{align}
    S(A) = \max_v \int_A \sqrt{h}\, n_{\mu}v^{\mu}, \quad |v| \leq \frac{1}{4G_N}.
\end{align}
This is equivalent to the Ryu-Takayanagi formula by the Riemannian max-flow min-cut theorem \cite{2018CQGra..35j5012H}. The analogy to the solution of random tensor networks arises because there we used the more common network version of max-flow min-cut. In particular, the bit threads on the graph are identified with augmenting paths. They are divergenceless by definition. In this identification, only the leading term in the entropy can be found. The bit threads are agnostic to the $O(1)$ terms. The solution on the graph is
\begin{align}
    S(A) = w_{\text{min-cut}}\log\left[ N\right] + O(1).
\end{align}

\begin{figure}
    \centering
    \includegraphics[width = .8\textwidth]{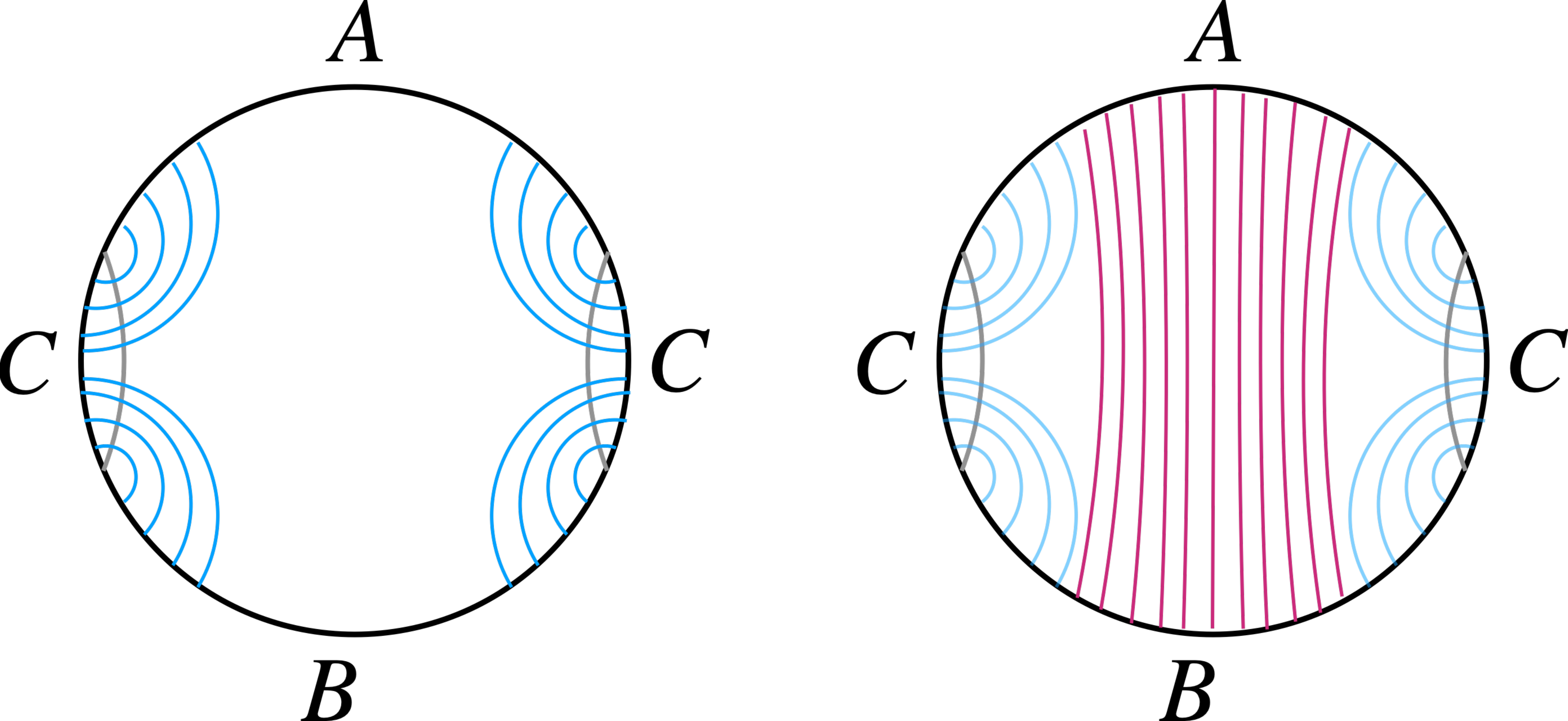}
    \caption{The left figure is the first step involving a maximal bit thread configuration flowing from $C$ to $A \cup B$. The gray lines represent the Ryu-Takayanagi surface. In the second step (right figure), the old flow $w$ (blue lines) remains and a new flow $v$ (magenta lines) must be constructed and optimized. It is this flow that computes the logarithmic negativity.}
    \label{bitthreads_fig}
\end{figure}

For logarithmic negativity in fixed-area states, we may follow the same logic. The bit threads, however, are more complicated as we need to consider two distinct flows, a specific example of a \textit{multicommodity flow}. For a boundary tripartition into $A$, $B$, and $C$, we first find a maximal flow, $w$, from $C$ to $A\cup B$. There may be many bit thread configurations that maximize the flux and we must consider all of them. Next, we must find the maximal flow of bit threads from $A$ to $B$. This flow is not independent of the initial bit threads; combined, they must not have norm larger than $\frac{1}{4G_N}$. The logarithmic negativity is given by the flux of this second flow conditioned on the first flow
\begin{align}
    \mathcal{E}(A,B) = \max_{v|w} \int_A \sqrt{h}\, n_{\mu}v^{\mu}, \quad |v|+|w| \leq \frac{1}{4G_N}.
\end{align}
The flow must be maximized over all configurations of both bit threads. We outline this two step bit thread procedure in Fig.~\ref{bitthreads_fig}.

We note that the union of the $v$ and $w$ flows represents a \textit{max thread configuration} from \cite{2019CMaPh.376..609C}. As noted in \cite{2019CMaPh.376..609C}, the number of threads connecting $A$ and $B$ in a max thread configuration is given by half the mutual information. Therefore, we confirm the proportionality of negativity and mutual information from bit threads.

\section{Islands in mixed state entanglement} \label{sec:islands}

In Section \ref{sec:west_coast_1_tensor} we discussed negativity in the model of \cite{2019arXiv191111977P} where the entanglement entropy of Hawking radiation was studied.
This raises naturally the question of the analog of an island formula for negativity.\footnote{We note that different island formulas for the negativity have been proposed in \cite{2020arXiv201203983B}.} Here, we will not give such a formula analogous to the quantum extremal surface formulas, but instead mention only the semi-classical approximation to it based on random tensor networks. We will then test it in a very simple example and find agreement.

\subsection{Doubly holographic systems and fixed generalized entropy}

In this section, we would like to analyze the negativity in systems that are coupled to gravity rather than between parts of a holographic system. The main motivation is to consider an evaporating black hole with Hawking radiation. In particular, we can think about dividing the Hawking radiation into two parts and considering the negativity between part of the radiation and the black hole, as we studied in Section \ref{sec:west_coast_1_tensor} for the model of \cite{2019arXiv191111977P}. Here we would like to learn lessons about more general situations.

\begin{figure}[]
\centering
\includegraphics[width=0.5\textwidth]{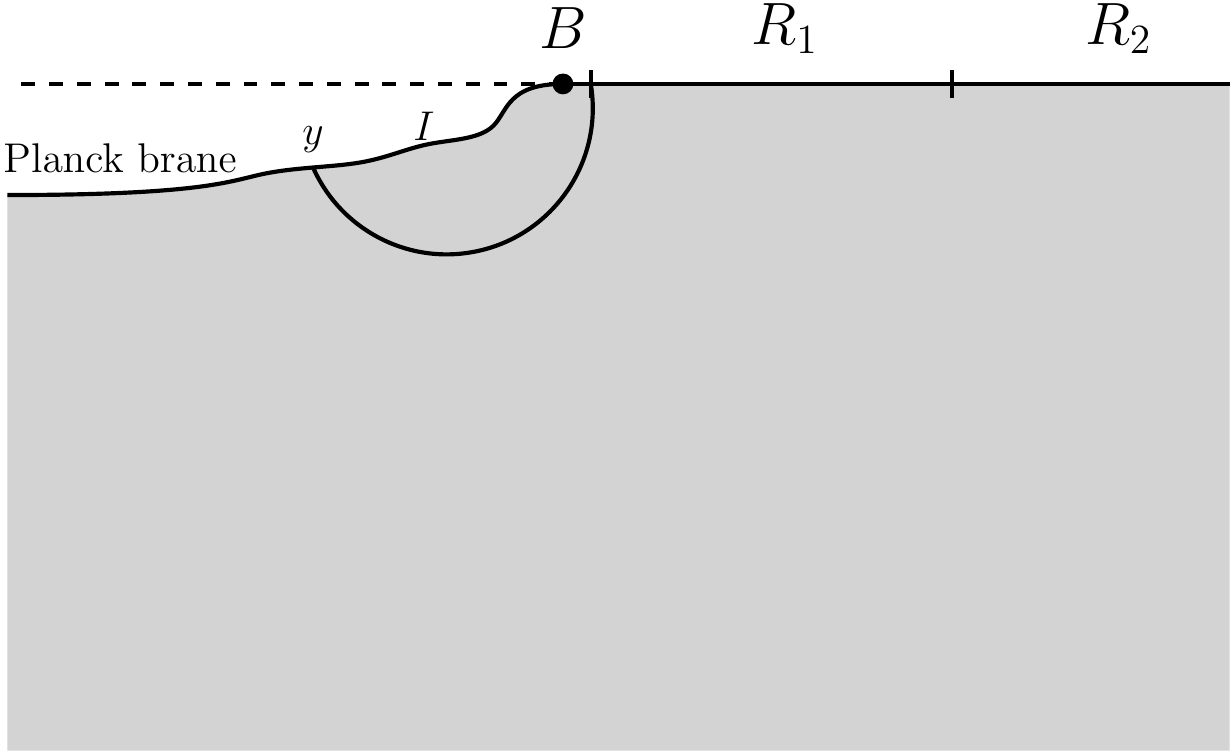}
\caption{A system with holographic matter.}
\label{fig:double_hol}
\end{figure}

In order to get an idea for how the negativity should behave, we will use the doubly-holographic idea of \cite{Almheiri:2019hni}. Let us consider a matter CFT through which the black hole can release radiation. That is, we have a gravitational theory where we can have a black hole coupled to a holographic CFT of large central charge. This system is coupled to the same CFT on a rigid space without gravity where we can measure the radiation. 
For convenience in the discussion, let us take the CFT to be two-dimensional though the analysis holds in any dimension. Without gravity, the CFT is dual to a three-dimensional gravitational system with the two-dimensional system on its boundary. In the non-gravitational part, we have the usual AdS/CFT boundary. In order to implement the 2d gravitational part of the action, the boundary in that region becomes dynamical, allowing a dynamical induced metric. This boundary is referred to as the Planck brane. For our purposes, we will divide the matter theory without gravity into two regions $R_1$ and $R_2$. This setting is shown in Fig.~\ref{fig:double_hol}. Therefore, in the bulk we have a 3d gravitational theory with a Planck brane. If we think about the boundary dual of this system, it is composed of the non-gravitational CFT made of $R_1$ and $R_2$ and we can think about the dual of the 2d gravity theory as a dual quantum mechanics system represented by a heavy quantum dot.

If we consider a subsystem of the boundary including part of the rigid CFT as well as the quantum mechanical system, its entropy can be calculated by the usual Ryu-Takayanagi formula in the 3d bulk. The RT surface is allowed to end on the Planck brane. The Planck brane contributes by the value of the 2d dilaton at that point. Indeed, recalling \cite{Almheiri:2019hni}, if we consider a region in the boundary denoted by $B$ including the quantum system, as shown in Fig.~\ref{fig:double_hol}, and naively apply the quantum extremal surface formula \cite{Engelhardt:2014gca} to it, we will need to extremize over the points $y$ on the 2d gravity dual the generalized entropy $\frac{\phi (y)}{4G_N^{(2)} } +S_{bulk}(I)$ where in 2d the RT term is given by the value of the dilaton $\phi $ and $I$ is the entanglement wedge ending on $y$. Because the CFT has a large central charge, $S_{bulk} (I)$ is dominated by the matter entropy, which in the 3d language is the area of the RT surface. At leading order, we can neglect the 3d bulk entropy.

In this 3d setting, we can find the negativity between $R_1$ and the black hole $B$. As we saw in Section \ref{sec:gen_network} for random tensor networks, semi-classically the negativity is given by half the holographic mutual information $\frac{S(B)+S(R_1)-S(R_2)}{2} $.
Holographically, we saw that random tensor networks describe fixed-area states. Therefore, we can think about the situation here as fixing the generalized entropy instead due to the Planck brane. 
$S(B)$ in 3d is given by the area plus the dilaton value on the Planck brane, which is the 3d value corresponding to the usual quantum extremal surface prescription in 2d
\begin{equation}
S(B)= \min \ext_{\gamma_B }  \left( \frac{A[\gamma_B ]}{4G_N} +S_{\text{eff}}(\Sigma )\right) 
\end{equation}
with $\gamma_B $ the RT surface and $\Sigma $ the entanglement wedge having $\gamma _B$ on its boundary, where from now on Newton's constant is always the 2d one. $S_{\text{eff}}$ is the 2d bulk entropy around a semi-classical geometry \cite{Faulkner:2013ana}.

\begin{figure}[]
\centering
\includegraphics[width=0.45\textwidth]{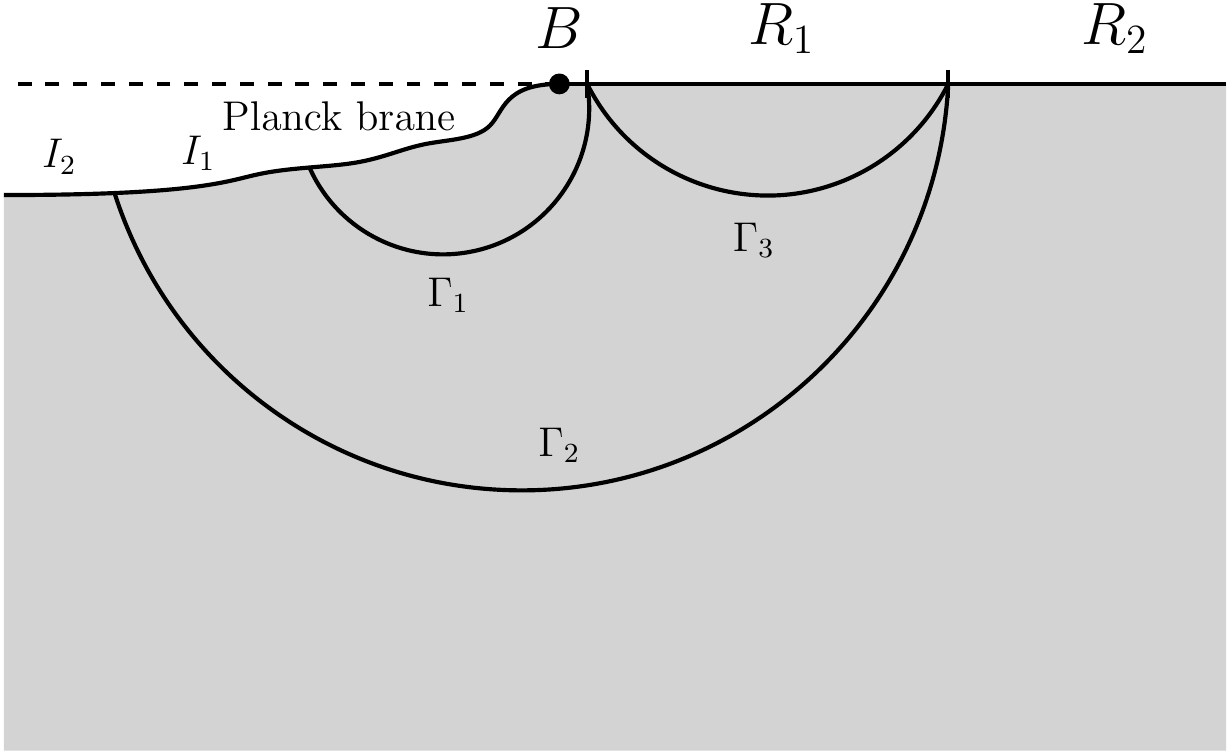} \hfill
\includegraphics[width=0.45\textwidth]{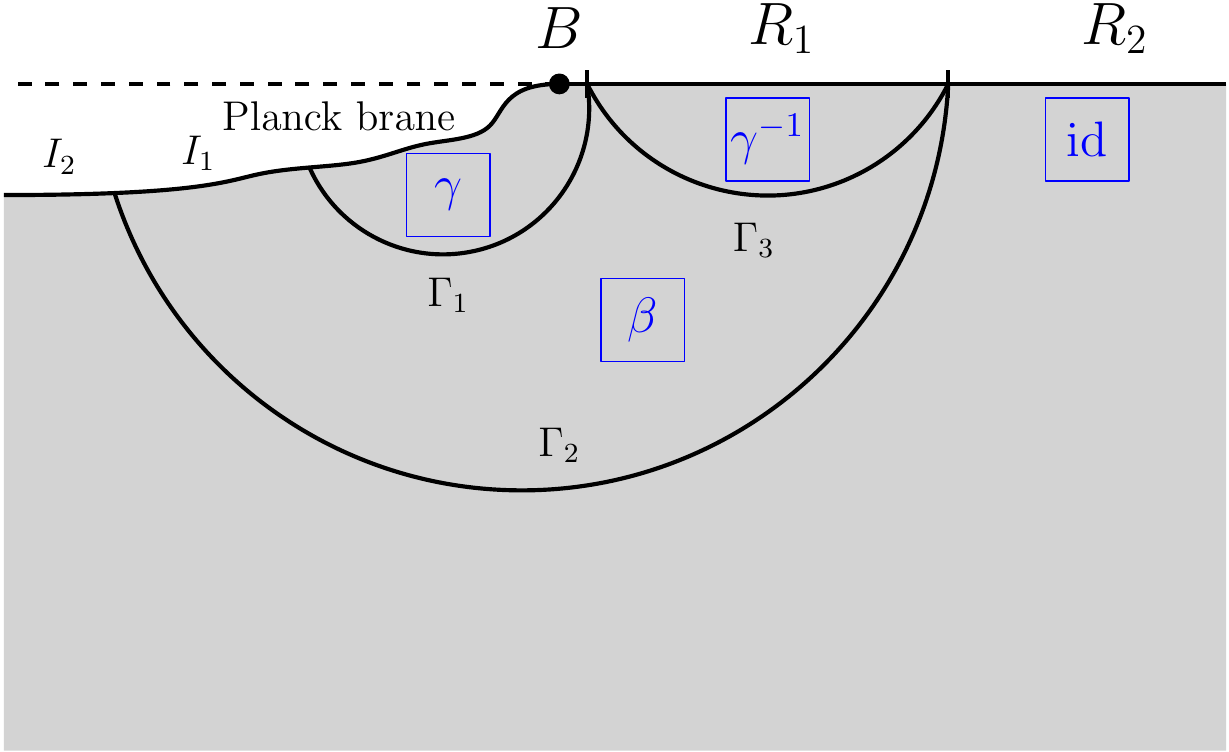}
\caption{Islands in the 3d picture. We show several options for extremal surfaces. On the RHS, an example of a corresponding tensor network is shown labeled by the permutations (this depends on what extremal surfaces correspond to each subsystem).}
\label{fig:double_hol_R1}
\end{figure}

More interestingly, moving on to $R_1$, the RT surface may consist of $\Gamma _1$ and $\Gamma _2$ (see Fig.~\ref{fig:double_hol_R1}). Its 3d entanglement wedge ends on an interval island $I_1$ on the Planck brane. The 3d RT prescription of $A[\Gamma _1]+A[\Gamma _2]$ corresponds in the $R_1$+$R_2$+Planck brane system to the entanglement entropy of the matter part, which is equivalent to the entanglement entropy $S_{\text{eff}}(I_1 \cup R_1)$.\footnote{$\Gamma _2$ corresponds to the leftmost part of the Planck brane together with $R_2$, while $\Gamma _1$ to the rightmost part of the Planck brane. The complement of these is $I_1 \cup R_1$ because the full system is pure.} The sum of the values of the dilaton on the end of $\Gamma _1$ and $\Gamma _2$ in the 2d language is the area of the boundary of the island. Here the island is fully isolated. We get
\begin{equation}
S(R_1) = \min \ext_{I_1}  \left( \frac{A[\partial I_1]}{4G_N} +S_{\text{eff}}(R_1 \cup  I_1) \right) .
\end{equation}
With the configuration of $R_1$ in the figure, we may also have $\Gamma _3$ as its RT surface. This just corresponds in the discussion above to having no island. In this case, we reduce to the naive expectation of $S_{\text{eff}}(R_1)$.

Lastly, $R_2$ may have an island of its own. This happens if in 3d its entanglement wedge is bounded by a surface such as $\Gamma _2$ in the figure, and the entanglement wedge ends on an island $I_2$ on the Planck brane. The area of $\Gamma _2$ corresponds to the entropy $S_{\text{eff}}(R_2 \cup I_2)$ and the dilaton term gives $A[\partial I_2]$, so in total
\begin{equation}
S(R_2) = \min \ext_{I_2}  \left( \frac{A[\partial I_2]}{4G_N} +S_{\text{eff}}(R_2 \cup  I_2) \right) .
\end{equation}

We are led to the following formula for the negativity expressing the entanglement between part $R_1$ of a system coupled to gravity and $B$, which is given in terms of two islands
\begin{equation} \label{eq:island_for_negativity}
\begin{split}
\cE(R_1, B) &= \frac{1}{2} \Bigg\{
\min \underset{\gamma_B}{\ext}  \left( \frac{A[\gamma_B ]}{4G_N} +S_{\text{eff}}(\Sigma )\right) +
\min \underset{I_1}{\ext}  \left( \frac{A[\partial I_1]}{4G_N} +S_{\text{eff}}(R_1 \cup  I_1) \right) - \\
& - \min \underset{I_2}{\ext}  \left( \frac{A[\partial I_2]}{4G_N} +S_{\text{eff}}(R_2 \cup  I_2) \right)
\Bigg\} .
\end{split}
\end{equation}
This formula is proportional to the mutual information and captures the semi-classical negativity. It is not meant to be an exact formula analogous to the quantum extremal surface prescription.
As mentioned, it is based on applying the random tensor network result to this setting, which we interpret as fixing the generalized entropy. In fact, in general, when we have an island, this should be the analog of fixed-area states. We cannot fix only the area term or the bulk entropy term, as these two balance each other. This is the setup where we expect such a formula to hold semi-classically. If we think about the entanglement entropy as an operator, it would not even be linear; the situation is much simpler in descriptions as the one above, where we can fix the extremal combination of the area and the dilaton value (or another area if we are in higher dimensions).

Fixing the generalized entropy is also necessary to avoid the tricky backreaction that is present in more honest calculations of holographic R\'enyi entropy and negativity. We expect that the final island formula for negativity will only include a global extremization instead of the linear combination of extremizations because the negativity is evaluated from a single density matrix, unlike the mutual information. Furthermore, we expect that in the final formula, the quantum corrections will not be in the form of bulk entropies, but bulk negativities (or some variant). This was, for example, observed in quantum error correcting codes in \cite{2019PhRvD..99j6014K}. With these important caveats, we progress.


\subsection{Islands from replica wormholes}

We have analyzed negativity in JT gravity including wormhole contributions. We shall now show that the result there matches to the formula \eqref{eq:island_for_negativity}. This provides an argument for the Euclidean wormhole origin of \eqref{eq:island_for_negativity} in a gravitational calculation.

Let us consider the three terms in the brackets of \eqref{eq:island_for_negativity} in JT with EOW branes. The first term is the entropy of the black hole or equivalently the radiation. In the language of the radiation, if there is no island, it is given by the entropy of the radiation. The state that we use is \eqref{eq:west_coast_state}, where for the purpose of the effective theory we consider the $|\psi_{i,a} \rangle _B$ states to be orthonormal. In this state, $S_{\text{eff}} (R_1)=N_1$, $S_{\text{eff}} (R_2)=N_2$, and $S_{\text{eff}} (B)=N_1+N_2$. Therefore, with no island we just get $N_1+N_2$. If there is an island, it includes in particular the EOW brane, so the bulk entropy vanishes and we just remain with extremization of the area term, giving $\Smicro$. Together, the first term gives
\begin{equation}
\min \underset{\gamma_B}{\ext}  \left( \frac{A[\gamma_B ]}{4G_N} +S_{\text{eff}}(\Sigma )\right) = S(R)= \min \begin{cases}
N_1+N_2 & \text{no island} \\
\Smicro & \text{with island}
\end{cases} .
\end{equation}
For the second term, if there is no island we just get $S_{\text{eff}} (R_1)=N_1$, but if there is an island which then includes the $B$ system, the bulk entropy term is actually $N_2$ but the minimal area is $\Smicro$. Together,
\begin{equation}
\min \underset{I_1}{\ext}  \left( \frac{A[\partial I_1]}{4G_N} +S_{\text{eff}}(R_1 \cup  I_1) \right) = \min \begin{cases}
N_1 & \text{no island} \\
\Smicro +N_2 & \text{with island}
\end{cases} 
\end{equation}
and similarly
\begin{equation}
\min \underset{I_2}{\ext}  \left( \frac{A[\partial I_2]}{4G_N} +S_{\text{eff}}(R_2 \cup  I_2) \right) = \min \begin{cases}
N_2 & \text{no island} \\
\Smicro +N_1 & \text{with island}
\end{cases} .
\end{equation}

We should now compare this to the result in Fig.~\ref{fig:west_coast_fixed_rad_amount}. Let us start with the intermediate region where $|\Smicro-N_1|<N_2<\Smicro+N_1$. In this range of parameters, there is an island only in the first term, giving $\frac{\Smicro+N_1-N_2}{2} $ just as in Fig.~\ref{fig:west_coast_fixed_rad_amount}.

In the first region where $N_2$ is smallest, we can consider two cases. If $\Smicro>N_1$, the range is $N_2<\Smicro-N_1$; there is no island in either term and we get $N_1$. If, on the other hand $\Smicro<N_1$, then $N_2<N_1-\Smicro$; there is an island in the first two terms and we get $\Smicro$. Together, these two cases give $\min(\Smicro,N_1)$ as in the figure. Lastly, when $N_2>\Smicro+N_1$, there is no island only in the second term, giving zero. This shows that Euclidean wormholes are indeed consistent with the formula \eqref{eq:island_for_negativity}.

\section{Discussion}

In this paper we have studied the negativity spectrum for general random tensor networks with large bond dimensions. We saw that a three-party flow network is useful in solving for the negativity spectrum in such networks. There are many possible negativity spectra that can be obtained using such random tensor networks and we mentioned several examples.

These random tensor networks are useful both in preparing states relevant to condensed matter many body systems, as well as in holography. We discussed two holographic situations described by random tensor networks, which are Jackiw-Teitelboim gravity with end-of-the-world branes and holographic fixed-area states. We also discussed island prescriptions for finding the negativity in systems coupled to gravity.

\paragraph{A complete holographic formula} Normal holographic states prepared by the Euclidean path integrals such as the vacuum state or thermal state are not fixed-areas states. Rather, they are superpositions of fixed-areas states with $O(\sqrt{G_N})$ fluctuations in the area. It would be interesting to analyze this sum over fixed-area states to include backreaction in the holographic formula for negativity. Due to the large backreaction between replicas in these normal states, it is not presently clear to us if the same qualitative behavior of negativity would remain i.e.~if proportionality with mutual information at leading order and the same topology of saddles dominate. This holographic formula would be a necessary first step to the correct island formula for negativity that incorporates quantum corrections which is clearly of great interest.
We believe this formula may be very intricate because of the replica symmetry breaking that appears to play an important role at leading order.

This brings us to a previous proposal for a holographic formula for negativity \cite{2019PhRvD..99j6014K}. In this proposal, the negativity was dual to the entanglement wedge cross section with backreaction included. Using 2d CFT arguments of twist operators, the negativity was related to a R\'enyi version of the reflected entropy, providing evidence for the proposal \cite{2019PhRvL.123m1603K}. A central assumption in \cite{2019PhRvL.123m1603K} was that a certain conformal block completely fixed the negativity at large central charge and that this block was identical to the conformal block for R\'enyi reflected entropy. This conformal block preserved replica symmetry and is analogous to taking $\tau_1 = \gamma^{-1}$ and $\tau_2 = \gamma$ in the two-tensor network. From random tensor networks and fixed-areas states, there is reason to believe 
that replica symmetry breaking saddles should be included in the CFT computation, though it is confusing from the boundary perspective as replica symmetry of the twist operator correlation functions is manifest. Understanding replica symmetry breaking from the CFT is an important open question that has implications far beyond negativity. Replica symmetry breaking saddles appear very generally in gravitational settings whenever there is more than one candidate extremal surface.

Here, we present a heuristic which we admit may be a red herring. The moments of the partially transposed density matrix for two disjoint intervals (with endpoints $(z_1, z_2)$ and $(z_3,z_4)$) in the vacuum are given by a four-point function of $\mathbb{Z}_n$ twist operators, $\sigma_n$ \cite{Calabrese:2012ew}
\begin{align}
    \tr \left(\rho^{T_A}\right)^n = \langle \sigma_n(z_1) \bar{\sigma}_n(z_2)\bar{\sigma}_n(z_3)\sigma_n(z_4)\rangle.
    \label{fourpoint_eq}
\end{align}
If we do the standard conformal block decomposition of the four-point function, when the intervals are close ($z_2 \sim z_3$), one usually expects the dominant channel to be an exchange of the double-twist field \cite{Calabrese:2012ew, 2014JHEP...09..010K}. This is the conformal block that agrees with the R\'enyi reflected entropy and preserves replica symmetry. To break the replica symmetry explicitly, we take inspiration from the random tensor networks in \cite{Dong:2021clv} where the domain walls split in two at the endpoints of the intervals. In CFT language, this is like a reverse OPE where the $\mathbb{Z}_n$ twist field splits into sum of pairs of $S_n$ twist fields
\begin{align}
    \sigma_n(x) \sim \lim_{x'\rightarrow x} \sum_{\tau \in S_n}\frac{C_{\sigma_n \tau}}{|x-x'|^{ \Delta_{\sigma_n}- \Delta_{\tau} - \Delta_{\sigma_n\circ \tau^{-1}}}} (\sigma_n \circ \tau^{-1})(x') \tau(x) + \dots ,
    \label{eight_point_eq}
\end{align}
where the $C$'s are numerical coefficients and the denominator is needed to absorb the scaling dimensions of the operators.
This changes \eqref{fourpoint_eq} to an eight-point function. What we gain are new OPE channels that break replica symmetry and resemble the gravitational saddles that compute negativity in fixed-area states. Of course, the sum over all of these channels retain replica symmetry just as in the gravitational calculations. It would be interesting to see if this approach to replica symmetry breaking in CFT could reproduce gravitational calculations. However, there is a key distinction due to the vanishing prefactor in the reverse OPE.

\paragraph{CFT dual of fixed-area states}

Fixed-area states have played a major role in this paper and many other papers on holographic entanglement. They vastly simplify gravitational path integrals and provide direct connections between tensor networks and gravity. One aspect of fixed-area states that is clearly lacking is their boundary description. How can we create a state in a CFT with an approximately flat entanglement spectrum and how different are these CFT states from their parent states without fixed areas? Recently, interesting progress has been made in \cite{2021JHEP...02..085G,2021arXiv210803346G} for the case of a single fixed-area surface. We would like to generalize this to multiple fixed-area surfaces. This is important to investigate further and we expect it to clarify various issues raised in the previous section.

\paragraph{Entanglement phase transitions}

Entanglement phase transitions have gained increasing attention in the condensed matter community. They have been studied in the context of many-body localization \cite{2015ARCMP...6...15N, 2019RvMP...91b1001A}, quantum circuits with measurement \cite{2019PhRvX...9c1009S, 2018PhRvB..98t5136L,2019PhRvB..99v4307C}, and random tensor networks \cite{PhysRevB.100.134203}. While in this work we have studied certain phase transitions in random tensor networks, it would be desirable to relate these to these other studies. This may involve new geometries of tensor networks and a backing away from the large-$N$ limit in the bond dimensions. We have found the negativity spectrum to be a robust diagnostic of entanglement phases, so it would be interesting to study it in quantum circuits and across the many-body localization transition.

\section*{Acknowledgments}

We thank Xi Dong, Xiaoliang Qi, Hassan Shapourian, and Michael Walter
for helpful discussions and comments. 
S.R.~is supported by the National Science Foundation under 
Award No.\ DMR-2001181, and by a Simons Investigator Grant from
the Simons Foundation (Award No.~566116).

\appendix

\section{Loop equations for partial transpose} \label{sec:loop_eq}

In this appendix, we show how to use the loop equations method for negativity. We will do this for the Haar random state, i.e., single tensor network.

Recall that in this case we have a single tensor $X_{ij\alpha } $ with probability distribution
\begin{equation}
P(X) = Z^{-1} \exp \left[ -L_A L_B L_C X_{ij\alpha } X^*_{ij\alpha } \right] ,
\end{equation}
where $Z$ is the partition function, that is, the Gaussian integral over $X$.
For the negativity between $A$ and $B$ (after tracing over $C$), we should consider traces of powers of the partially transposed density matrix, which is the matrix
\begin{equation}
H_{i_1j_1,i_2j_2} = \sum _{\alpha }  X_{i_1j_2\alpha } X^*_{i_2j_1\alpha } .
\end{equation}

We can obtain a loop equation for these moments $m_k = \mean{ \tr H^k} $ by considering
\begin{equation}
\begin{split}
& Z^{-1}  \int dX dX^{^{\dagger} } \frac{\partial }{\partial X^*_{i_2j_1\alpha } } \left( X^*_{i_1j_2\alpha } H^k_{i_1j_1,i_2j_2} e^{-L_A L_B L_C X_{ij\alpha } X^*_{ij\alpha } } \right)  = 0.
\end{split}
\end{equation}
This gives us
\begin{equation} \label{eq:Wishart_SD_eq}
 -L_AL_BL_C \mean{  \tr H^{k+1} } + \sum _{p=1} ^k \mean{ \tr \left[ \tr _{B} \left( H^{p-1} \right)  H \tr _{A} \left( H^{k-p} \right) \right] } + L_C \mean{  \tr H^k }  = 0.
\end{equation}
The first term is obtained by acting on the exponent, the second term comes from acting on the $H^k$, while the last one is just the derivative of the explicit appearance of $X^*$.

This equation is exact. The first and third terms are just the moments we are interested in, but the second term is more involved. One could continue and write Schwinger-Dyson equations for higher traces, such as the one in the second term.

\begin{figure}[]
\centering
\includegraphics[width=0.9\textwidth]{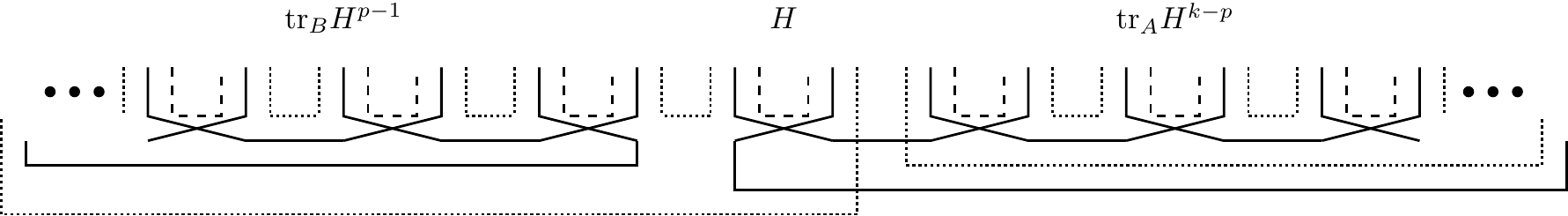}
\caption{The representation of the mixed trace term in the diagrammatic notation. Dotted lines represent system $A$, solid lines stand for system $B$, and dashed lines are for system $C$.}
\label{fig:Wishart_triple_tr}
\end{figure}

However, this equation is already sufficient in the regime we describe now. In general, this second term
\begin{equation} \label{eq:Wishart_second_term}
 \sum _{p=1} ^k \mean{ \tr \left[ \tr _{B} \left( H^{p-1} \right)  H \tr _{A} \left( H^{k-p} \right) \right] } 
\end{equation}
is represented in the diagrammatic notation used before as the sum over contractions of the diagram in Fig.~\ref{fig:Wishart_triple_tr}. Contractions are allowed to connect the different traces.

Because of the separation of the different lines, we expect that when the $L$'s are large, the averaging factorizes. This should be the case unless one of the system sizes is excessively larger than the others (e.g., larger than the product of the other two). In such a case, we would have a higher weight for a specific kind of line and so different diagrams will be preferred. That is, this should be valid in the bulk of the phase diagram of Fig.~\ref{fig:one_tensor_phase_diagram}, i.e., in region III.

\begin{figure}[]
\centering
\includegraphics[width=0.9\textwidth]{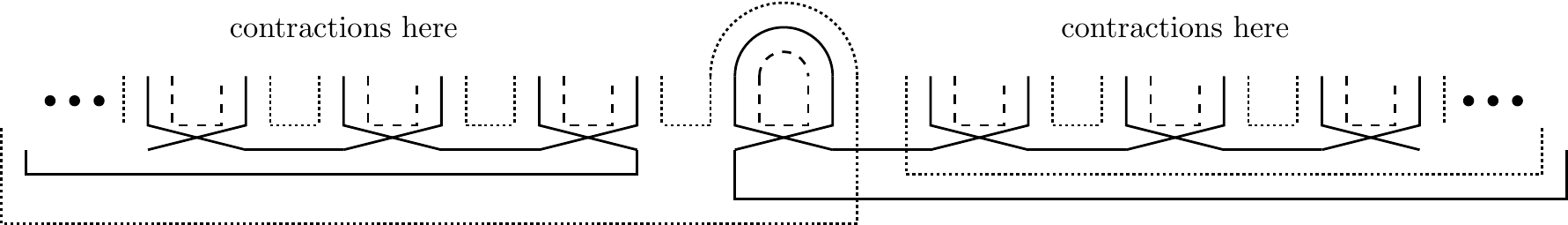}
\caption{A factorized form of Fig.~\ref{fig:Wishart_triple_tr}.}
\label{fig:Wishart_triple_tr_factorize}
\end{figure}

In this case, each of the three pieces of the trace are contracted among themselves as shown in Fig.~\ref{fig:Wishart_triple_tr_factorize}. Then the \eqref{eq:Wishart_second_term} term is approximated by
\begin{equation}
\begin{split}
& \sum _{p=1} ^k \tr \left[\mean{  \tr _{B}  H^{p-1} } \cdot\mean{ H} \cdot \mean{ \tr _{A}  H^{k-p} }  \right] = 
\frac{1}{L_A L_B} \sum _{p=1} ^k \mean{ \tr H^{p-1} } \cdot \mean{  \tr H^{k-p} } ,
\end{split}
\end{equation}
where the factor of $1/(L_A L_B)$ enters since the $1/(L_{A}L_{B})$ propagator of the middle $H$ is not compensated by free index lines because they are shared with the other traces. Note that $\mean{ \tr H} =1$, as can be seen for instance in the only diagram with a single $H$. The traces on the RHS are in the full $AB$ system.

For the resolvent, we use the convention of \cite{Shapourian:2020mkc} in order to compare to their result
\begin{equation}
G(z)= \frac{1}{L_A L_B} \sum _{k=0} ^{\infty } \frac{L_C^k m_k}{z^{k+1} } .
\end{equation}

Multiplying \eqref{eq:Wishart_SD_eq} by $(L_C/z)^{k+1} $ and summing over $k$, we get\footnote{We use $m_0=L_A L_B$.}
\begin{equation}
\begin{split}
& -L_AL_BL_C \sum _{k=0} ^{\infty } \frac{m_{k+1} L_C^{k+1} }{z^{k+1} } + \frac{1}{L_AL_B} \sum _{k=1} ^{\infty } \sum _{p=1} ^k \frac{m_{p-1} m_{k-p} L_C^{k+1} }{z^{k+1} } + L_C \sum _{k=0} ^{\infty } \frac{m_k L_C^{k+1} }{z^{k+1} } = \\
& \qquad =  L_A^2L_B^2 L_C - z L_A^2L_B^2 L_C G(z) +\frac{1}{L_AL_B} \sum _{p_1,p_2 \ge 0} \frac{m_{p_1} m_{p_2} L_C^{p_1+p_2+2} }{z^{p_1+p_2+2} } +L_AL_B L_C^2 G(z) = \\
& \qquad =  L_A^2L_B^2 L_C - z L_A^2L_B^2 L_C G(z) +L_AL_B L_C^2 G(z)^2 +L_AL_B L_C^2 G(z) = 0 .
\end{split}
\end{equation}
Note that in the second term above the sum over $k$ starts at 1 as clearly this term is missing in the derivation when $k=0$. With the definition $q=L_C/(L_AL_B)$ this equation is
\begin{equation}
1-zG(z)+qG(z) + qG(z)^2=0 .
\end{equation}
This indeed agrees with the Schwinger-Dyson equation (5.13) of \cite{Shapourian:2020mkc}.



\section{Phases of the two-tensor network} \label{sec:2_tensor_phases}

In this appendix, we will classify the phases of the two-tensor network. In order to do that, we will use methods that were developed in Section \ref{sec:gen_network}. We will see that this also provides us with several non-trivial checks on the results there.
The way that tensor networks are solved there is by finding the maximal flow in a corresponding flow network. In fact, our approach here will be the reverse. We shall classify the different phases rigorously by considering all the options for a flow network and determining in which cases such a flow is a maximal flow.

There are five possible flow lines that go between the three external nodes:
\begin{equation}
\vcenter{\hbox{\includegraphics[scale=0.8]{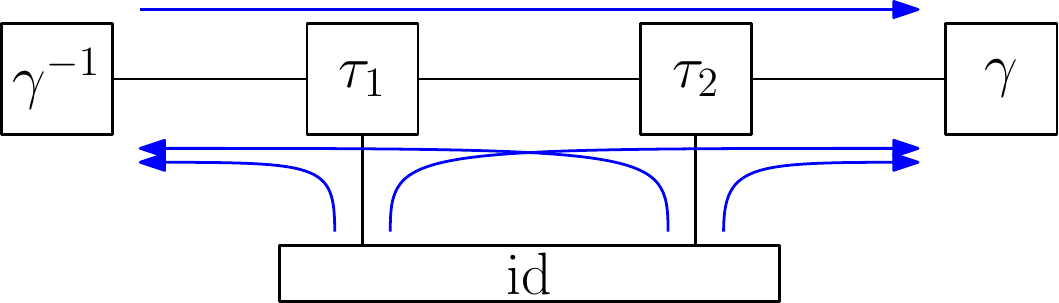}}} 
\end{equation}
and we should consider all the subsets of them as a flow. We may ignore cases where two flow lines appear together that go along the same edge in two opposite directions because this can be replaced by other flows, as explained in Fig.~\ref{fig:flow_reudction}. There is one such pair of flow lines in this case.

We will assume that all the bond dimensions are non-vanishing and that there are no special relations between them. Fine-tuned relations correspond to phase boundaries\footnote{This is the case as long as we verify that on both sides of the transition there are two distinct phases. We will find several ranges for phase 1 of Table \ref{tab:2_tensor_phases} and so any value in between them still belongs to the same phase.}
like those analyzed for the one-tensor network in the main text.

We can also compare each phase to the phases appearing in the one-tensor network (Fig.~\ref{fig:one_tensor_phase_diagram}). The idea is that once we find the leading permutations, we know that the moments are $m_n = \prod _i L_i^{-|\beta _i^{-1} \circ \beta _{i'} |} $ (from Section \ref{sec:gen_network}). Thus, we can compare this to the expression that one gets in the one tensor network.\footnote{In the one-tensor network, the dominant permutation in region I is the identity, in region II it is the cyclic or anti-cyclic permutation depending on whether $A$ or $B$ is smaller, and in region III the $NC_2$ permutations dominate.} When reducing to a one-tensor network, we denote the subsystem that we trace out in the one-tensor network by $C_{\text{one-tensor}} $ and the remaining two subsystems by $A_{\text{one-tensor}} $ and $B_{\text{one-tensor}} $. In region I, the negativity vanishes, so the moments depend only on the product of the two subsystems $A_{\text{one-tensor}}$ and $B_{\text{one-tensor}} $. In region II, the moments depend only on the smaller system between $A_{\text{one-tensor}}$ and $B_{\text{one-tensor}} $, and on $C_{\text{one-tensor}}$.

We follow two simple rules in each case. Firstly, each flow should be maximal in the sense that we cannot add any more flow either in existing flow lines or by adding a new line. Secondly, there is no flow $i \to X \to j$ such that in the residual network $X$ is connected to $k$, where $\{i,j,k\}=\{\text{id},\gamma ,\gamma ^{-1} \}$ and $X$ is some internal node.

We have the following options for maximal flows:

\begin{enumerate}
\item 
\begin{equation}
\vcenter{\hbox{\includegraphics[scale=0.8]{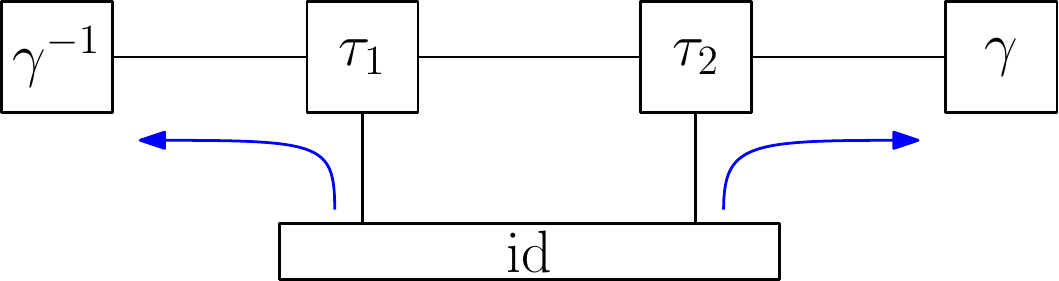}}} 
\end{equation}
Since $\tau _1$ is in a path $\text{id} \to \gamma ^{-1} $, it cannot be connected to $\gamma $, so $w_B<w_{C_2} $ and the right flow line is of weight $w_B$. The same reasoning holds for the other flow line. Therefore, this is a maximal flow in the cases where
\begin{equation} \label{eq:2_tensor_phases_range_1}
\begin{split}
& w_A<w_{C_1},\\
& w_B<w_{C_2} .
\end{split}
\end{equation}
In the residual network $\tau _1$ and $\tau_2$ are in the connected component of id, so
$\tau _1=\tau _2=\text{id}$.

The negativity vanishes, matching phase I of the one-tensor network with $L_A^{\text{one-tensor}}L_B^{\text{one-tensor}}=L_AL_B$.
The range \eqref{eq:2_tensor_phases_range_1} is a special case of \eqref{eq:2_tensor_phases_range_9} and \eqref{eq:2_tensor_phases_range_10} below.

\item 
\begin{equation}
\vcenter{\hbox{\includegraphics[scale=0.8]{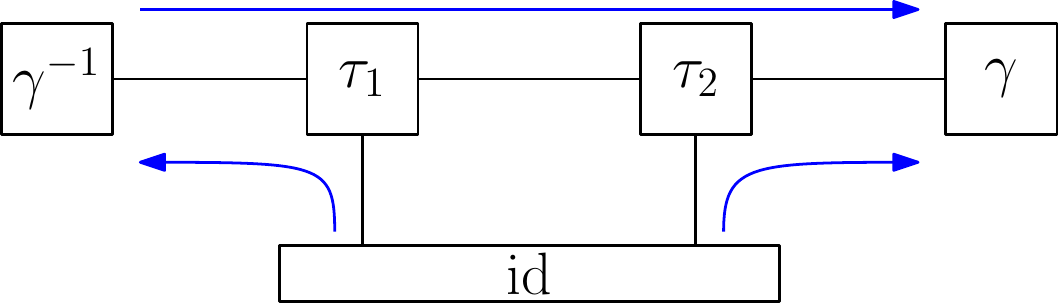}}} 
\end{equation}
The two nodes cannot remain connected to id, so the two flows at the bottom of the diagram are of weights $w_{C_1} $ and $w_{C_2} $ respectively. In order to allow the flow line $\gamma ^{-1} \to \gamma $, we should have that $w_{E_W} $ is less than the remaining weights in $w_A$ and $w_B$. In the other option where this flow line maximizes the remaining weight on $w_A$ or $w_B$, we will violate the second point in the rules above unless both $w_A$ and $w_B$ are saturated, but this requires the weights to satisfy a linear relation. So, away from phase boundaries, this is a maximal flow when
\begin{equation}
\begin{split}
& w_A> w_{C_1}+w_{E_W} ,\\
& w_B > w_{C_2} + w_{E_W} .
\end{split}
\end{equation}
In this case,
$\tau _1=\gamma ^{-1} $, $\tau _2=\gamma $, matching phase II of the one-tensor network with the smaller subsystem of size $L_{E_W} $ and $L_C^{\text{one-tensor}}=L_C:=L_{C_1} L_{C_2} $.

\item 
\begin{equation} \label{eq:2_tensor_phases_3}
\vcenter{\hbox{\includegraphics[scale=0.8]{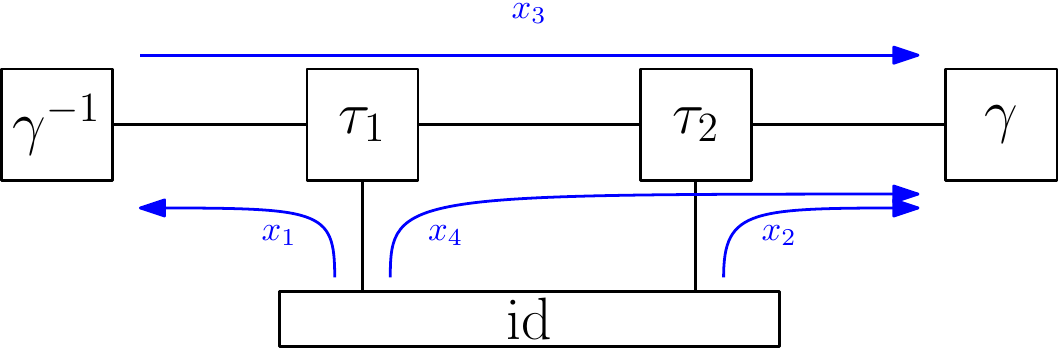}}} 
\end{equation}
Here, because of the second rule, $w_{C_1}$ and $w_A$ must be saturated by the flow and similarly either $w_{E_W} $ or $w_B$ should be as well. Let us first consider the case where $w_{E_W} $ is saturated by the flow. This means that
\begin{equation}
x_1+x_3=w_A,\qquad x_4+x_3=w_{E_W} ,\qquad x_1+x_4=w_{C_1} 
\end{equation}
and because of the flow $\gamma ^{-1} \to \gamma $ we must also have $x_2=w_{C_2} $. The conditions are then $x_i >0$ for $i=1,\cdots,4$ and that $x_2+x_3+x_4 < w_B$.  Therefore, this is a maximal flow when
\begin{equation}
\begin{split}
& w_{E_W} <w_A+w_{C_1} \\
&w_{C_1} <w_A+w_{E_W} \\
& w_A < w_{C_1} +w_{E_W} \\
& w_B>w_{C_2} +w_{E_W}
\end{split}
\end{equation}
$\tau _1$ appears in flows between all the nodes, so it is non-crossing with only 2-cycles (and a single 1-cycle when $n$ is odd), as found around \eqref{eq:NC2}, and these permutations are denoted by $NC_2$. $\tau _2$ remains connected to $\gamma $. Thus,
$\tau _1 \in NC_2$, $\tau _2=\gamma $.

This matches a rescaled phase III, by $L_{C_2} $, of the one-tensor network with $L_A^{\text{one-tensor}}=L_A$, $L_B^{\text{one-tensor}}=L_{E_W}$, and $L_C^{\text{one-tensor}}=L_{C_1}$.\footnote{In order to see this, note that the moments are the same as those in the one-tensor network in phase III, except that there is an additional factor of $L_{C_2} ^{-(n-1)} $. This factor corresponds to rescaling the negativity spectrum by $\ns '(\lambda )=L_{C_2}^2 \ns (L_{C_2} \lambda )$.}

\item 
We have the same flow network with $A \leftrightarrow B$ (by which we mean exchanging the two sides, so that also $C_1 \leftrightarrow C_2$).
This gives a maximal flow when
\begin{equation}
\begin{split}
& w_{E_W} <w_B+w_{C_2} \\
&w_{C_2} <w_B+w_{E_W} \\
& w_B < w_{C_2} +w_{E_W} \\
& w_A>w_{C_1} +w_{E_W} 
\end{split}
\end{equation}
and in this case
$\tau _1=\gamma ^{-1} $, $\tau _2 \in NC_2$.

This matches a rescaled phase III, by $L_{C_1} $, of the one-tensor network with $L_A^{\text{one-tensor}}=L_{E_W}$, $L_B^{\text{one-tensor}}=L_{B}$, and $L_C^{\text{one-tensor}}=L_{C_2}$.

\item 
Going back to the flow network \eqref{eq:2_tensor_phases_3}, as mentioned the second case is when $w_B$ is saturated by the flow. In this case the flow is
\begin{equation}
x_1+x_3=w_A,\qquad x_1+x_4=w_{C_1},\qquad x_2+x_4+x_3=w_{B} 
\end{equation}
and still $x_2=w_{C_2} $. The conditions are $x_i>0$ and $x_3+x_4<w_{E_W} $, which become
\begin{equation}
\begin{split}
& |w_{A} -w_{C_1}| <w_B-w_{C_2} <w_A+w_{C_1} \\
& w_{E_W} >w_B-w_{C_2} 
\end{split}
\end{equation}
and in this case $\tau _1$ and $\tau _2$ are in their own connected component, so that
$\tau _1=\tau _2 \in NC_2$ .

Since $\tau _1=\tau _2$, this is just the same as phase III of the one-tensor network with $L_A^{\text{one-tensor}}=L_A$, $L_B^{\text{one-tensor}}=L_{B}$, and $L_C^{\text{one-tensor}}=L_{C}$.

\item 
Again, we have the same for $A \leftrightarrow B$. This happens when
\begin{equation}
\begin{split}
& |w_{B} -w_{C_2} |<w_A-w_{C_1} <w_B+w_{C_2} \\
& w_{E_W} >w_A-w_{C_1} 
\end{split}
\end{equation}
and in this case
$\tau _1=\tau _2 \in NC_2$.

Again, this is the same as phase III of the one-tensor network with $L_A^{\text{one-tensor}}=L_A$, $L_B^{\text{one-tensor}}=L_{B}$, and $L_C^{\text{one-tensor}}=L_{C}$.

\item 
\begin{equation} \label{eq:2_tensor_phases_7}
\vcenter{\hbox{\includegraphics[scale=0.8]{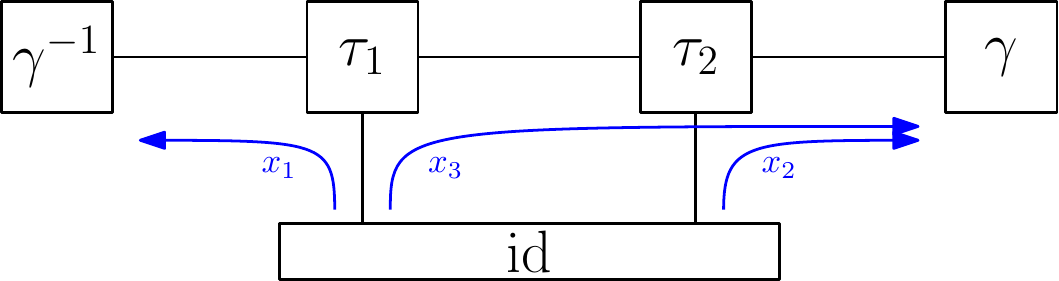}}} 
\end{equation}
By the second rule $\tau _1$ cannot remain connected to $\gamma ^{-1} $, so anyway $x _1=w_A$. But it also cannot be connected to $\gamma $ in the residual network. So we have two cases depending on which edge $x_2,x_3$ saturate. In the first case $x_3=w_{E_W} $ and $x_2=w_{C_2} $. Then, this is a valid maximal flow when
\begin{equation}
\begin{split}
& w_{C_1} > w_A+w_{E_W} \\
& w_B>w_{C_2} +w_{E_W} 
\end{split}
\end{equation}
and in this case
$\tau _1=\text{id}$, $\tau _2=\gamma $.

The negativity here vanishes, and this matches phase I of the one-tensor network with $L_A^{\text{one-tensor}}L_B^{\text{one-tensor}}=L_AL_{E_W}L_{C_2} $.

\item 
The $A \leftrightarrow B$ of this is
\begin{equation}
\begin{split}
& w_{C_2} > w_B+w_{E_W} \\
& w_A>w_{C_1} +w_{E_W} 
\end{split}
\end{equation}
and in this case
$\tau _1=\gamma ^{-1}$, $\tau _2=\text{id} $.

The negativity here vanishes, and this matches phase I of the one-tensor network with $L_A^{\text{one-tensor}}L_B^{\text{one-tensor}}=L_{C_1}L_{E_W}L_{B} $.

\item 
The second option for \eqref{eq:2_tensor_phases_7} is when $\tau _1$ is disconnected from $\gamma $ because $w_B$ is saturated. In this case, $x_2+x_3 = w_B$. The constraints are that $x_1=w_A<w_{C_1} $, $x_3<w_{E_W} ,w_{C_1} -w_A$, and $x_2<w_{C_2} $.
This happens when
\begin{equation} \label{eq:2_tensor_phases_range_9}
\begin{split}
& w_A<w_{C_1} \\
&w_B<w_{C_2} +w_{E_W} \\
&w_A+w_B < w_{C_1} +w_{C_2} 
\end{split}
\end{equation}
and in this case
$\tau _1=\tau _2=\text{id}$.

The negativity here vanishes, and this just matches phase I of the one-tensor network with $L_A^{\text{one-tensor}}L_B^{\text{one-tensor}}=L_{A}L_{B} $.

\item 
The $A \leftrightarrow B$ of this is
\begin{equation} \label{eq:2_tensor_phases_range_10}
\begin{split}
& w_B<w_{C_2} \\
&w_A<w_{C_1} +w_{E_W} \\
&w_A+w_B < w_{C_1} +w_{C_2} 
\end{split}
\end{equation}
and in this case
$\tau _1=\tau _2=\text{id}$.

Again, the negativity here vanishes, and this matches phase I of the one-tensor network with $L_A^{\text{one-tensor}}L_B^{\text{one-tensor}}=L_{A}L_{B} $.

\item 
\begin{equation}
\vcenter{\hbox{\includegraphics[scale=0.8]{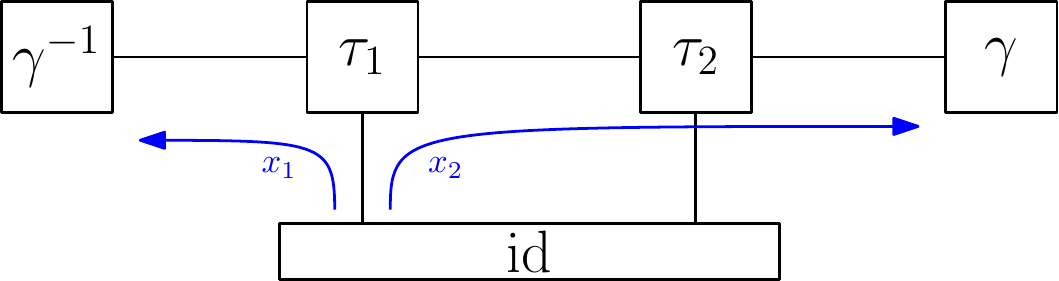}}} 
\end{equation}
Here again $\tau _1$ cannot remain connected to $\gamma ^{-1} ,\gamma $. So $x_1=w_A$. Note that it cannot be made disconnected from $\gamma $ by saturating $w_{E_W} $ since then we will have a remaining possibility for a flow $\text{id} \to \gamma $ by going along $w_{C_2} $ and then $w_B$, and so this would not be a maximal flow, violating the first rule. Therefore, $x_2=w_B$. The conditions are then $x_2<w_{E_W} $ and $x_1+x_2<w_{C_1}$. So this happens when
\begin{equation} \label{eq:2_tensor_phases_range_11}
\begin{split}
& w_{C_1} > w_A+w_B \\
&w_B<w_{E_W} 
\end{split}
\end{equation}
and in this case
$\tau _1=\tau _2=\text{id}$. However we can see that this flow is a special case of the flow that lead to \eqref{eq:2_tensor_phases_range_9}. In particular, indeed the conditions \eqref{eq:2_tensor_phases_range_11} imply those of \eqref{eq:2_tensor_phases_range_9}.

\item 
The $A \leftrightarrow B$ of this is
\begin{equation}
\begin{split}
& w_{C_2} > w_A+w_B \\
&w_A<w_{E_W} 
\end{split}
\end{equation}
and
$\tau _1=\tau _2=\text{id}$. Again, this is a special case of \eqref{eq:2_tensor_phases_range_10}.

\item 
\begin{equation}
\vcenter{\hbox{\includegraphics[scale=0.8]{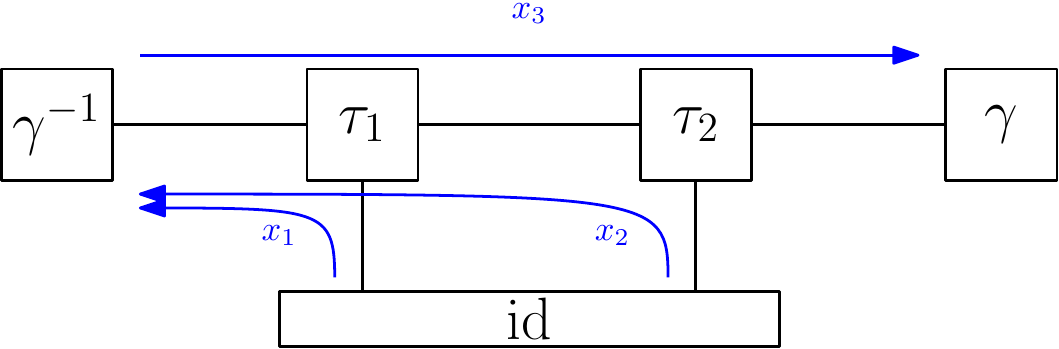}}} 
\end{equation}
By the second rule, we must have $x_2=w_{C_2} $, $x_1=w_{C_1} $, and $x_3=w_B$ that are uniquely fixed. The conditions are $x_2+x_3<w_{E_W} $ and $x_1+x_2+x_3<w_A$, so that
\begin{equation}
\begin{split}
& w_{E_W} >w_B+w_{C_2} \\
& w_A > w_B+w_{C_1} +w_{C_2} 
\end{split}
\end{equation}
and since $\tau _1,\tau _2$ remain in the connected component of $\gamma ^{-1} $,
we have $\tau _1=\tau _2=\gamma ^{-1} $.

This matches phase II of the one-tensor network with the smaller subsystem of size $L_{B} $ and $L_C^{\text{one-tensor}}=L_C$.

\item 
The $A \leftrightarrow B$ of this is
\begin{equation}
\begin{split}
& w_{E_W} >w_A+w_{C_1} \\
& w_B > w_A+w_{C_1} +w_{C_2} 
\end{split}
\end{equation}
and then
$\tau _1=\tau _2=\gamma $.

This matches phase II of the one-tensor network with the smaller subsystem of size $L_{A} $ and $L_C^{\text{one-tensor}}=L_C$.

\end{enumerate}

It is immediate to see using the rules above that all the other flows are not possible as maximal flows. Let us drop the three cases that are special cases of other ones. For any two cases that give a different result for $\tau _1,\tau _2$, we should get that the ranges where they apply must not overlap. This is indeed the case, and can be checked explicitly by looking at the ranges. There are two cases giving $\tau _1=\tau _2 \in NC_2$, but actually away from phase transitions they do not overlap. Therefore, if we group the two cases that give $\tau _1=\tau _2=\text{id}$ into one, we remain with regimes that are distinct. This analysis provides a non-trivial check on the methods of Section \ref{sec:gen_network}.

\bibliography{main}
\bibliographystyle{JHEP}
\end{document}